\documentclass[[authoryear,preprint,review,12pt]{wiley-article}

\usepackage[authoryear]{natbib}

\usepackage{amssymb}
\usepackage{amsmath}

\usepackage{graphics}
\usepackage{lineno}
\usepackage{siunitx}
\usepackage{lscape}
\usepackage{caption}
\usepackage{subcaption}
\usepackage{url}
\usepackage{float}
\floatstyle{plaintop}

\restylefloat{table}

\papertype{Original Article}
\paperfield{Manuscript in Preparation}

\title{The effect of air purifiers and curtains on aerosol dispersion and removal in multi-patient hospital rooms}

\author[1]{Steven N. Rogak}
\author[2]{Adam Rysanek}
\author[1]{Jim Myungjik Lee}
\author[1]{Surya Venkatesh Dhulipala}
\author[1]{Naomi Zimmerman}
\author[3]{Martin Wright}
\author[3]{Mitch Weimer}

\affil[1]{Department of Mechanical Engineering, University of British Columbia, Canada}
\affil[2]{School of Architecture and Landscape Architecture, University of British Columbia, Canada}
\affil[3]{Fraser Health Authority, British Columbia, Canada}

\corraddress{Steven N. Rogak\\ Department of Mechanical Engineering,\\ University of British Columbia, 6250 Applied Sciences Lane, Vancouver, BC, Canada, V6T 1Z4}
\corremail{rogak@mech.ubc.ca}

\fundinginfo{Funding was provided by the Mitacs Accelerate internship program and Fraser Health Authority.}

\runningauthor{Rogak et al.}

\begin{document}
\raggedbottom

\cleardoublepage

\begin{frontmatter}
\maketitle

\begin{abstract}
\textbf{Abstract} Airborne transmission of disease is of concern in many indoor spaces. Here, aerosol dispersion and removal in an unoccupied 4-bed hospital room was characterized using a transient aerosol tracer experiment  for 38 experiments covering 4 configurations of air purifiers and 3 configurations of curtains.  NaCl particle (mass mean aerodynamic diameter $\sim 3 \mu m$) concentrations were measured around the room following an aerosol release. Particle transport across the room was 1.5 - 4 minutes which overlaps with the characteristic times for significant viral deactivation and gravitational settling of larger particles. Concentrations were close to spatially uniform except very near the source.  Short curtains had no consistent effects on  concentrations at the non-source patient locations  while floor-length curtains reduced concentrations slightly depending on the purifier configuration. The aerosol decay rate was in most cases higher than expected from the clean air delivery rate, but the reduction in steady-state concentrations resulting from air purifiers was less than suggested by the decay rates.   Apparently a substantial (and configuration-dependent) fraction of the aerosol is removed immediately and this effect is not captured by the decay rate.  Overall, the combination of curtains and purifiers is likely to reduce disease transmission in multipatient hospital rooms.
\end{abstract}

\end{frontmatter}

\begin{figure}[h!]
\centering
\includegraphics[width=14cm]{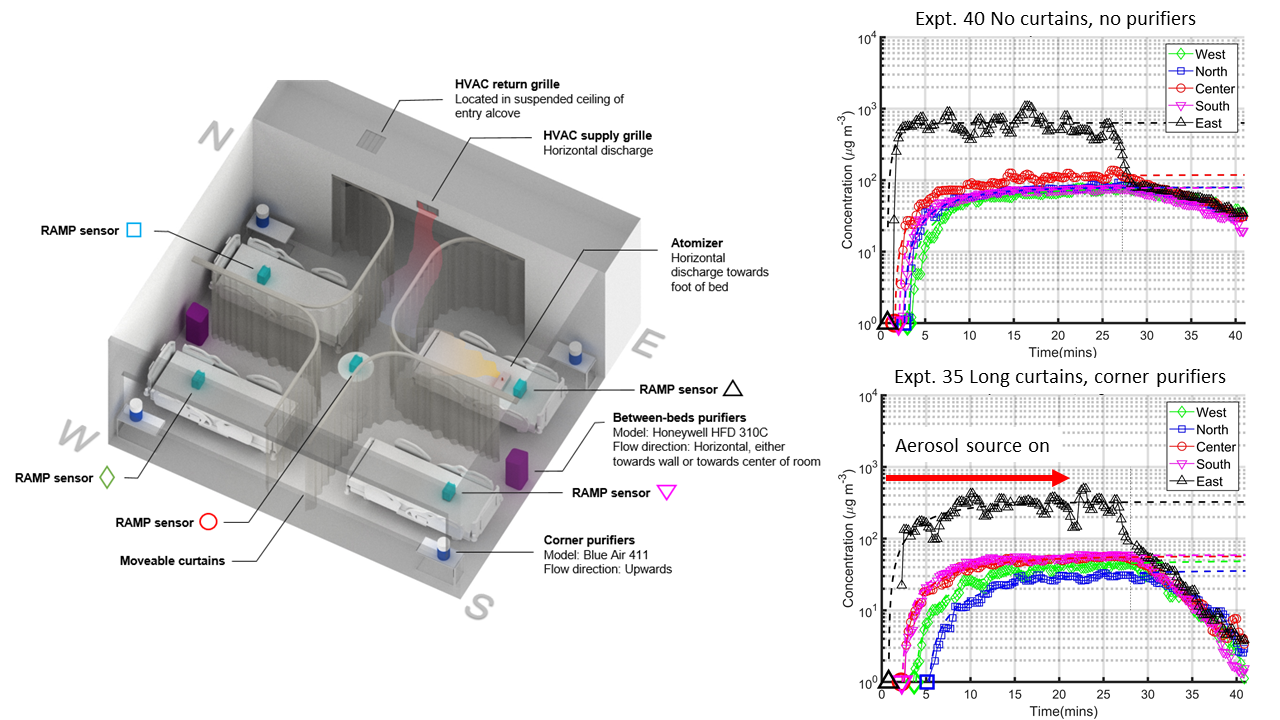}
\caption*{Graphical Abstract}
\label{abstract}
\end{figure}

\section{Introduction}

COVID-19 has spread globally mainly via inhalation of virus-containing particles produced via coughing, sneezing, vocalizations and even normal breathing \citep{Jayaweera2020Transmission, Zhang2020Identifying,Morawska2020Airborne,Bazant2021guideline}. The particle size most critical to transmission is unclear  \citep{Morawska2020How,Tang2021Dismantling}. Particles larger than $5\mu$m in diameter are commonly referred to as "respiratory droplets"; smaller particles are interchangeably referred to as aerosols, droplet nuclei and airborne particles\citep{WorldHealthOrganization2014Infection}; these smaller particles appear to carry most of the airborne viruses  \citep{Leung2020}. Airborne particles $ > 1\mu$m are easily filtered by almost any building air handling unit but will be distributed by the gentle air currents found in most indoor settings. Larger droplets ($\sim 30\mu$m) will  settle to the ground in less than a minute, while a 10 $\mu$m particle will settle out of a room in about 10 minutes \citep{Hinds1999Aerosol}.  These factors suggest that aerosols are most important, but recent analysis of a restaurant outbreak suggests that larger droplets may results in transmission if there are strong air currents (> 1m/s)  \citep{Kwon2021Evidence}. Recent work \citep{Reid2022} shows that very rapid viral deactivation can be expected within the first minute of exhalation, then much more gradual deactivation would occur on timescales of 5 minutes to an hour. Those authors attribute the rapid deactivation to the physical transformations from drying (increased at lower humidity) and change in pH as the exhaled plume entrains air with a low CO$_2$ concentration. This underlines the importance of the transport time from the source to the receptor and may affect the particle size that is most important to transmission.

Masks reduce airborne transmission \citep{Macintyre2020rapid}, and both patients and workers in hospitals are expected to wear them. However, the pleated over-the-ear surgical masks often fit poorly and patients in a hospital may not always wear their masks correctly. Further, in public spaces such as bars and restaurants, mask wearing is counter to the purpose of the space. Thus, there is a need for a layer of protection on top of mask wearing \citep{Morawska2020How}.

There is much prior research on the use of portable air filtration systems in buildings.  Experiments and modeling of aerosol transport and removal in a house \citep{Novoselac2009Impact} outline the basic behavior we study here for the semi-partitioned space in a hospital room. A simplified mass balance model \citep{Ward2005effectiveness} was used to assess the use of air purifiers in the event of an external bioterrorism attack. A recent study motivated by COVID-19 considered the particle removal efficiency of for different filter grades in dental clinics \citep{Zhao2020Using} but did not consider air movement within rooms. Another study focused on effectiveness of air purifiers in a single patient hospital room \citep{Lee2021Effectiveness} but did not consider other interventions such as curtains. Experiments conducted in a 6-bed hospital ward  \citep{Qian2010Particle} showed that a single HEPA filter removed particles at a rate within 30$\%$ of the theoretical value, and that the exhaust, aimed to the room center, produced a global circulation that dominated the overall air flow pattern. The experiment was not set up to quantify particle transport from one bed to another. A few recent studies address the effectiveness of air purifiers in minimizing COVID19 transmission. Firstly, computer simulations were used to evaluate different configurations of purifiers added to a classroom  \citep{He2021, Burgmann2021}.  Secondly, an experimental study of an occupied classroom determined the concentration decay rate produced by air purifiers and the degree of concentration uniformity in a large room without any ventilation \citep{Curtius2021}. Similarly, another experimental study determined the effectiveness of air purifiers in an occupied conference room  \citep{lindsley2021efficacy}.

Plastic barriers have been ubiquitous in the workplace since the start of the pandemic.  At the extreme end, sealed partitions can create separate rooms with their own filtration units, reducing the migration of aerosols from the patient space to the neighboring corridor by 98$\%$ \citep{Mousavi2020Performance}.   Local extraction systems using a tightly-sealed enclosure around the patient  \citep{Mead2004evaluation,Mick2020Aerosolgenerating,Sommer2020Recommendations,Francois2020SimulationBased, Cottrell2020operative} provide excellent patient isolation but would be a serious impediment to provision of health care. Plastic barriers of the kind meant to protect cashiers or office workers are less restrictive, but evidence for their effectiveness is quite conflicting \citep{Eykelbosh2021}. 

As summarized in Table \ref{Table_1}, the key factors related to the effectiveness of air purifiers in mitigating COVID-19 transmission have been considered in one or more studies, for particular real or idealized spaces, but there is a dearth of studies simultaneously considering the interaction of real-world ventilation conditions with air purifiers, barriers, and within-room aerosol transport.  Also, the effect of transport time within the room, critical to droplet transmission and viral deactivation, has not been studied experimentally before.

\begin{table}[h!]
\centering
\large
\begin{tabular}{lllllllll}
                              & \multicolumn{8}{c}{\textbf{Features of the Study}}                                          \\
\textbf{Study}    & \rotatebox{90}{Modeling}& \rotatebox{90}{Experimental}& \rotatebox{90}{Within-room variations}& \rotatebox{90}{Filter config.} & \rotatebox{90}{Barriers/curtains} & \rotatebox{90}{Particle size}  & \rotatebox{90}{Transport time}&\rotatebox{90}{Baseline ventilation}\\

\citep{He2021}                          & X  &    & X  & X  &    &  X &  & X\\
\citep{Ward2005effectiveness}          &  X  &    &    & X  &    &  X  &  & X\\ 
\citep{Abuhegazy2020}          &  X  &    &    &    &  X &  X  &  & X\\ 
\citep{Burgmann2021}                    & X  & X  & X  &    &    &  X & X & \\
\citep{Curtius2021}                     &    & X  &    &    &    &  X  &  &  \\
\citep{Mousavi2020Performance}          &    & X  &    &    &  X &  X  &  & \\     
\citep{Mead2004evaluation}              &    & X  &   &     &  X  &    &     & \\     
\citep{Lee2021Effectiveness}            &    & X  &    & X  &    &     &  & X\\ 
\citep{Novoselac2009Impact}            &  X  & X  &    & X  &    &  X  &  & X\\ 
\citep{Buising2021}            &    & X  &    & X  & X  &     &  & X\\  
\citep{Qian2010Particle}                &    & X  & X  &    &    &  X  &  & X\\    
\citep{lindsley2021efficacy}                &    & X  & X  &    & X  &    &  & X\\  
\textbf{This Work}                      &    & \textbf{X}  & \textbf{X}  & \textbf{X}  & \textbf{X}  &    & \textbf{X}& \textbf{X}\\

\end{tabular}
\caption{Summary of literature on the effect of air purifiers on aerosol transport and removal}

\label{Table_1}
\end{table}


\section{Methods}
We conducted the experiment in a real hospital room because the objective was to inform the purchase of purifiers by a local health authority. However, given access restrictions during the pandemic, and the large number of uncontrolled factors in an occupied room, we conducted experiments in an unoccupied room. Nevertheless, the natural variability of the setting (from weather and ventilation system operation) undoubtedly influenced the results. We use a novel transient tracer experiment that allows us to assess the transport delay time, effective decay rate, and spatial inhomogeneity of steady-state concentrations. Our approach resembles the pulse tracer technique \citep{Persily1990} that has been used for gaseous tracers. 

\subsection{Test Room Layout}
\begin{figure}[h!]
\centering
\includegraphics[width=13cm]{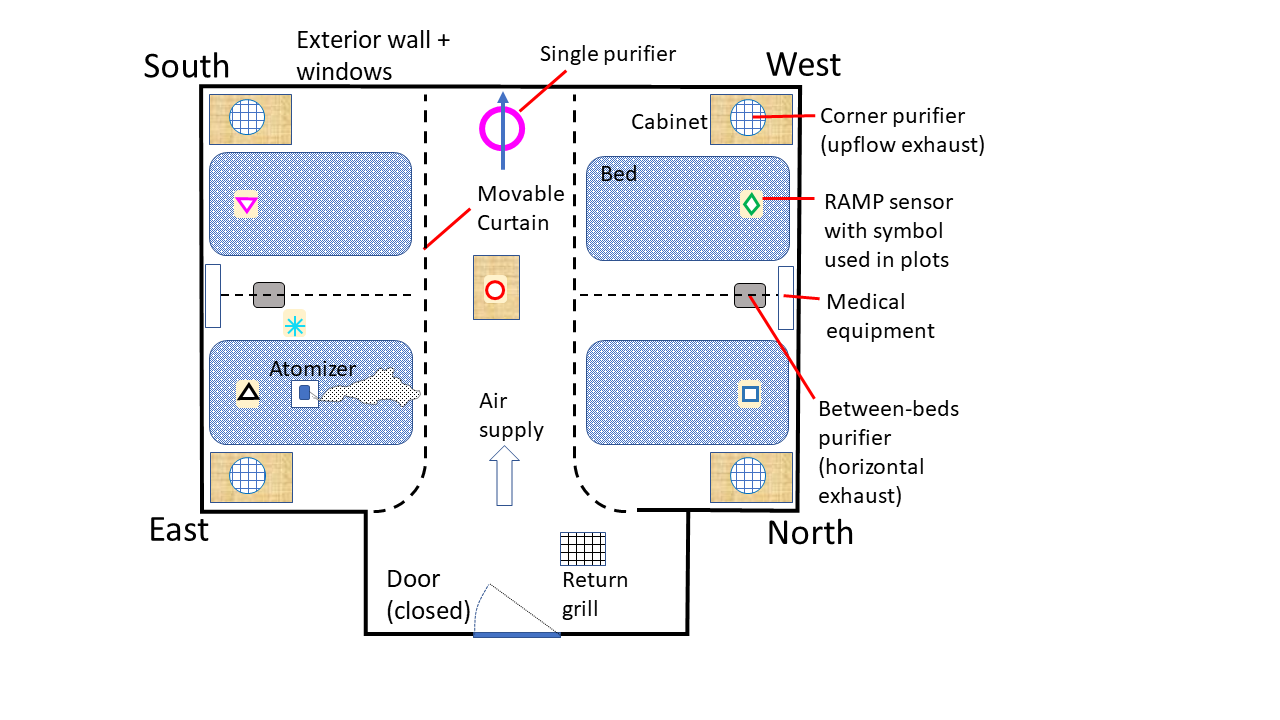}
\caption{Plan of test room. Air purifiers were either cylindrical upflow units in the corners or horizontal towers on the floor, exhausting to the room center or the walls. Particle sensors (RAMPS) were placed where a patient's head would be or where a healthcare worker would attend the sick patient. Supply (inlet) grill is mounted on the edge of the alcove drop ceiling and directs air to the SW. Return (outlet) grill is mounted on the drop ceiling. Room width=6.5 m; volume=54.09 m$^3$.}
\label{layout}
\end{figure}

The test room (RM 213 of Delta Hospital, Figure \ref{layout}) is intended for 4 patients and has an area of 398 sf (37 m$^{2}$) including a 48 sf (4.5 m$^{2}$) entry alcove. The door to the washroom was closed and this small room is not included in the totals above. The room supply air is located on the alcove drop ceiling and directs air along the center-line towards the exterior wall. The return air grill is located directly above the room entry door. Thus, the normal ventilation system creates a tumbling pattern of air moving towards the windows at ceiling height and returning along the floor to the entry door. Flow rates are likely variable, but measurements before the present test program suggests that it provides 3.5 ACH with 206 cfm (350 m³/h). 

Each bed space was fitted with sliding curtains. These curtains had gaps of approximately 50 cm at the top and bottom. In addition, we retrofitted curtain extensions to the floor ("Long" curtain configuration). There is nothing to control wrinkling or the foot location of the curtains precisely- as expected for real curtains - and this could contribute to experimental variability.

Because tests were conducted in a working hospital applying COVID-19 protocols, the researchers conducting the tests stayed in the room during all tests, wearing masks and remaining in the SE or NW sections of the room. Undoubtedly this had some small effect on air flows, but the level of activity was much less than the normal activity in an occupied hospital room.

\subsection{Aerosol Source and Test Sequence}
A sodium chloride (NaCl) solution was dispersed using a Sonair MedPro ultrasonic nebulizer.  The concentration of the solution was set to produce dried aerosol particles with a mass median diameter of approximately 3 µm as measured by at TSI Model 3330 OPS. For 3 µm particles, filtration efficiency would be nearly 100$\%$ for all purifiers tested. For this particle size, gravitational settling reduces concentrations at a rate equivalent to 0.5 ACH, on top of the true air exchange or cleaning rate. We expect the true particle losses to room surfaces to be about 30\% higher than computed from sedimentation, based on previously reported experiments \citep{Bivolarova2017}.  The aerosol leaving the nebulizer was not neutralized, so static charges on room surfaces would have increased removal rates (slightly) for all experiments.   The nebulizer was placed near the middle of one bed to simulate the cloud of particles produced by an infected patient. The nebulizer produced a jet of particles with a “throw” of 30-50 cm directed towards the foot of the bed, away from the sensor in that space.   From a  collocation experiment (Supplemental Information A) in a sealed room with a fan but no filtration, the source strength is inferred to be 1970+/-100 $\mu$g/min.

Experiments 1-27 were conducted using aerosol generation for 10 min 15 s set by the nebulizer.  Experiments 28-41 used 25 minutes of aerosol generation to provide more accurate estimates of the steady-state concentrations.  After the generator was turned off, there was 15 minutes in the specified configuration to determine the decay rate, followed by 10 minutes of clearing with all purifiers on.  By the end of this sequence, PM$_{10}$ concentrations were consistently reduced to less than 1 $\mu$g m$^{-3}$ – about 1$\%$ of the peak values recorded during the experiment. This manuscript focuses on the configurations that were replicated at least once (and in most cases 3-4 times); data from all 41 experiments is contained in the Supplemental Information sections B and C.

\subsection{Particle Measurements}
We used the Remote Affordable Multi-Pollutant Sensor (RAMP, SENSIT Technologies), which includes low-cost sensor modules for measuring Particulate Matter (PM$_1$, PM$_{2.5}$, PM$_{10}$), NO, NO$_2$, CO, Ozone(O$_3$) and CO$_2$ \citep{Zimmerman2018machine}. The RAMP uses laser scattering to measure PM. Data from the RAMP sensors are uploaded via cellular networks to an online server enabling remote monitoring. The RAMP records data every 15 seconds. A running three-element median filter was applied to the raw data. The sensors are located as shown in Figure \ref{layout}. 
Many factors control the accuracy of low-cost optical particle sensors; we consider the major issues in the Supplemental Information section A.  From collocation experiments, examination of humidity and temperature variations, and from experiments in the hospital room with good mixing, it appears that calibration drift should be less than 3 \% over an experiment and sensors read consistently with each other to better than 10\%.

\subsection{Air Purifier Configurations }
In addition to the baseline configuration without added air filtration, we tested configurations with 1, 2 or 4 purifiers in the room, in locations that would cause minimal disruption to the activities of patients and health care workers. 

\subsubsection{Single purifier by far wall}
One experiment was conducted with a single large purifier (Atmosphere Sky true HEPA CADR 300 CFM).  This was located on the center of the SW wall with exhaust directed upwards towards that wall. For completeness the data from this experiment is included (and identified) in the summary plots to follow.

\subsubsection{Corner upflow}
In this configuration, a vertical exhaust purifier was placed in the room corner near the head of the bed.  The intake was located 1.2 meters above the floor.  For the corner upflow configuration we used the Blue Air 411 purifiers (except for experiments 3 and 4 which used 2 Honeywell towers on their sides, in addition to 2 Blue Air units).  The BlueAir411 uses radial inflow through cylindrical filters with a nominal efficiency of 99$\%$ for PM$_{2.5}$. Our tests of the filter material suggest that the actual performance is approximately 99$\%$ filtration at 1 $\mu$m. These units use a vertical exhaust.  Each is rated for 5 ACH in a 161 sf (15 m$^{2}$) room, equivalent of 103 cfm (175 m$^{3}$h$^{-1}$). Thus, 4 of these units should clean 7.1 room volumes per hour. 

\subsubsection{Between-beds horizontal flow}
In this configuration, two horizontal axis towers were placed between pairs of beds. We used a Honeywell HFD 310C (Air Genius 4) for the west wall. It uses a true HEPA filter and is rated for a 250 square feet room with 161 cfm (274 m$^{3}$h$^{-1}$). A unit of this type would clean at a rate equivalent to 2.7 ACH.  For the east wall (near the aerosol source in most tests), we used a Honeywell HFD 122qc. The filter has a nominal efficiency of 99$\%$ at 0.3 $\mu$m. The unit is rated for 170 sf (16 m$^{2}$) with 109 cfm (185 m$^{3}$h$^{-1}$). There are two important variants of the between-beds configuration: exhaust towards the foot of the beds, or exhaust towards the head of the beds (or walls).

\subsection{Data Analysis}
Evaluation of the air purifier and barrier configurations was based on concentration decay time, transport time from source to another "patient", and estimated steady-state concentrations. 

Approximately 5 minutes after the aerosol source is turned off, concentrations at all sensors decay at a similar rate. By fitting an exponential we obtained the first order decay time (\emph{t$_{decay}$}), the inverse of the “effective air changes per hour” which is the ventilation rate for a well-mixed room where the only particle removal mechanism is air exchange. This parameter controls the steady-state  concentrations for steady aerosol generation. 

We were also concerned about the potential for air purifiers to create drafts that could rapidly transport large droplets from an infected person to other people in the room. As a surrogate for this risk, we extracted the time delay  (\emph{t$_{trans}$}) between the start of aerosol generation and the detection of particles at the 4 sensors away from the aerosol source (as a model for 3 other patients and a health care worker in the center of the room). This transport time is indicative of the time available for particles to settle to the ground before the air parcel reaches another person. Practically, in order to extract the delay time, a line was fit to the first few points at concentrations exceeding a threshold (typically 0.5 $\mu g/m^3$); this line was extrapolated to find the time to cross the 1 $\mu g/m^3$ level.  This provided robust and consistent (if arbitrary) estimates of the delay time.  Almost certainly this time is less than the \emph{average} transport time from one location to another, but we expect it to be useful in comparing alternative interventions. 

Although the experiments were transient, concentrations approached a plateau before the source was turned off, and assuming an exponential approach (accounting for the delay time), an estimate of the steady-state concentration was obtained for each sensor.  The uncertainty in these steady- state concentration estimates includes the statistical uncertainty from the fitting process.  For the shorter experiments, the 95\% confidence intervals on the estimates is about 5\%, except for the west sensor, where the confidence interval is often as large as the mean value. However, the total uncertainty in this fit is certainly larger as the true functional form might be more complex than assumed, and from inspection of the graphs, the uncertainty might be 50\% for the short-duration experiments. Later we will present medians for 4 sensors, which would thus have uncertainties of (conservatively) 25 \%. However, in all of the longer experiments, the 95\% confidence interval of the steady-state estimates is always less than 2\% and the fit appeared to be excellent.

Given the steady-state concentrations (for a hypothetical continuous source) and the concentration decay rate, the source strength can be estimated for a well-mixed room. The discrepancy between this source estimate and true source rate is an indication of the importance of the direct removal of aerosols by the exhaust or filter, before the aerosols are distributed through the room.

\section{Results and Discussion}

We consider first the concentration time-series of experimental runs with different purifier and curtain configurations. After this, results from the complete experimental set are considered.  In total, 41 experiments were conducted over 5 days and the complete results are included in the Supplemental Information.  The nebulizer was placed in the West corner for 3 experiments; in the main manuscript we only consider runs with the source in the East corner.  

\subsection{Evolution of concentrations in typical experimental runs}
\begin{figure}[h!]
\centering
\includegraphics[width=13cm]{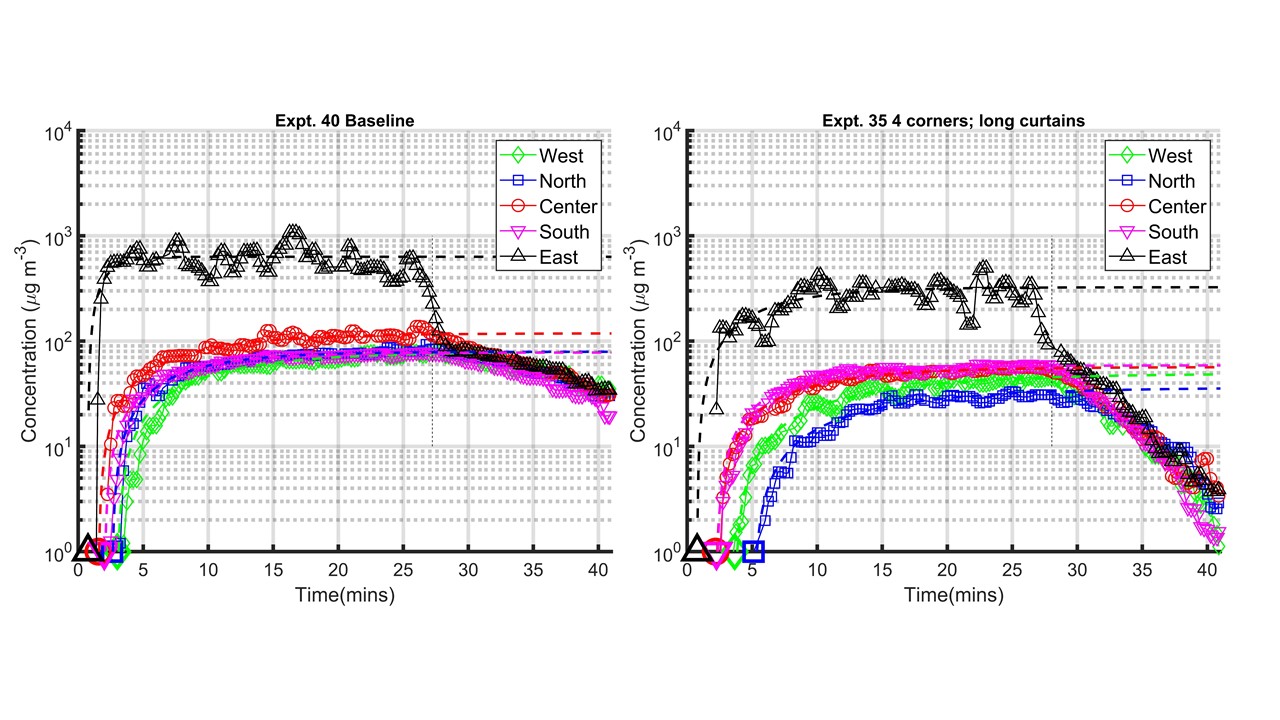}
\caption{Representative experiments for baseline (Expt. 40, no purifiers or curtains) and corner purifiers with long curtains (Expt. 35).  Dashed lines indicate the fitted curves used to estimate steady-state concentrations. Aerosol injection started at t = 0.}
\label{baseline_4corners}
\end{figure}

Figure \ref{baseline_4corners} shows the progress of a baseline experiment and one with upflow corner purifiers with long curtains. Although the atomizer is directed away from the nearest sensor (East bed), circulation in the room brings the particles back to that sensor in approximately 2 minutes. As expected, the concentrations at the 4 sensors away from the atomizer are much lower, and show a delayed response indicating a finite travel time (2-4 minutes).  Generally, the concentrations at the East sensor, nearest the source, show the largest fluctuations, about a factor of 3.  This is consistent with a concentrated aerosol plume wafted intermittently towards the sensor. Already at ~2.5 meters away, the sensors are reading roughly room averaged concentrations, to within a factor of 2-3. This is entirely consistent with detailed simulations of similar rooms \citep{Burgmann2021} and experimental studies \citep{Curtius2021, Bivolarova2017}.   

After the nebulizer switches off, an exponential fit to the concentrations provides an estimate of the decay rate or effective air changes per hour; a fine dashed vertical line shows the start of that fitted decay curve (about 27 minutes for the longer experiments).  Dashed curves indicate the exponential fit used to estimate the steady-state concentrations at each sensor for the period that the atomizer is on.  

\begin{figure}[h!]
\centering
\includegraphics[width=13cm]{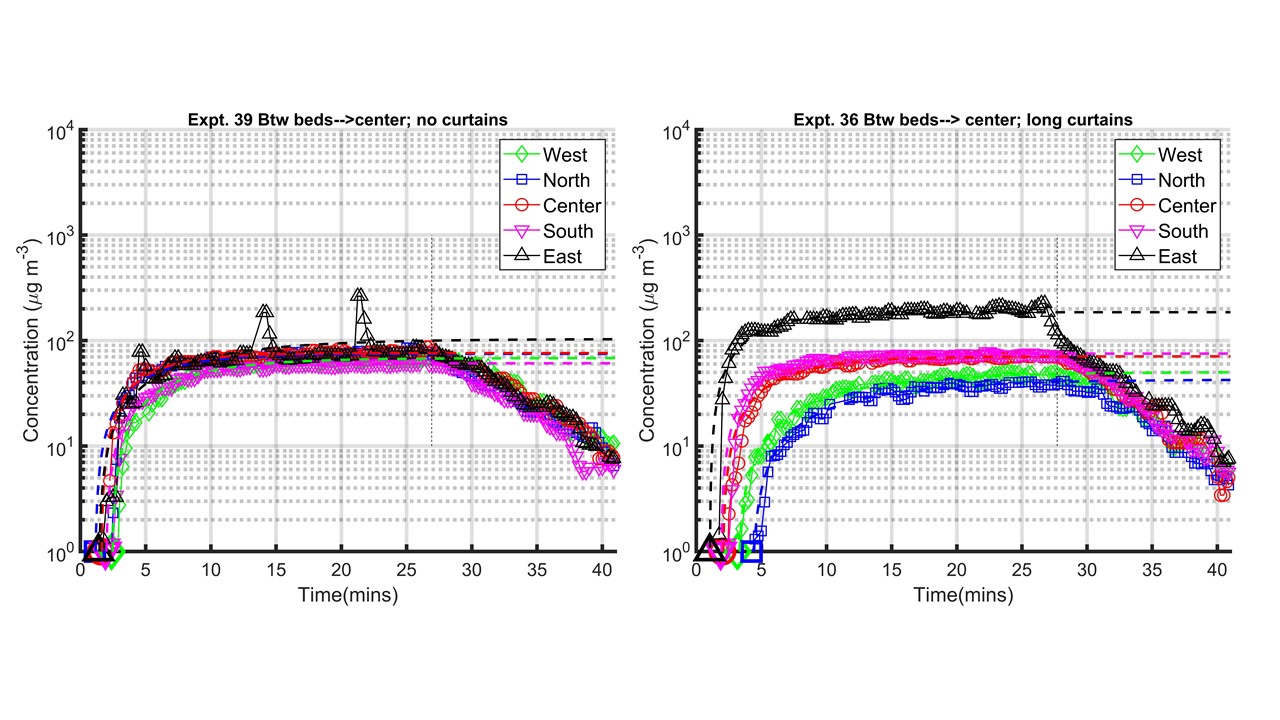}
\caption{Representative experiments using 2 purifiers between beds, exhausting towards the room center. Left panel is without curtains (Expt. 39) and right is with long curtains (Expt. 36). }
\label{BTWbedstocenter}
\end{figure}

Distinctly different concentration patterns are produced by air purifiers placed between beds with horizontal exhausts.  When directed towards the room center (Figure \ref{BTWbedstocenter}), the exhausts produce a high degree of mixing, especially when there are no curtains. Long curtains result less uniform concentrations.  When the purifier is turned 180 degrees (Supplemental Information Figures C12, 13, 23, 26), concentrations are markedly less uniform, especially with long curtains. 

\clearpage

\subsection{Effect of purifiers and curtains on whole-room metrics}
\begin{figure}[h!]
\centering
\includegraphics[width=11cm]{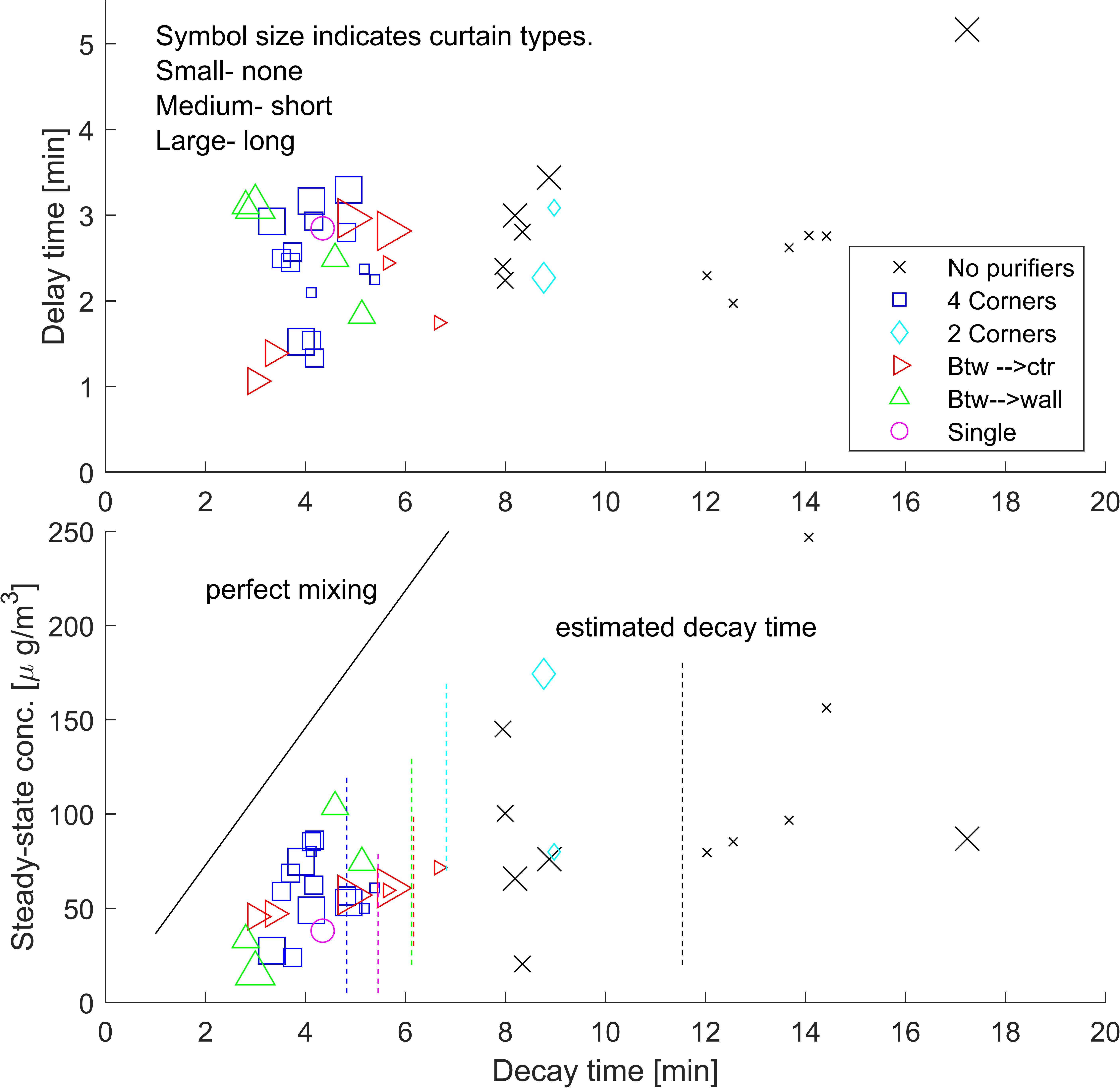}
\caption{Delay time (top) and estimated steady-state concentrations (bottom) as a function of the particle concentration decay time. Vertical dashed lines indicated the estimated decay time for each purifier configuration.  The "perfect mixing" line uses the measured decay times and aerosol generation rate to predict the steady-state concentration. }
\label{Timescales}
\end{figure}

Figure \ref{Timescales} compiles some key metrics for all experiments with the source on the East corner bed.  The decay times are the medians for all sensors, while the delay times are the averages for the 4 sensors away from the source bed area (i.e. South, West, North, Center sensors). The steady-state concentrations plotted in the figure are the medians for the 4 sensors away from the source. As expected, the concentration decays faster with purifiers (3-7 minute decay time) than without (8-18 minutes). From the nominal air supply flow and an estimate of particle losses (settling and additional 30\% for non-vertical surfaces and turbulence, see section 2.2), the baseline case should have a decay rate equivalent to 4.15 air changes per hour (ACH), which is equivalent to 14.5 minutes decay time. However, the average of the no-purifier data is 11.5 minutes (5.2 ACH); shown as a vertical dashed black line in the figure. For the cases with the purifiers, we expect the effective ACH to be increased by the Clean Air Delivery Rate (CADR) and thus the decay time can be estimated for each configuration. These estimates are shown as vertical dashed lines with the appropriate colour.  Concentrations decay faster in nearly all cases than this estimated time.  Several factors may contribute to this bias. Dead spaces (under beds, in entry vestibule, in furniture) make the effective room volume smaller than used in the decay time estimate. Also, aerosol losses to surfaces could be larger than assumed. Decay time is more consistent when purifiers are used, particularly for the 4 corners and between-bed-to-wall configuration. We speculate that for the between-beds-to-center configuration, results may be sensitive the precise alignment of the exhaust jet with respect to curtains (which are pinned back about 2 feed to avoid interference with the exhaust jet).  
 
 Our results are broadly consistent with the literature.  Curtius et al find that the incremental effect of air purifiers is consistently 80\% of that expected from the CADR.  In those experiments, the two measurement locations were in corners of the room, far from the purifiers, where ventilation might indeed have been lower. Meanwhile, Lindsley et al \citep{lindsley2021efficacy} find that effect of air purifiers varies with the location of air purifiers in the room; it is lower or higher than that expected from the CADR. 

The delay time (top panel of Figure \ref{Timescales}) is relevant to disease transmission because it would affect viral deactivation and the settling of large droplets (were we to consider a source with broader size range).  Long curtains (large symbols) tend to result in larger delay times but here too there is large variability.  Note that the RAMPs record 15 second averages (which are then smoothed using a 3-element median filter), and further that there could be errors of up to 10 seconds in the startup of the atomizer: practically the resolution of delay time will be about 1/2 minute.  Typically the delays were under 3 minutes, implying air velocities fast enough to carry droplets of ~20 microns across the room.  The range of delay times is potentially significant: for SARS-CoV2 it could change the viable virus concentration by more than a factor of 2 \citep{Reid2022}. 

Steady-state concentration estimates (lower panel of Figure \ref{Timescales}) are lower for the configurations with purifiers but there is again large variability between experiments. If the particles in the room were well mixed, we would expect the steady-state concentrations $C$ to depend only on the decay time $\tau$, particle generation rate $S$ and room volume $V$ as $C=S\tau/V$ as shown as the "perfect mixing" line in Figure \ref{Timescales}. All configurations produce lower concentrations than expected from perfect mixing.  This implies that the apparent aerosol source rate is lower than the true source rate.  Furthermore, the discrepancy is greater for the baseline cases with no purifiers- and yet the same atomizer and salt concentration is used in all cases.  The most likely explanation is that a portion of the aerosol is carried directly to either the room exhaust or one of the purifiers. By chance, this portion is larger for the baseline configuration, and smallest for the 4 corner purifiers.  This is consistent with the observed circulation created by the room air supply ejecting air southwest along the ceiling and returning northeast near the floor- this puts the aerosol source just upstream of the return grill. This pattern cannot be generalized to all rooms or even all source locations within the test room, but generally there would be special locations with more local source capture.

\clearpage
\subsection{Spatial variability}
Although the well-mixed model is often used to interpret indoor air concentrations of pollutants, we see some important deviations from this idealization.  As noted above, near the source, concentrations are highly non-uniform and this will influence the fluctuations experienced by the nearest sensor and also the extent to which nearby filters might capture particles immediately. Beyond this, the timeseries shown in Figures \ref{baseline_4corners} and \ref{BTWbedstocenter} suggest that people in zones away from the source might experience different levels of risk. 

\begin{figure}[h!]
\centering
\includegraphics[width=15cm]{"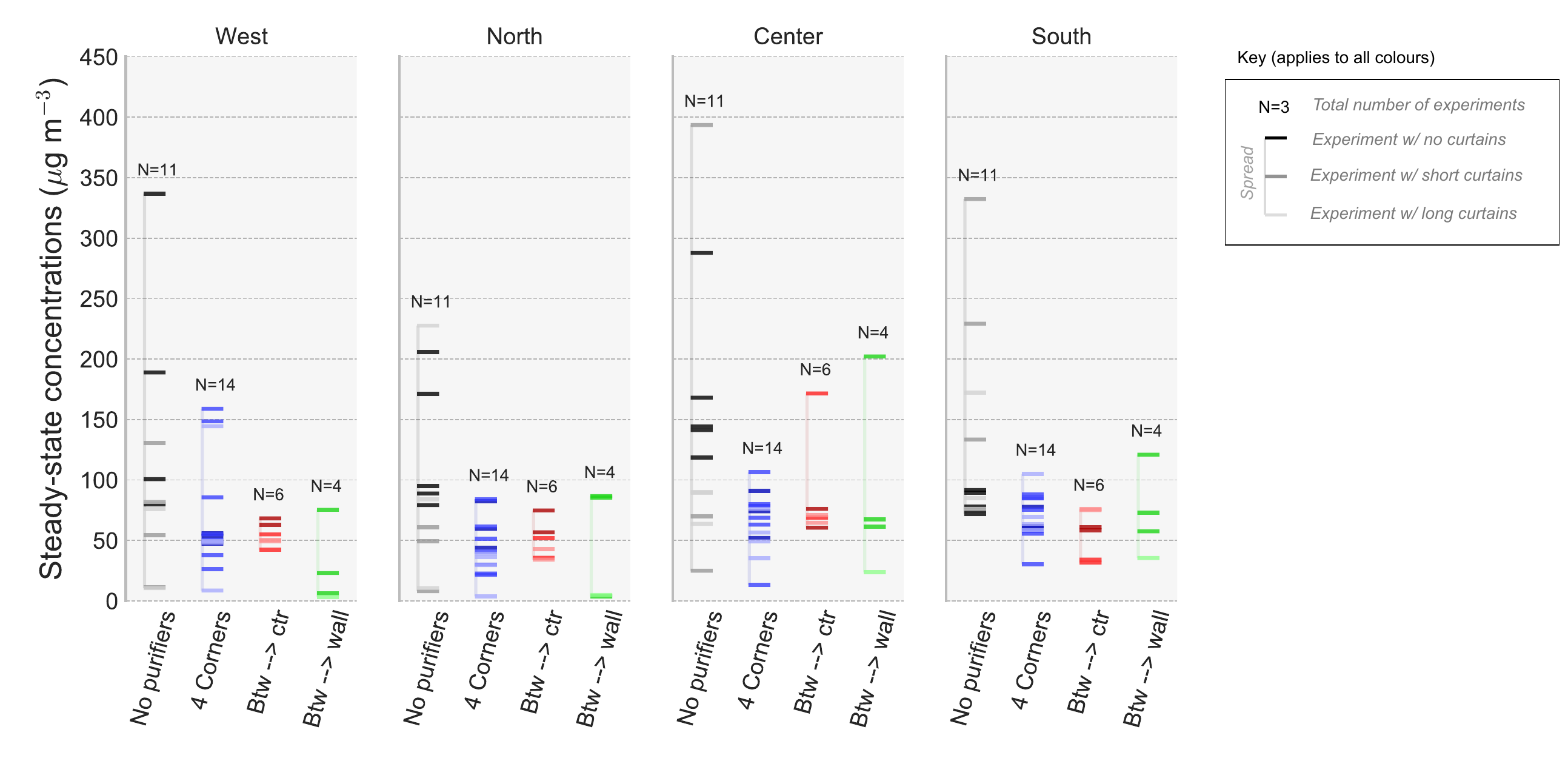"}
\caption{Spread of steady-state concentrations for the 4 sensors away from the aerosol source zone. The number of experiments per configuration is indicated as "N." }
\label{sensors_SSConly}
\end{figure}

In Figure \ref{sensors_SSConly}, the steady-state concentrations for individual sensors away from the source are illustrated independent of delay time and decay time. There are few clear trends on the effect of curtains on aerosol dispersion. While it may be the case that, in the no-purifier scenario, curtains reduce steady-state concentrations in the Center or North space, they may do the opposite in the South. With purifiers, there is less of a spread between experiments and less of an indication that the configuration of curtains has a strong role in reducing concentrations.

\begin{figure}[h!]
\centering
\includegraphics[width=11cm]{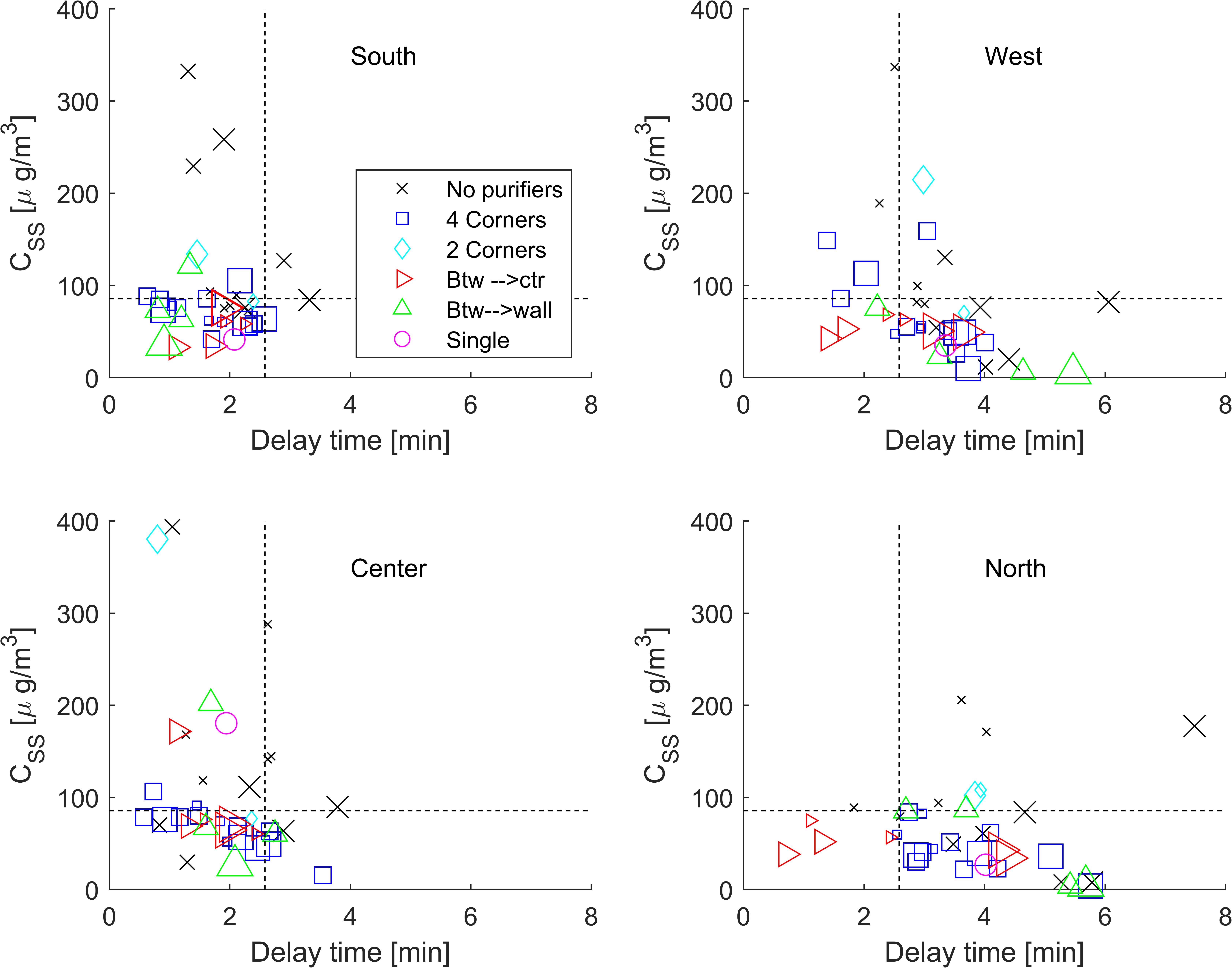}
\caption{Steady-state concentrations and delay times for the 4 sensors away from the aerosol source zone. Dashed lines indicate the averages for the full dataset; these lines divide the plot into quadrants.  The lower right quadrant (long delay and low concentration) is expected to indicate lower than average risk of infection.}
\label{sensors}
\end{figure}

Figure \ref{sensors} shows the steady-state concentrations and delay times for individual sensors away from the source.  The averages for all sensors and experiments are superimposed on the plots as a reference (dashed lines).  The risk of infection from a sick patient would be increased with short delay times and higher concentration, that is the upper left quadrant of each plot.  In a configuration without purifiers, it is apparent that the patient in the south bed or a worker in the center zone would be at the highest risk; the patients in the west and north beds would have the lowest risk. The configurations with curtains tend to have lower risk but the effect of curtains is less pronounced and less spatially uniform than the effect of purifiers. Regardless of the configuration of curtains, or even the configuration/location of the purifiers, the use of purifiers appears to mitigate risks of exposure within all spaces of the room adjacent to the source.

\clearpage

\section{Conclusions and Outlook}
Transient injections of sodium chloride aerosols were used to assess the ventilation patterns of a 4-bed hospital room using various configurations of air purifiers and curtains around beds. These experiments allowed us to estimate the aerosol concentration decay rate resulting from the purifiers, the transport delay time across the room, steady-state concentrations, and the portion of aerosol that is immediately removed from the room.  We believe that transient release experiments can provide metrics critical to assessing the effectiveness of infection control measures, but can recommend some improvements in future applications of the method.  Delay times will often be on the order of several minutes, so it would be useful to have sensor time response of only a few seconds.  On the other hand, concentrations will approach steady-state within 2 or 3 air change times, so experiments should run over 4-6 air change times in order to recover all the  features relevant to particle distribution and removal. 

Across 38 experiments, we found that the addition of stand-alone air purifiers increased the concentration decay rate by at least as much as expected from the purifier clean air delivery rate (CADR).   On the other hand, the change in steady-state concentrations was generally smaller than one would expect based on CADR.  This apparent paradox is resolved by noting that a large portion of the aerosol source is immediately removed from the air and not subject to the well-mixed removal behavior characterized by the concentration decay portion of the experiment.  Furthermore, this immediately-removed portion of the aerosol is altered by the air purifier configuration, and shows that there is scope for optimizing the location of purifiers for particular rooms and source-receptor positioning.  

The particle transport time from one bed to another was only a few minutes, which would provide ample time for 15-30 $\mu$m droplets to disperse widely within the room before settling out. Also, given the low velocities in the room ($<<1 m/s$) and large dimensions of the obstacles, inertial impaction of particles is expected to be negligible. Our observations here imply that the plastic barriers used in bars, restaurants and stores will be ineffective for the smaller droplets and aerosols believed to be largely responsible for transmission of COVID-19.  Indeed, curtains that surround the beds and extend from the floor to within 50 cm of the ceiling reduce concentrations by less than a factor of 2 and more commonly less than 20 {\%}.   
The surprisingly small effect of the barriers might be because the purifiers introduced air circulations larger than the zone around each bed, and this provided relatively efficient transport past the curtains.  Nevertheless, the delay times measured here \emph{are} large enough to affect the degree of viral deactivation that occurs due to the physical and chemical changes that occur on drying.

The effect of partial barriers on COVID-19 transmission has been reviewed recently \citep{Eykelbosh2021}.  Several CFD studies showed potentially very large impacts at close range, while the epidemiological evidence was very mixed. For example, CFD modelling \citep{Abuhegazy2020} predicted that small barriers would reduce concentrations by over 90\% for the student nearest the "source" student in a conventional classroom.  In light of our experiments, this appears possible for a perfectly aimed source under steady conditions.  However, this would not be representative of ensemble average conditions for real rooms, in which variations in large scale air motions would tend to move the concentrated aerosol plumes around the room.  This would reduce (on average) the very highest concentrations, even without barriers, and it would make it all but impossible to place small barriers in a useful location.  

In summary, curtains and air purifiers should substantially reduce airborne disease transmission in hospital rooms, but the sensitivity of room air flows to variable large scale air motions will make it challenging to optimize their use. Computational fluid dynamics could have an important role in designing these interventions, but they should consider variations in boundary conditions representative of real rooms, and experimentally validated under realistic conditions.

\section*{Conflict of Interest}
The authors declare no conflict of interest. 

\section*{Acknowledgements}
The authors would like to acknowledge Fraser Health Authority for allowing use of an empty patient room for experiment days. Funding was provided by the Fraser Health Authority and the Mitacs Accelerate program.

\section*{Supporting Information}
Supporting Information is available online. 

\bibliographystyle{elsarticle-harv}
\bibliography{References.bib}

\appendix

\clearpage
\appendix
\counterwithin{figure}{section}
\counterwithin{table}{section}
Supplemental Information
\section{Sensor Comparisons}
The accuracy of low-cost optical PM sensors signal is affected by (i) the omission of particles smaller than minimum detectable size ($<300 nm$); (ii) influence of temperature and (iii) Relative Humidity (RH) \citep{Malings2019Development}. Calibration or collocation alongside scientific grade instruments is needed for the PM sensors and gas-phase sensors \citep{Zimmerman2018machine,Malings2019Development,Malings2020Fine}.   However, we were interested in the spatial differences in concentrations and concentration decay times, thus we require only linearity and consistency between sensors.  

\subsection{Collocation}
A collocation test was conducted prior to Day 3. All 6 RAMP sensors used in this study were placed in an unventilated room (10'x8'x23') with uniform mixing created by running a BlueAir 411 without the filter in place. The atomizer and solution were the same as used in the hospital room tests.

\begin{figure}[h!]
\centering
\includegraphics[width=11cm]{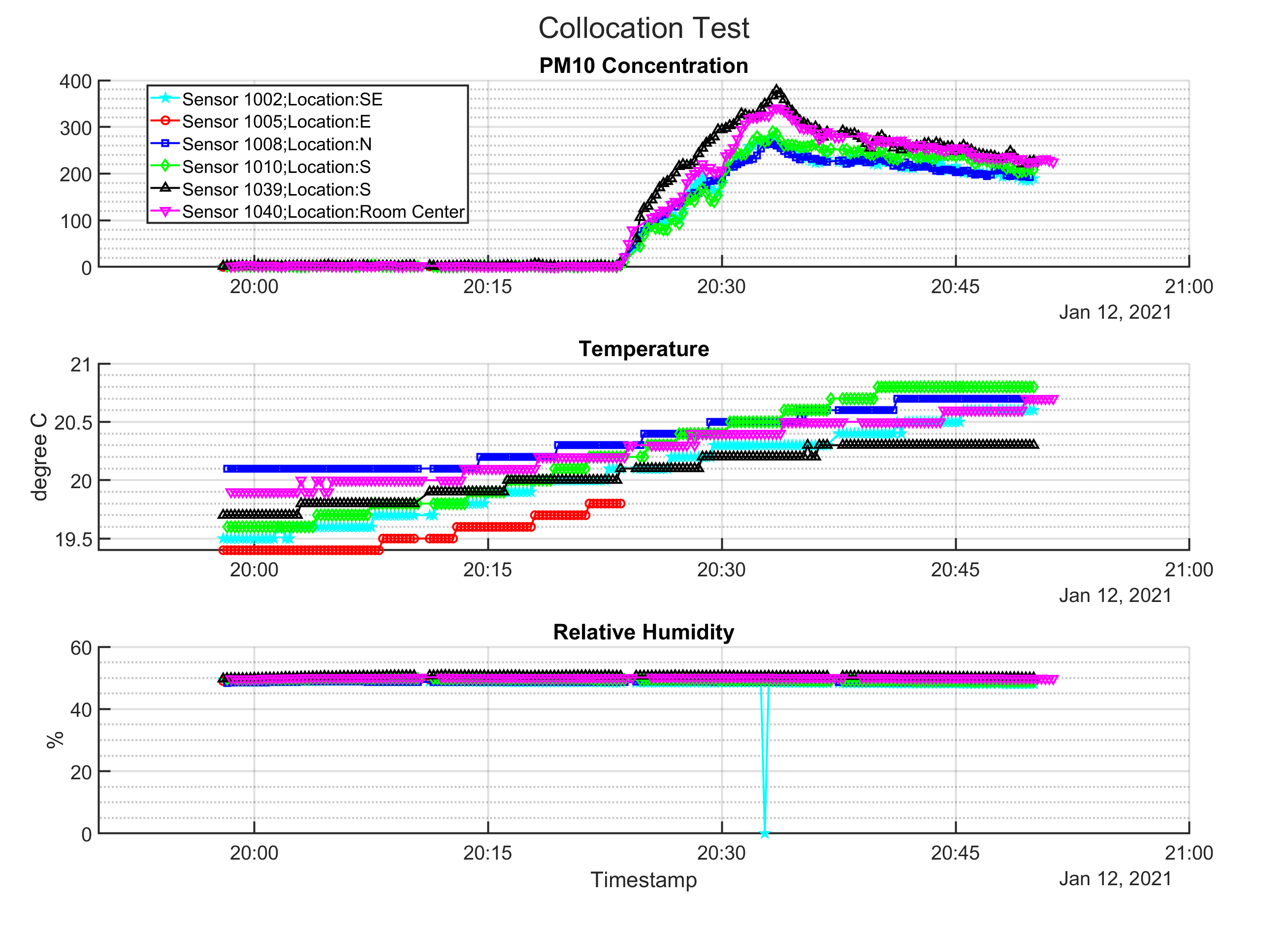}
\caption{PM10, Temperature and Relative Humidity during collocation}
\label{TimeSeries}
\end{figure}

NaCl particles were then introduced diagonally across from the sensors. All sensors picked up the PM10 signal at the same time. The temperature and Relative Humidity (RH) were constant as well. Figure \ref{TimeSeries} shows the time-series plots for the collocation test. The difference in magnitude of sensor concentrations should not affect the results presented in this study as we focus on the decay rate (to calculate effective ACH) and transport time.  Pearson correlation between sensors is good (Table \ref{Pearson}).

We expect that the true difference in calibration between the sensors is actually less than indicated here, however.  In this experiment, the sensors were stacked on a table and there may haven been a small gradient away from the surface.  This effect is quite evident in the first part of the experiment because the discrepancies are larger and variable.

The aerosol source rate can be estimated from the initial slope of the concentration curves (sensor-averaged) knowing the room volume.  Naturally there are large uncertainties in the true mass of the particles (as determined optically), but because the same sensors are used in the hospital experiments, we expect that this source estimate can be used to interpret the hospital measurements.

\begin{table}[h!]
\centering
\includegraphics[width=13cm]{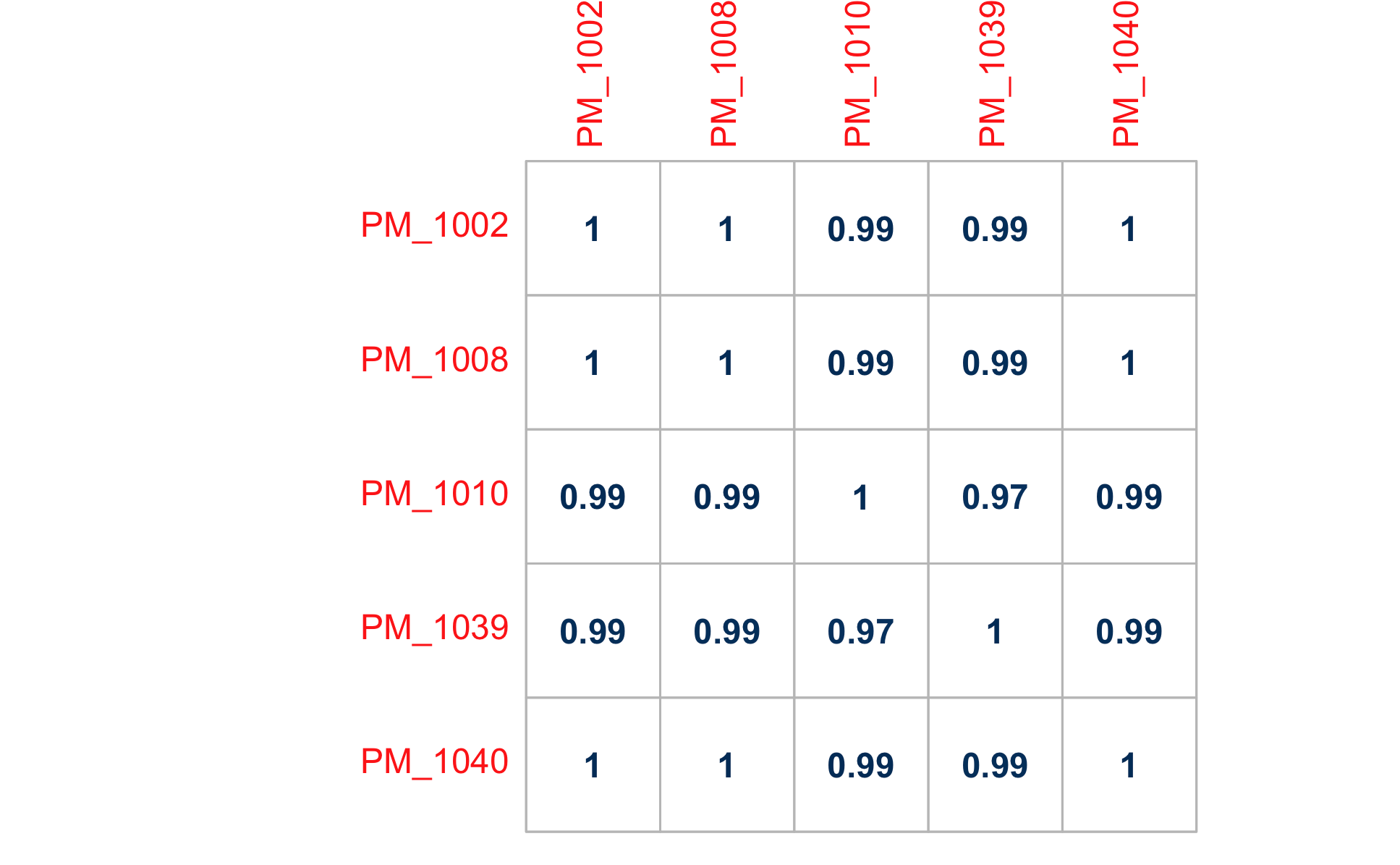}
\caption{Pearson correlation coefficients during collocation.}
\label{Pearson}
\end{table}

\subsection{Relative humidity variations}
\begin{figure}[ht!]
\centering
\includegraphics[width=13cm]{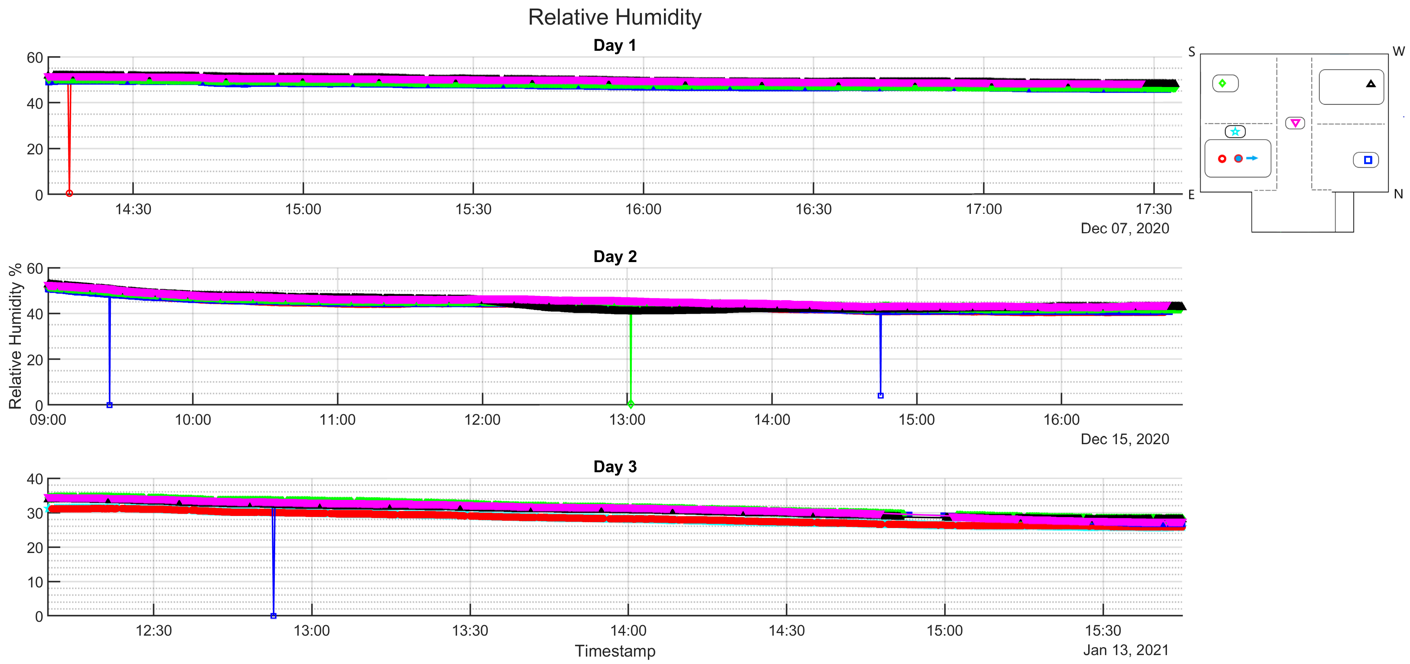}
\caption{Relative humidity by sensor for test days 1-3.}
\label{humidity}
\end{figure}

The relative humidity in the hospital room was not controlled but only varied a small amount between test days (30-55\%) and by much smaller amounts on any given test day (Figure \ref{humidity}). Data for test days 3 and 4 are not shown here but patterns are similar, but with RH ~52 to 56\%.

\subsection{Temperature variations}
\begin{figure}[ht!]
\centering
\includegraphics[width=13cm]{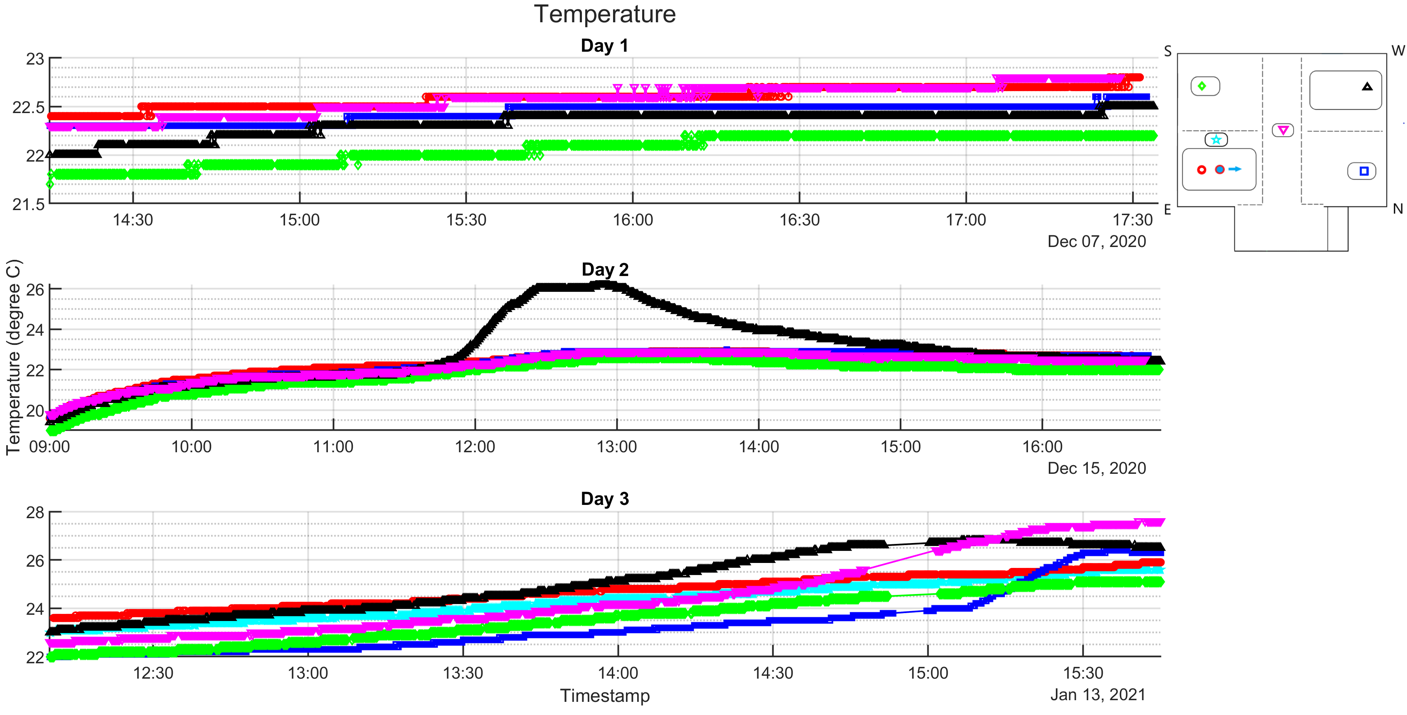}
\caption{Temperature by sensor for test days 1-3.}
\label{temperature}
\end{figure}

Temperature of the hospital is controlled at room level through the supply of conditioned air, but there are relatively large temperature variations (Figure \ref{temperature}). Data for test days 3 and 4 are not shown here but patterns are similar, with T increasing from 19 to 23 \textdegree C over the course of the day.

 Using the correction factors previously published \citep{Malings2020Fine} and using sample RH (50$\%$) and T (22°C), it is estimated that raw PM$_{2.5}$ measurements of 10 $\mu$g m$^{-3}$ and 100 $\mu$g m$^{-3}$ correspond to a corrected PM$_{2.5}$ of 9.6 $\mu$g m$^{-3}$ and 71.2 $\mu$g m$^{-3}$. A different RH (30$\%$) and same T(22°C) would result in similar values: 10 $\mu$g m$^{-3}$ (8.57 $\mu$g m$^{-3}$) and 100 $\mu$g m$^{-3}$ (65.88 $\mu$g m$^{-3}$) where values in parenthesis indicate corrected PM$_{2.5}$ concentrations.  Fortunately, variations in RH during a given test day were <3 \%  at most and would therefore not affect the comparisons made later.  The Mailings correction is non-linear at low concentrations, especially when the contributions of sub-300nm particles is important, but this is not likely the case for the NaCl that dominates the concentrations in our experiments.  We use the PM10 channel in all that follows.


\clearpage

\section{Overview of all experiments}
A summary of all experiments is included in Table \ref{summary ta}. The delay time is the average [minutes] time delay between the atomizer being turned on and the concentration reaching 1 microgram per cubic meter.  The decay time [tau, minutes] is obtained by fitting an exponential to the latter portion of the experiment, over which all sensors show a consistent decay rate (this can be assessed in the plots in the next section).  For the whole-room metric, we have taken the median values rather than the average because in some experiments with long curtains, the north sensor exhibited a very long delay time as a result of never truly exhibiting the log-linear decay.  The steady-state concentration (SSsource, $\mu g/m^3$) is estimated assuming an exponential approach to the steady state value.  Similarly, SSaway is for the 4 sensors in the other zones.  The coefficient of vatiation reported here is the standard deviation of the SSaway values divided by the mean value.

The description includes some minor configuration variations that have been grouped into major categories reported in the main manuscript.  For example, there was a small window-mounted exhaust fan operated in Experiment 13, off but unsealed in experiments 1-12, 14-19, and sealed from experiments 20-41.  There is no detectable impact from this on any of configurations repeated through the 5 days. 
\begin{table}[ht!]
\centering
\includegraphics[width=14cm]{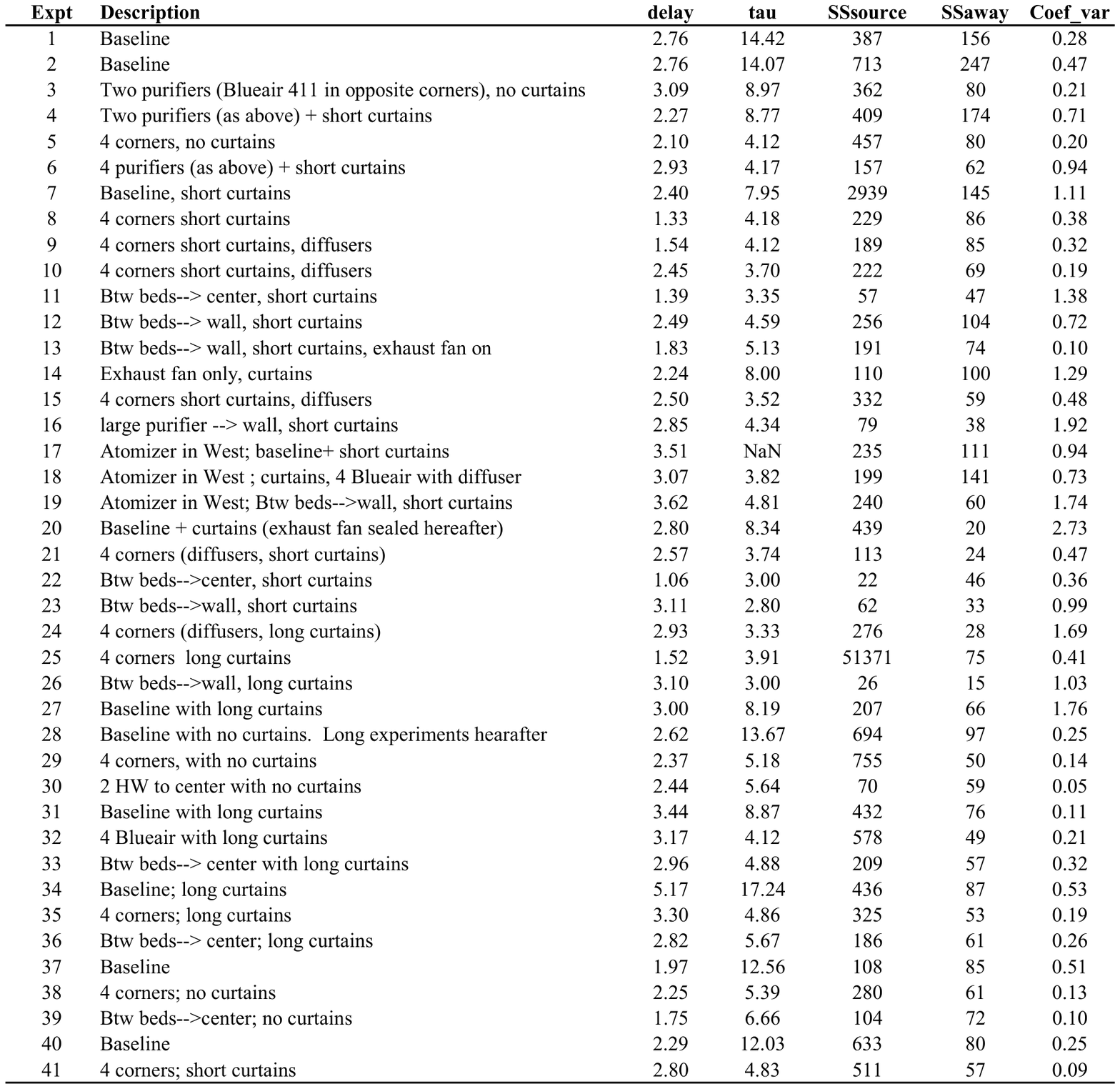}
\caption{Summary of all experiments}
\label{summary ta}
\end{table}{}

\pagebreak
\section{Summary plots for all experiments}
RAMP PM10 measurements are shown in each plot with the symbols.  Locations in the legend are described in the main manuscript.  To determine the decay time (after the vertical dashed line), exponentials were fit to the data and the resulting fits are shown on the plots (appearing as straight lines on the log-linear plots). The fit was performed for times after \~ 14 minutes for day 1-3 data and after approximately 28 minutes for day 4-5 data. 

Also shown are the exponential approaches to steady state determined from the fit to the first portion of each experiment.  The large symbols at 1 microgram per cubic meter indicate the "delay time" reported in the manuscript.  Numerical data for each experiment is given in SI section B.

\begin{figure}[h!]
\centering
\includegraphics[width=10cm]{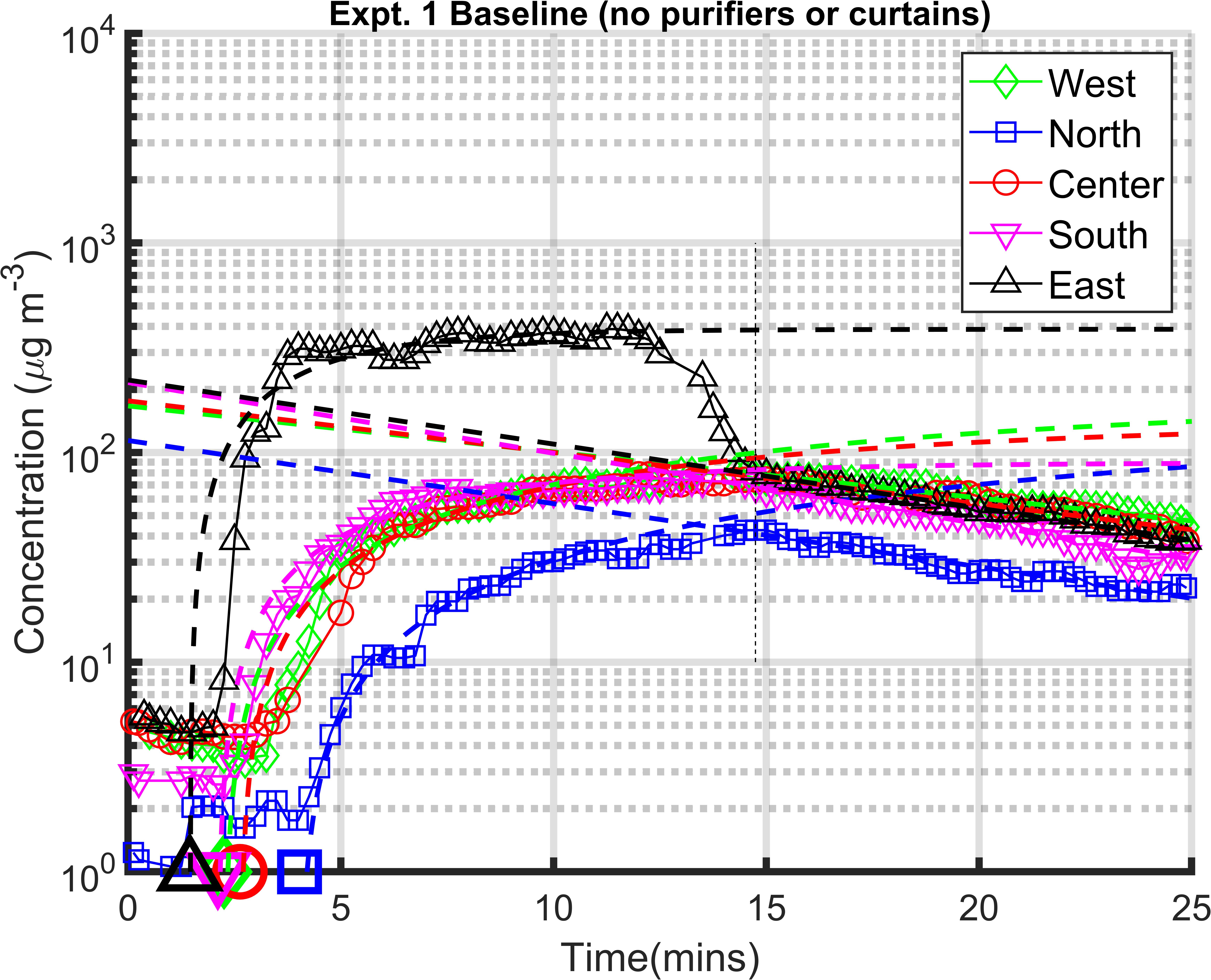}
\caption{Experiment 1. Day1}
\label{Expt1}
\end{figure}

\begin{figure}[h!]
\centering
\includegraphics[width=10cm]{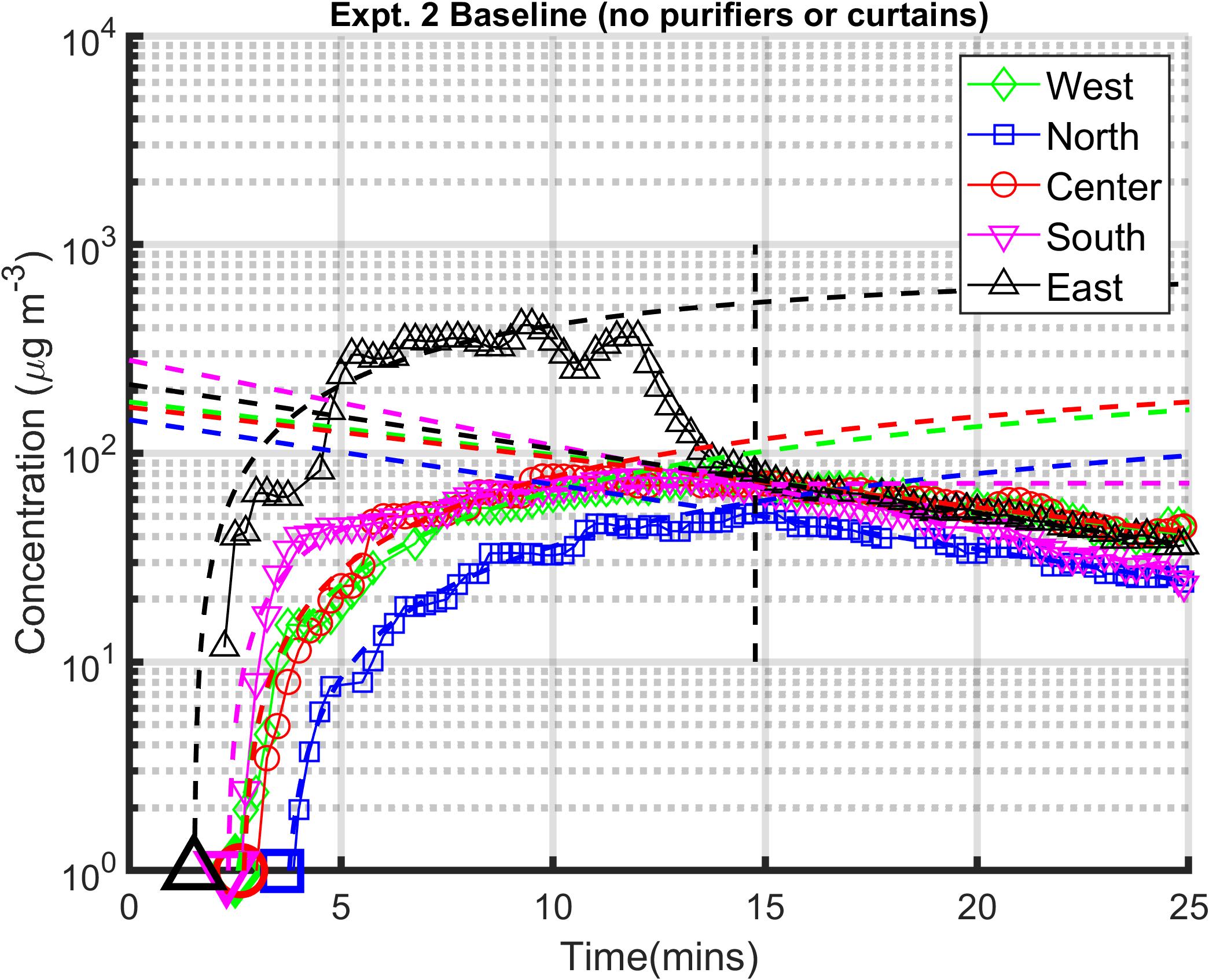}
\caption{Experiment 2. Day1}
\label{Expt2}
\end{figure}

\begin{figure}[h!]
\centering
\includegraphics[width=10cm]{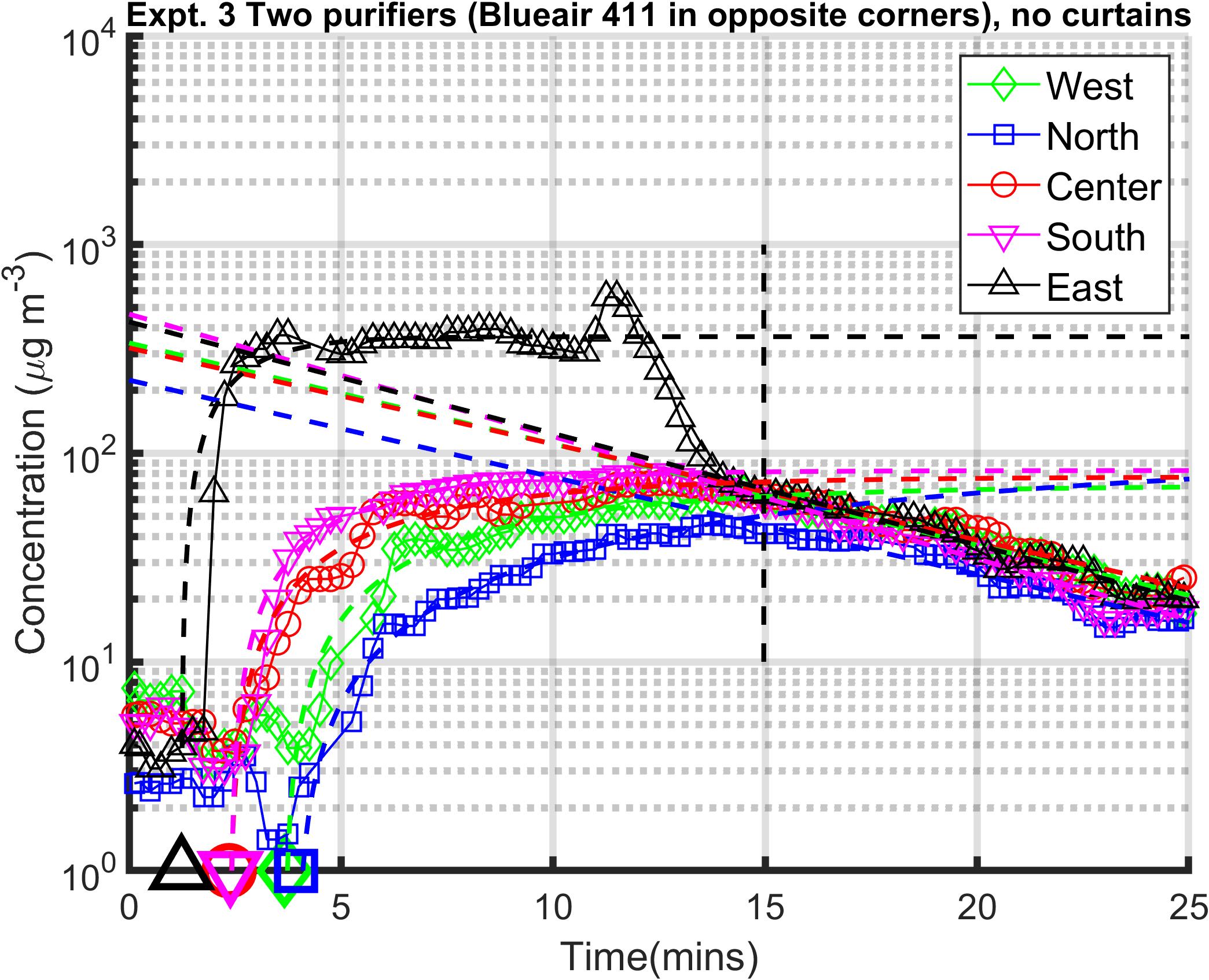}
\caption{Experiment 3. Day 1}
\label{Expt3}
\end{figure}

\begin{figure}[h!]
\centering
\includegraphics[width=10cm]{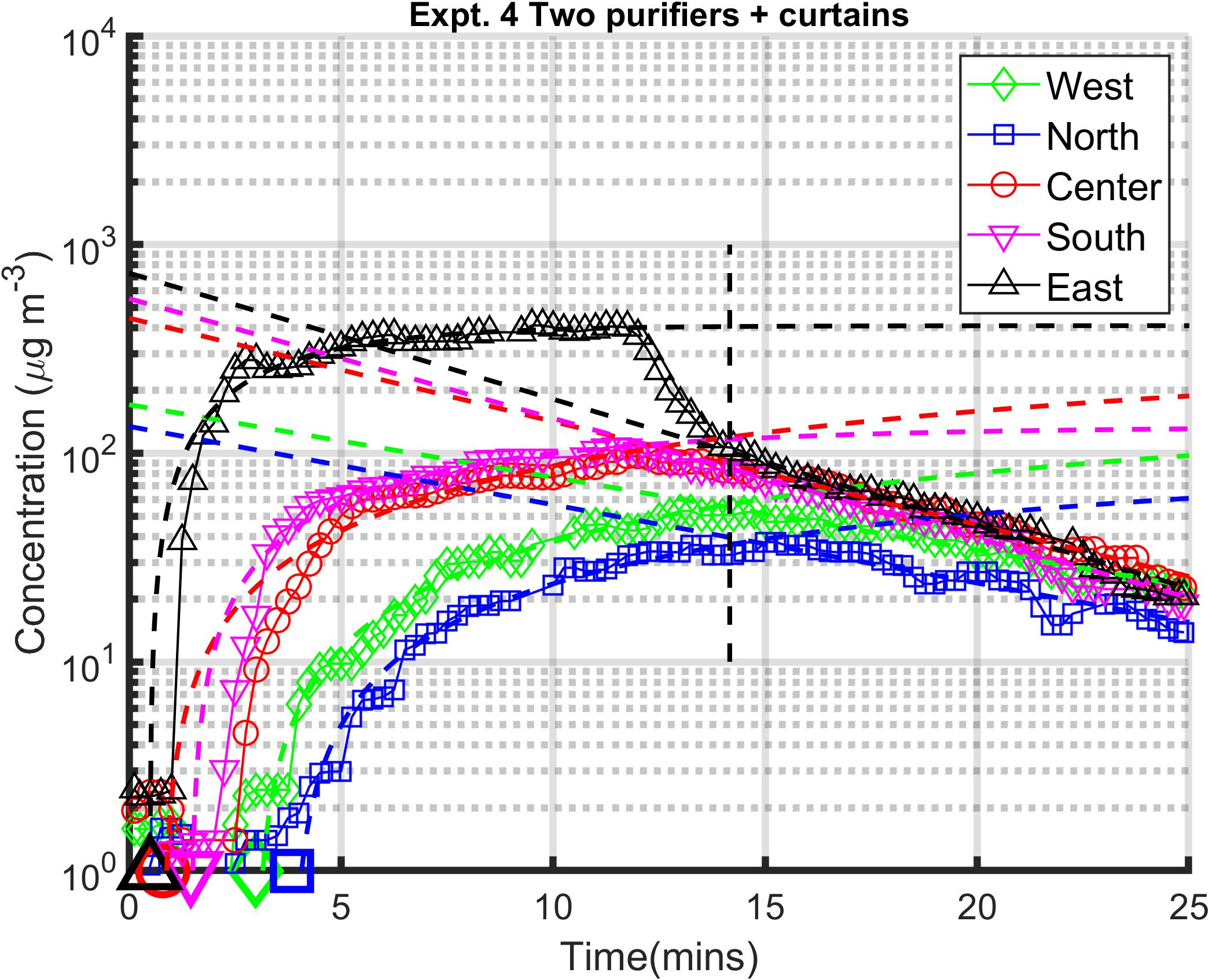}
\caption{Experiment 4. Day 1.  Purifiers in corners as Expt. 3. Short curtains.}
\label{Expt4}
\end{figure}

\begin{figure}[h!]
\centering
\includegraphics[width=10cm]{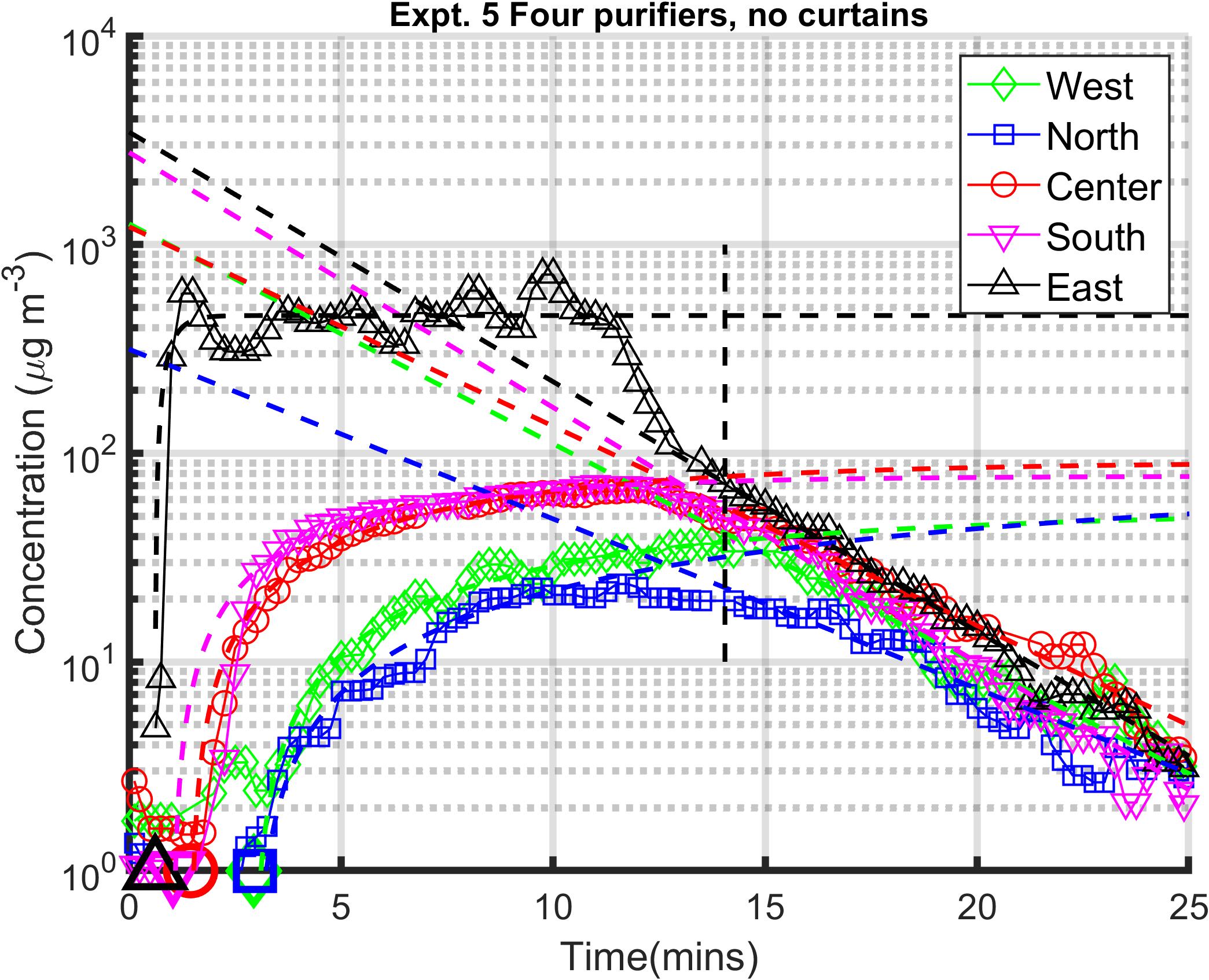}
\caption{Experiment 5. Day 1}
\label{Expt5}
\end{figure}

\begin{figure}[h!]
\centering
\includegraphics[width=10cm]{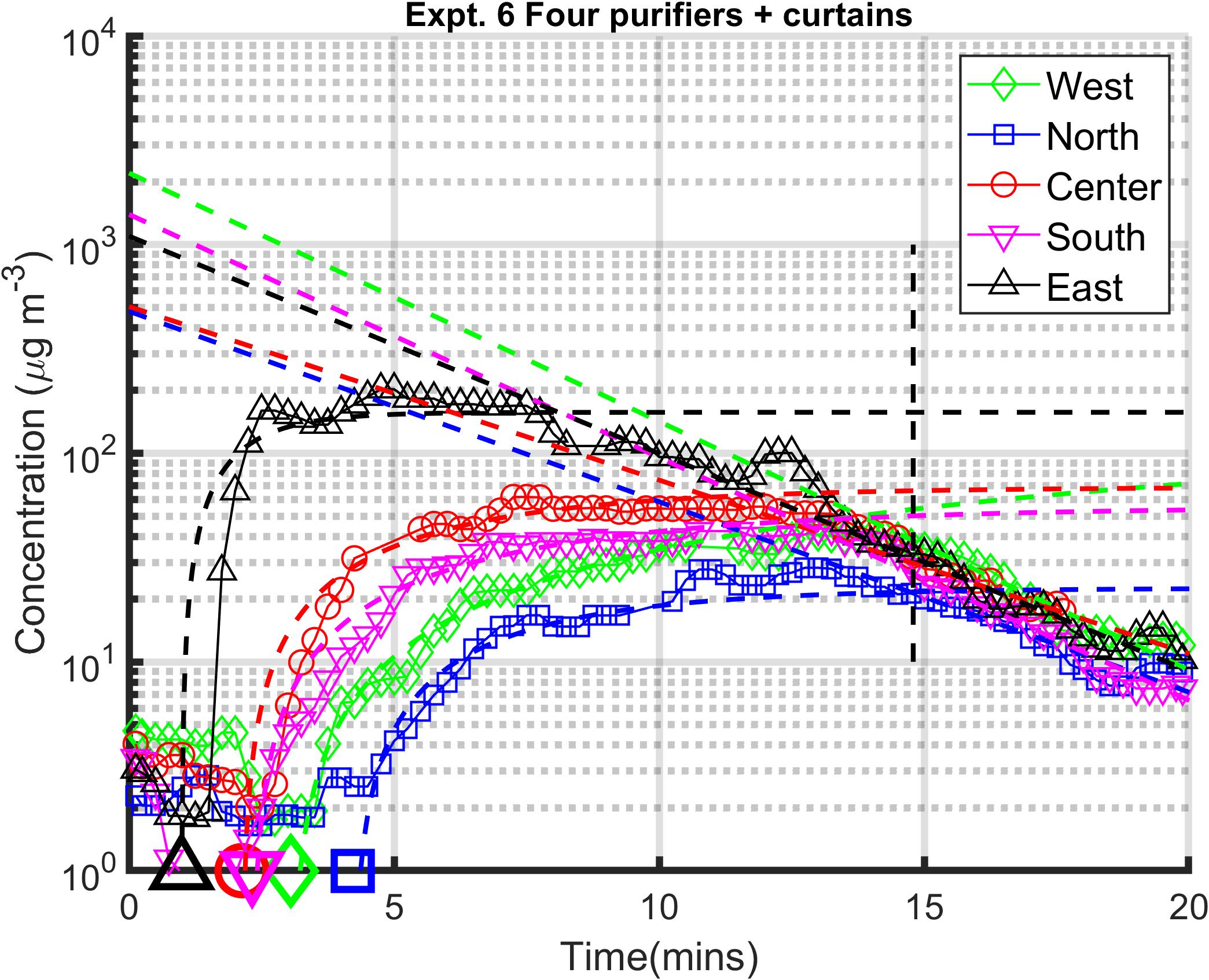}
\caption{Experiment 6.  Day1 Purifiers in corners; short curtains.}
\label{Expt6}
\end{figure}

\begin{figure}[h!]
\centering
\includegraphics[width=10cm]{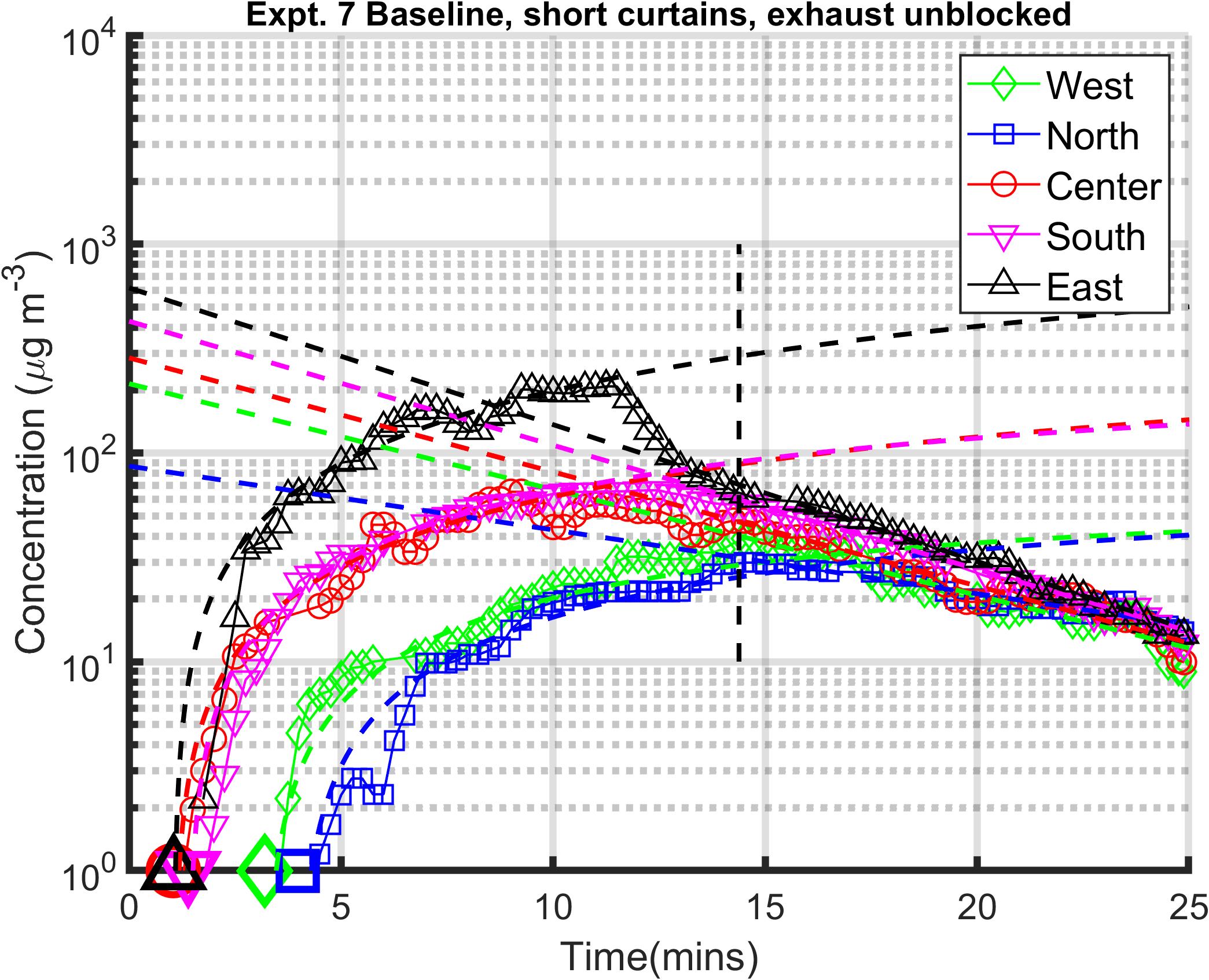}
\caption{Experiment 7. Day 2 Window exhaust fan not sealed (as in Day 1).}
\label{Expt7}
\end{figure}

\begin{figure}[h!]
\centering
\includegraphics[width=10cm]{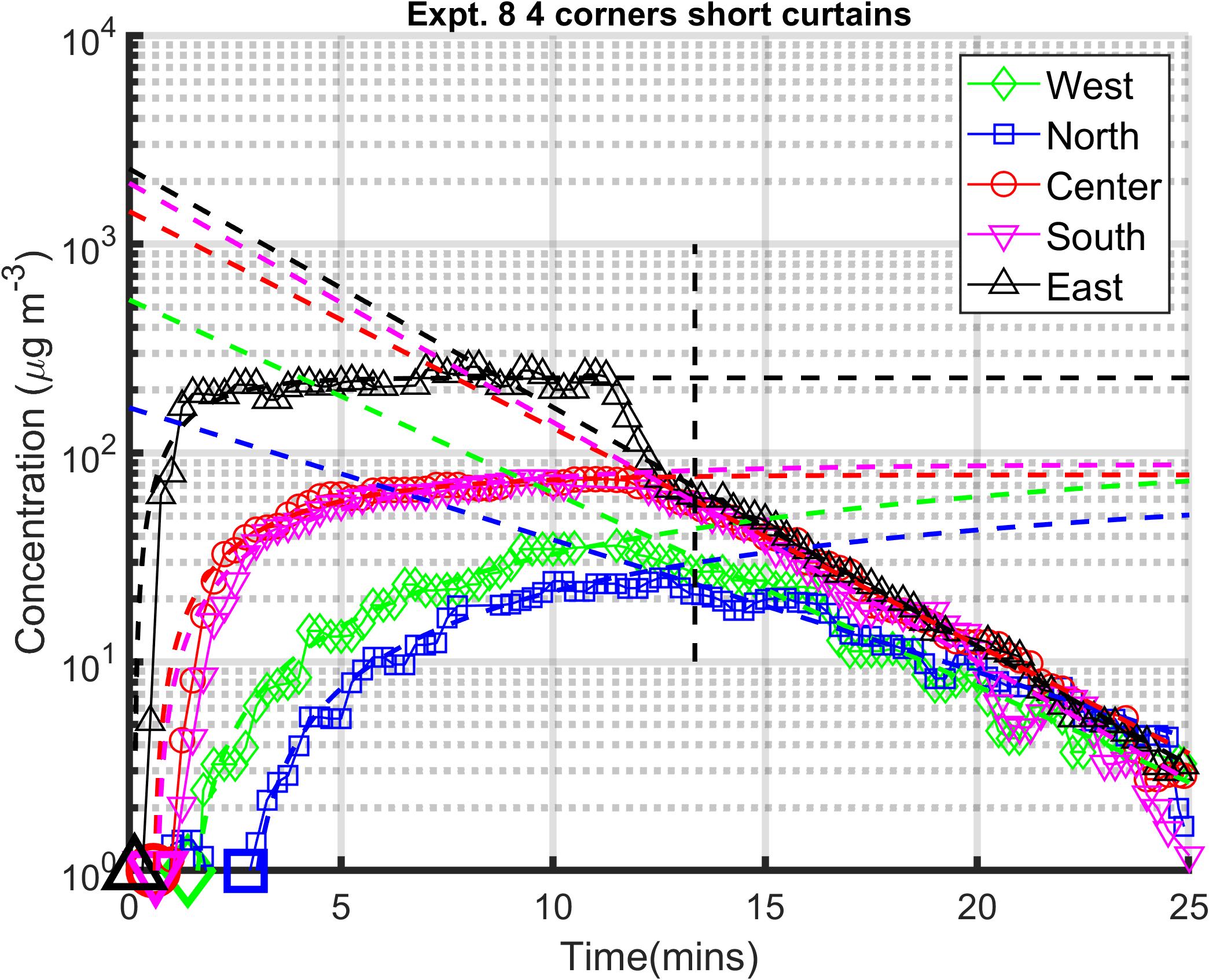}
\caption{Experiment 8.Day 2}
\label{Expt8}
\end{figure}

\begin{figure}[h!]
\centering
\includegraphics[width=10cm]{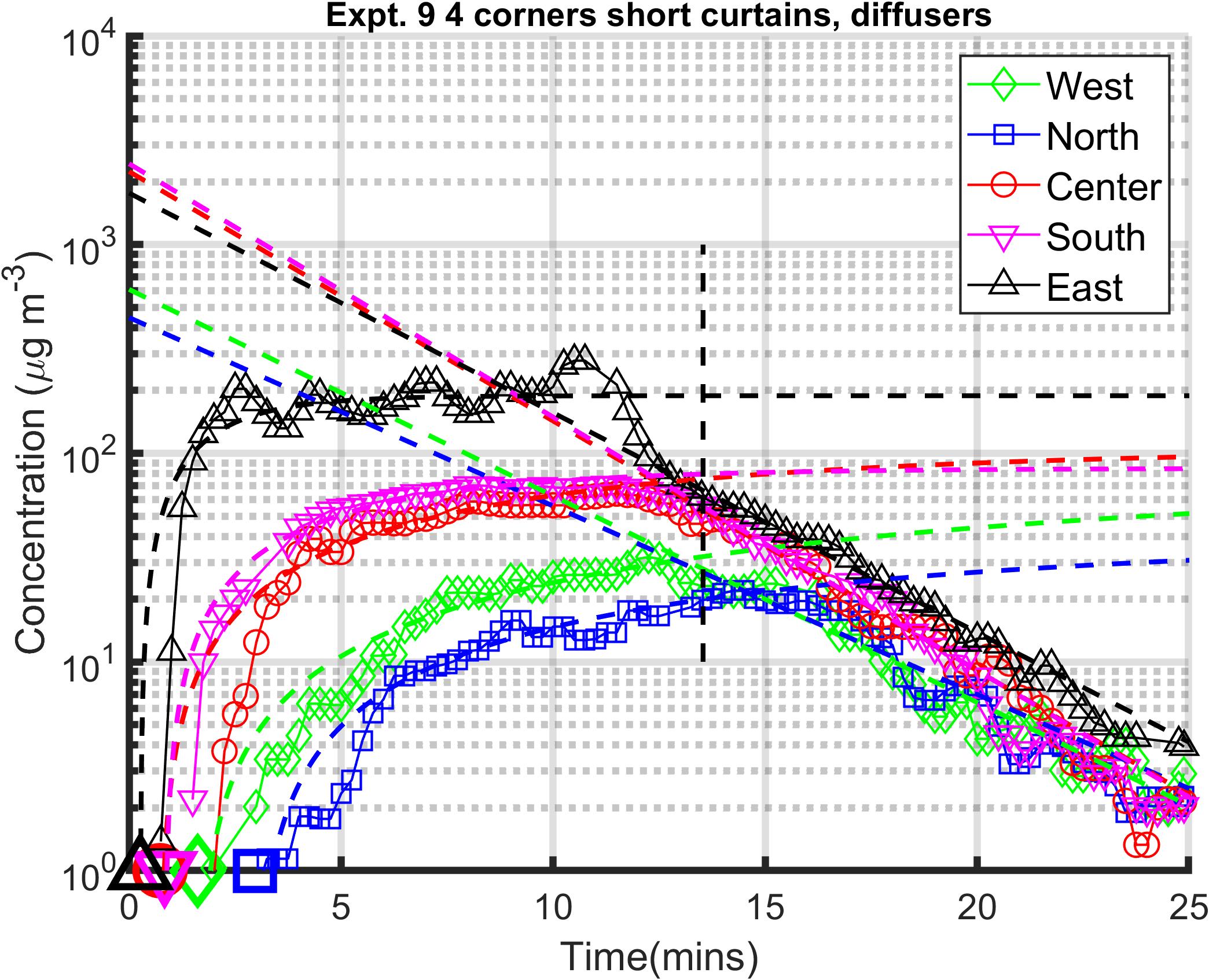}
\caption{Experiment 9. Day 2 Diffusers (mesh cones at purifier exhaust) had no detectable effect of flow, decay time or delay times.}
\label{Expt9}
\end{figure}

\begin{figure}[h!]
\centering
\includegraphics[width=10cm]{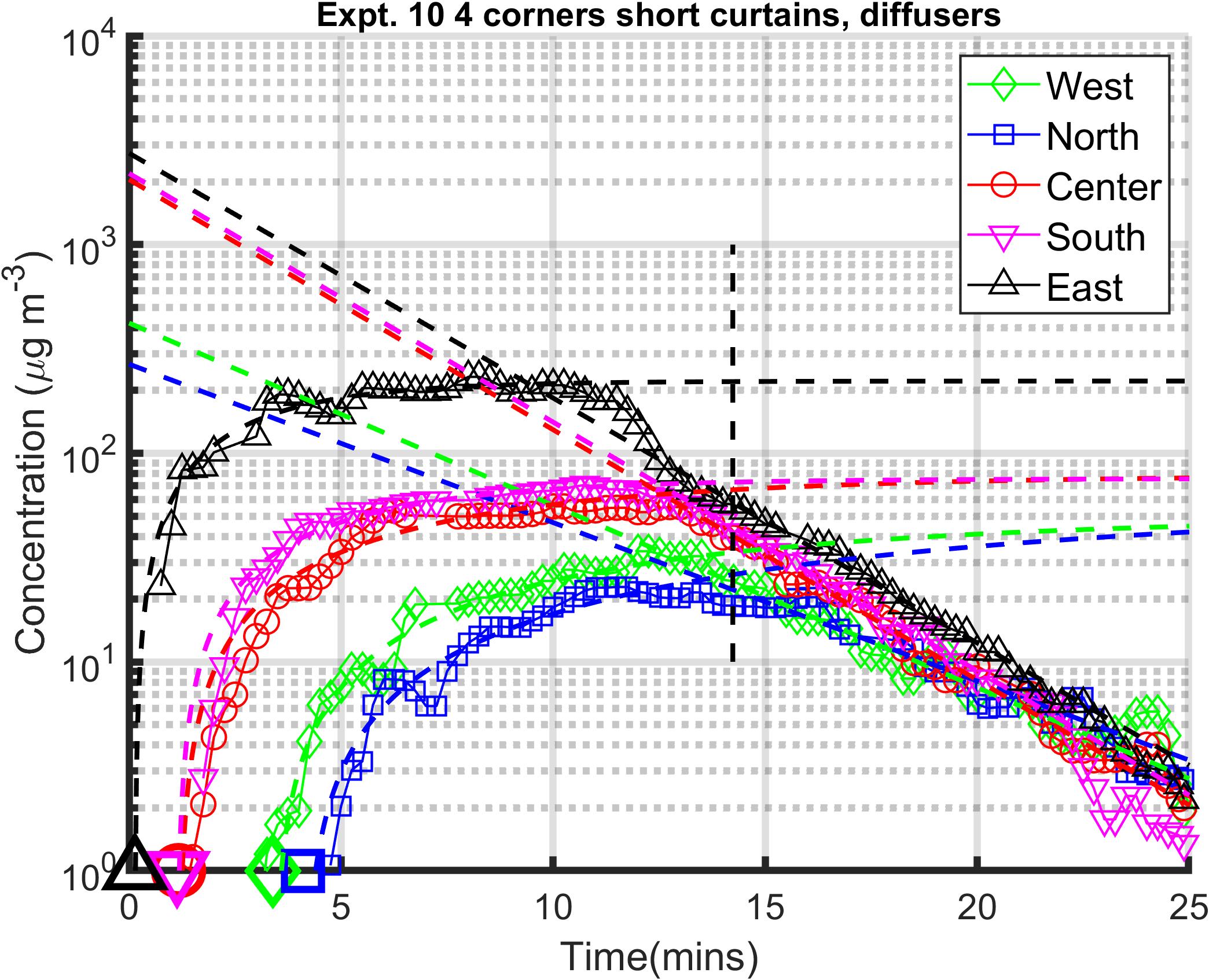}
\caption{Experiment 10. Day 2 }
\label{Expt10}
\end{figure}

\begin{figure}[h!]
\centering
\includegraphics[width=10cm]{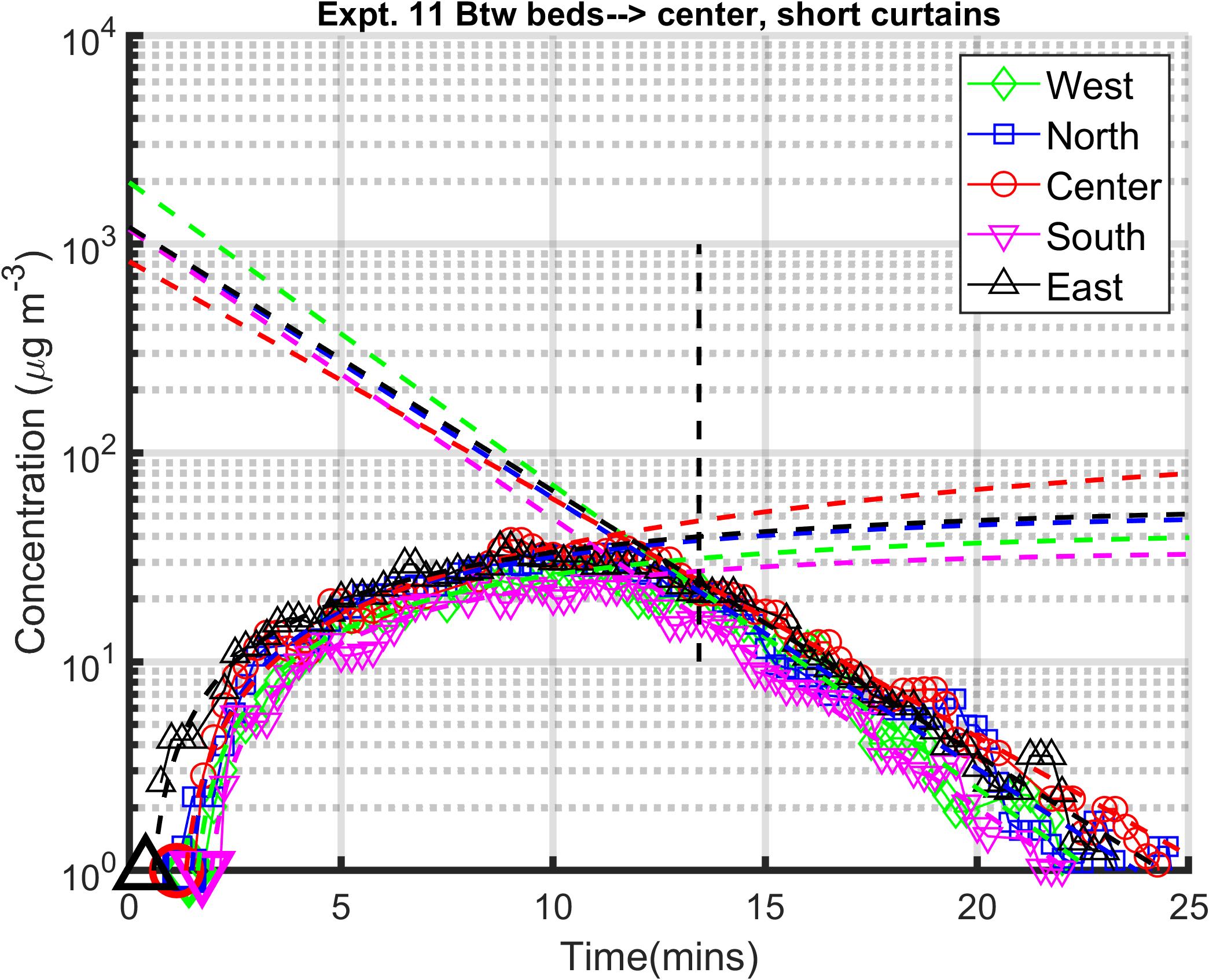}
\caption{Experiment 11. Day 2 Two Honeywell tower purifiers exhausting towards room center.}
\label{Expt11}
\end{figure}

\begin{figure}[h!]
\centering
\includegraphics[width=10cm]{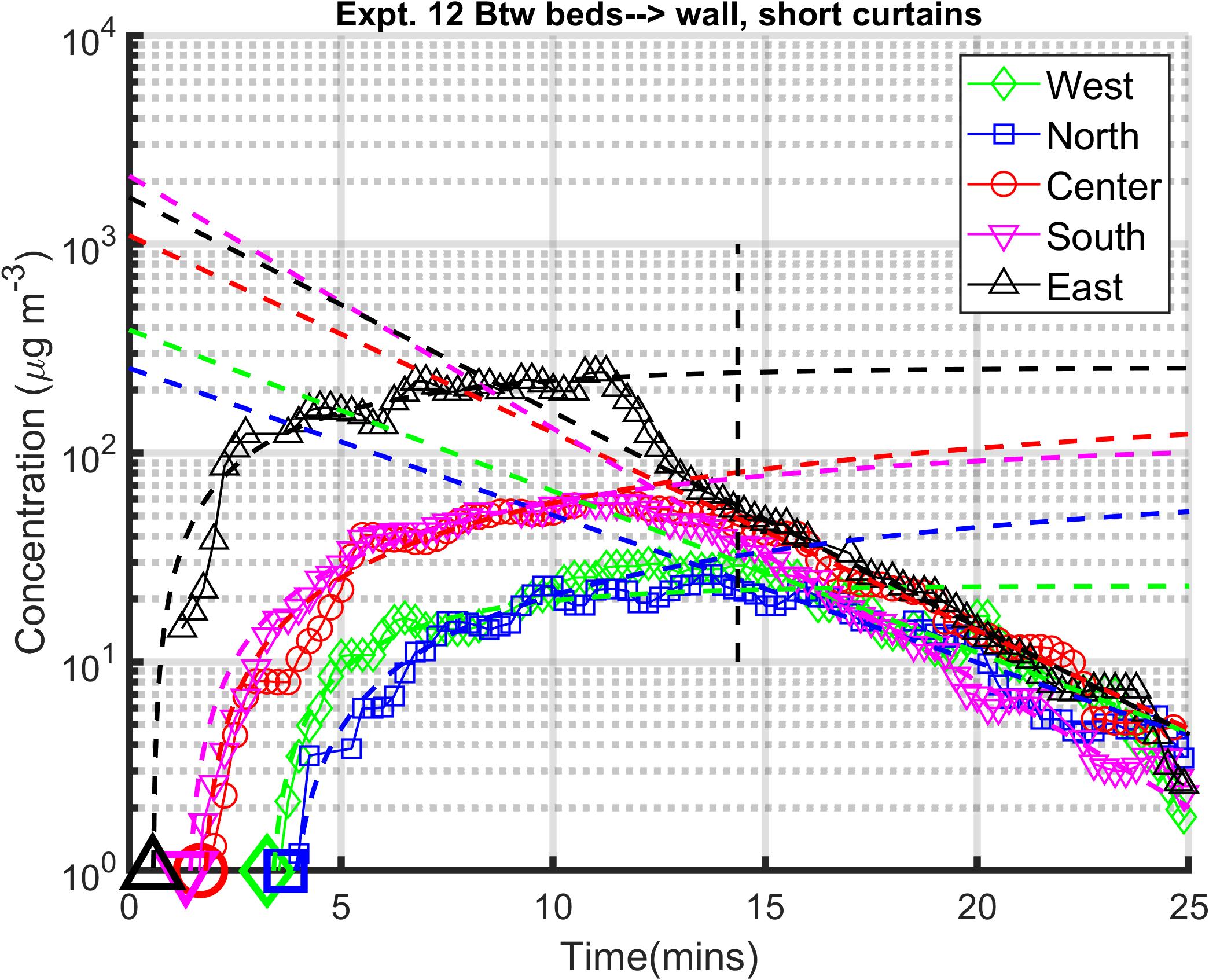}
\caption{Experiment 12. Day 2 Tower purifiers; exhaust direction reversed compared to Experiment 11.}
\label{Expt12}
\end{figure}

\begin{figure}[h!]
\centering
\includegraphics[width=10cm]{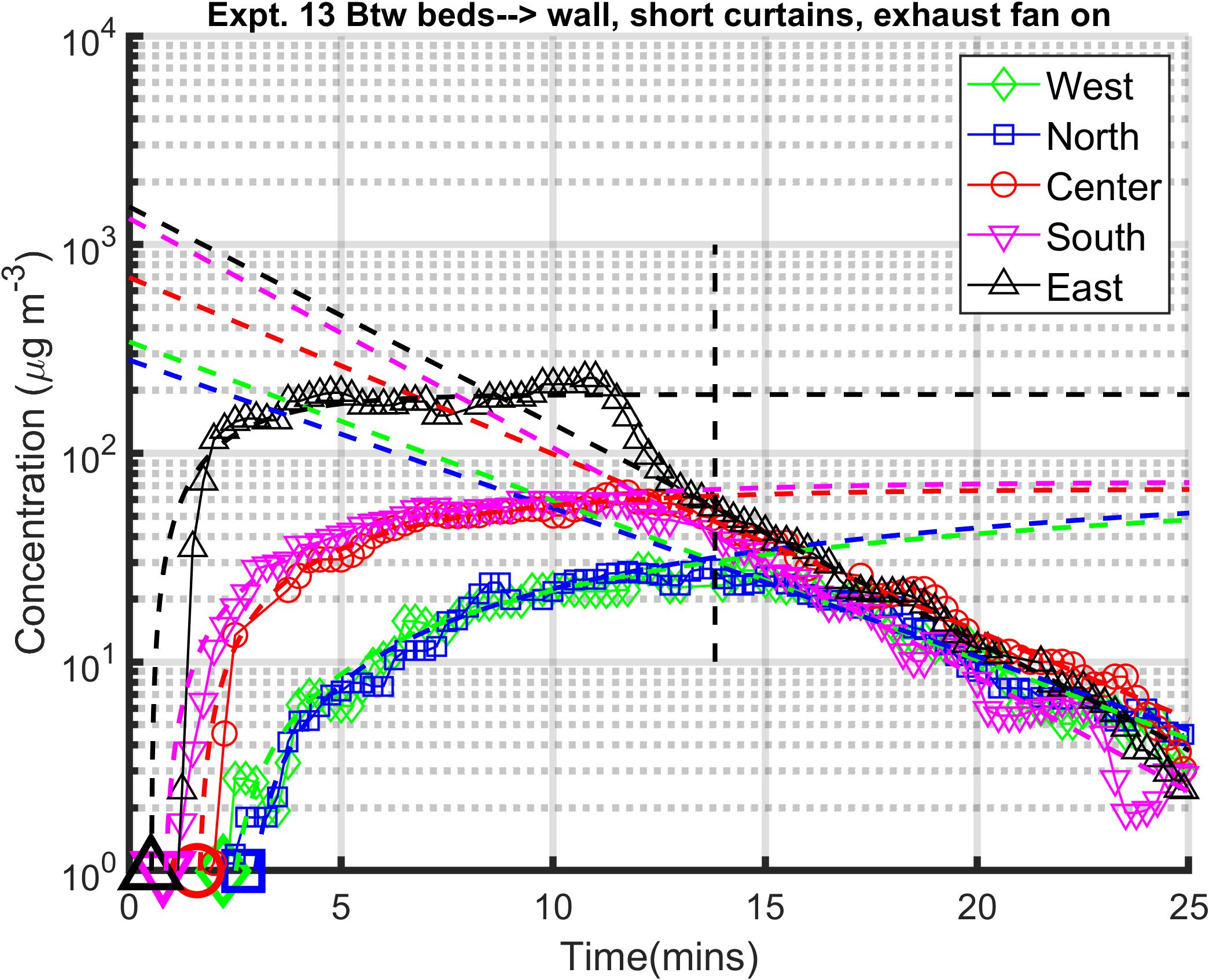}
\caption{Experiment 13. Day 2 Identical to Expt. 12 but window-mounted fan on.}
\label{Expt13}
\end{figure}

\begin{figure}[h!]
\centering
\includegraphics[width=10cm]{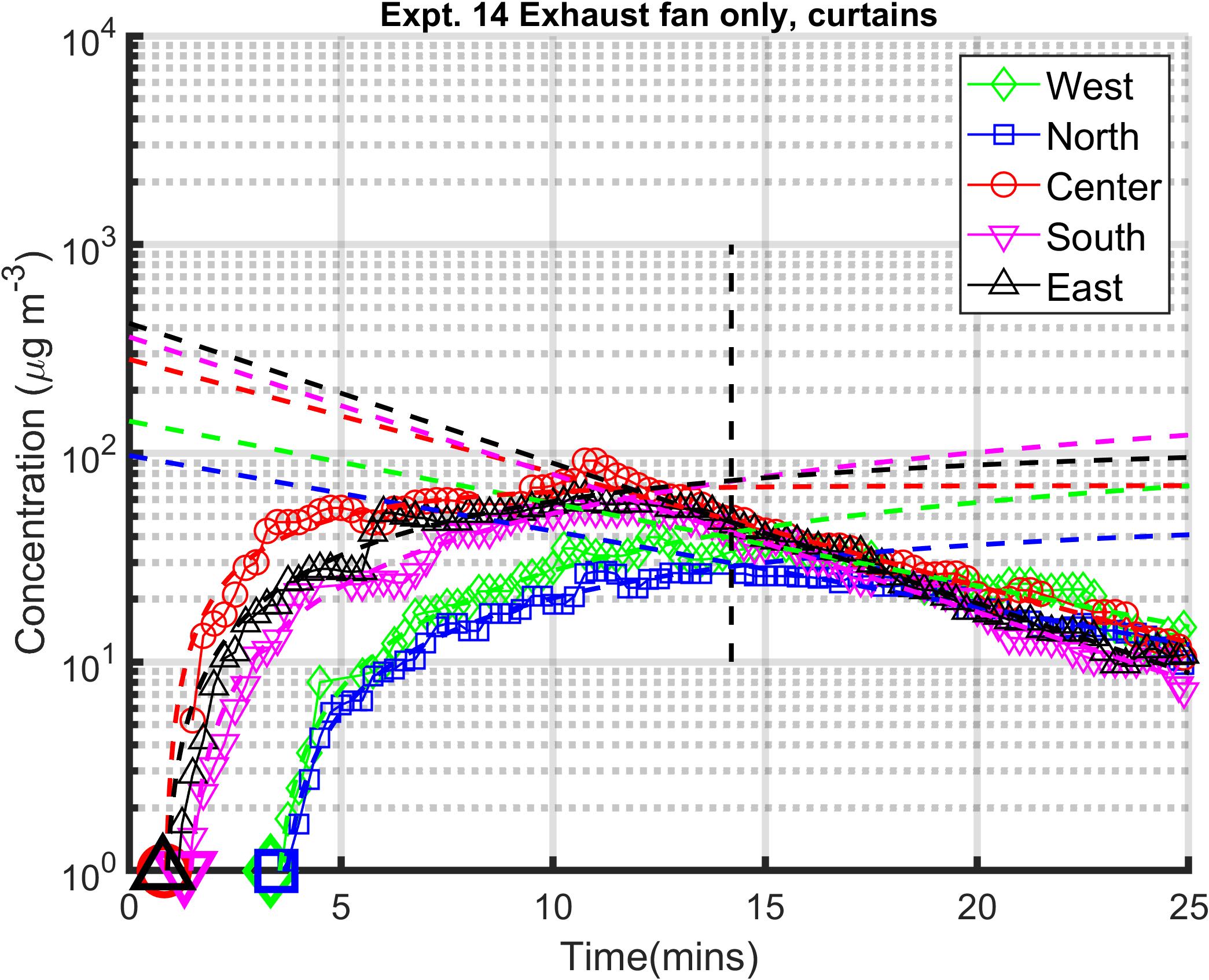}
\caption{Experiment 14. Day 2 No purifiers; short curtains; exhaust fan on.}
\label{Expt14}
\end{figure}

\begin{figure}[h!]
\centering
\includegraphics[width=10cm]{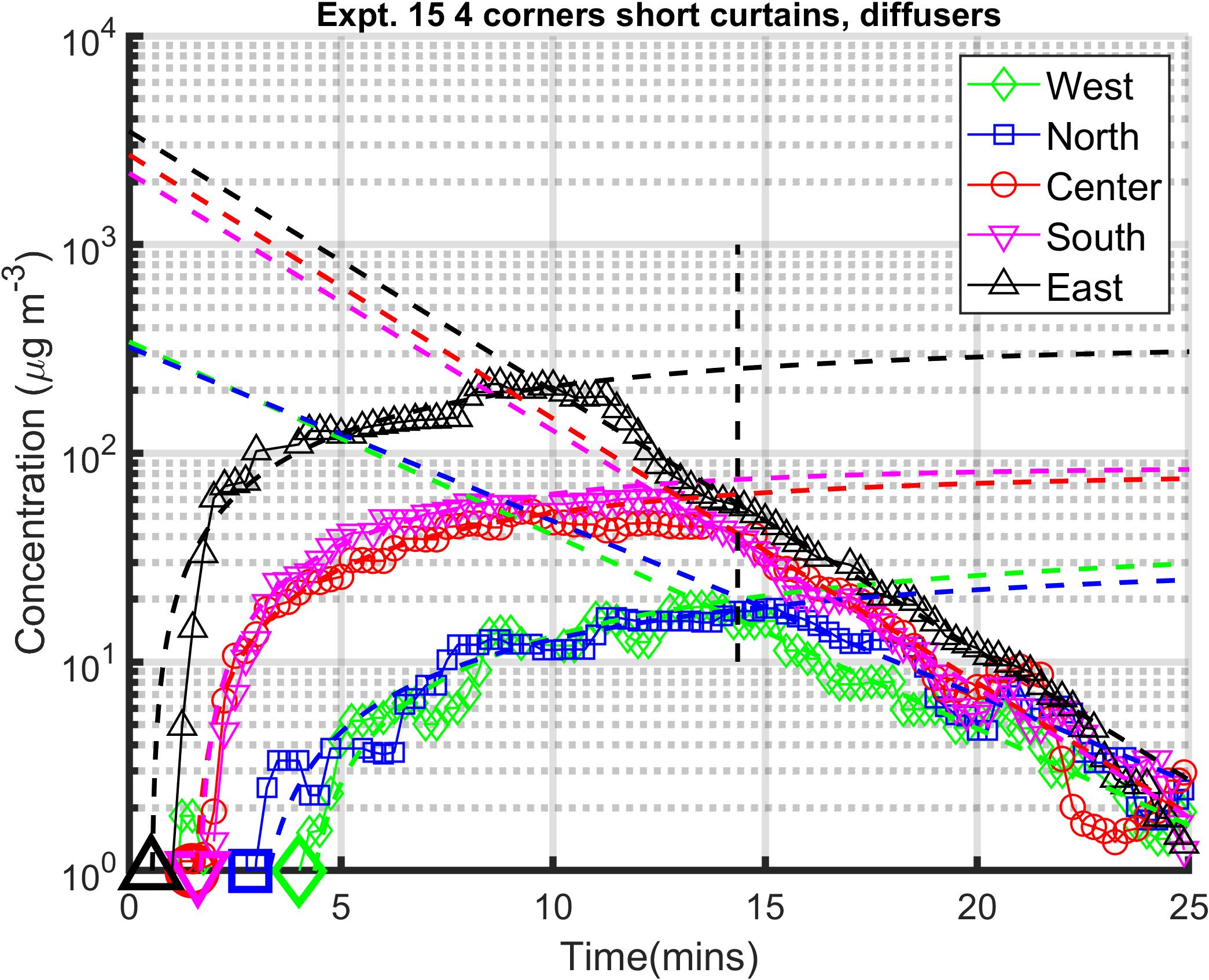}
\caption{Experiment 15. Day 2}
\label{Expt15}
\end{figure}

\begin{figure}[h!]
\centering
\includegraphics[width=10cm]{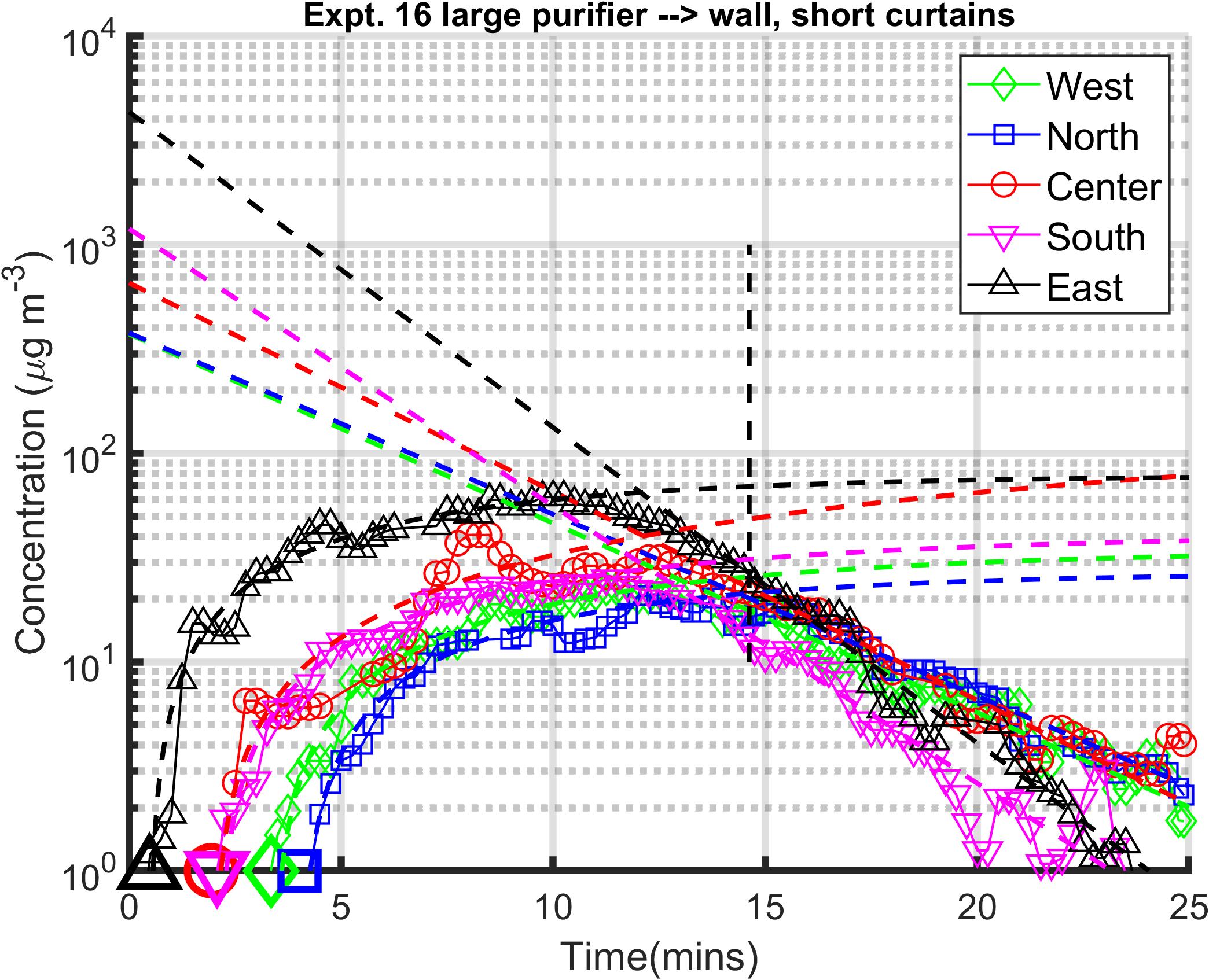}
\caption{Experiment 16. Day 2 "Atmosphere" purifier in aisle, centered on SW wall, exhausting towards wall.}
\label{Expt16}
\end{figure}

\begin{figure}[h!]
\centering
\includegraphics[width=10cm]{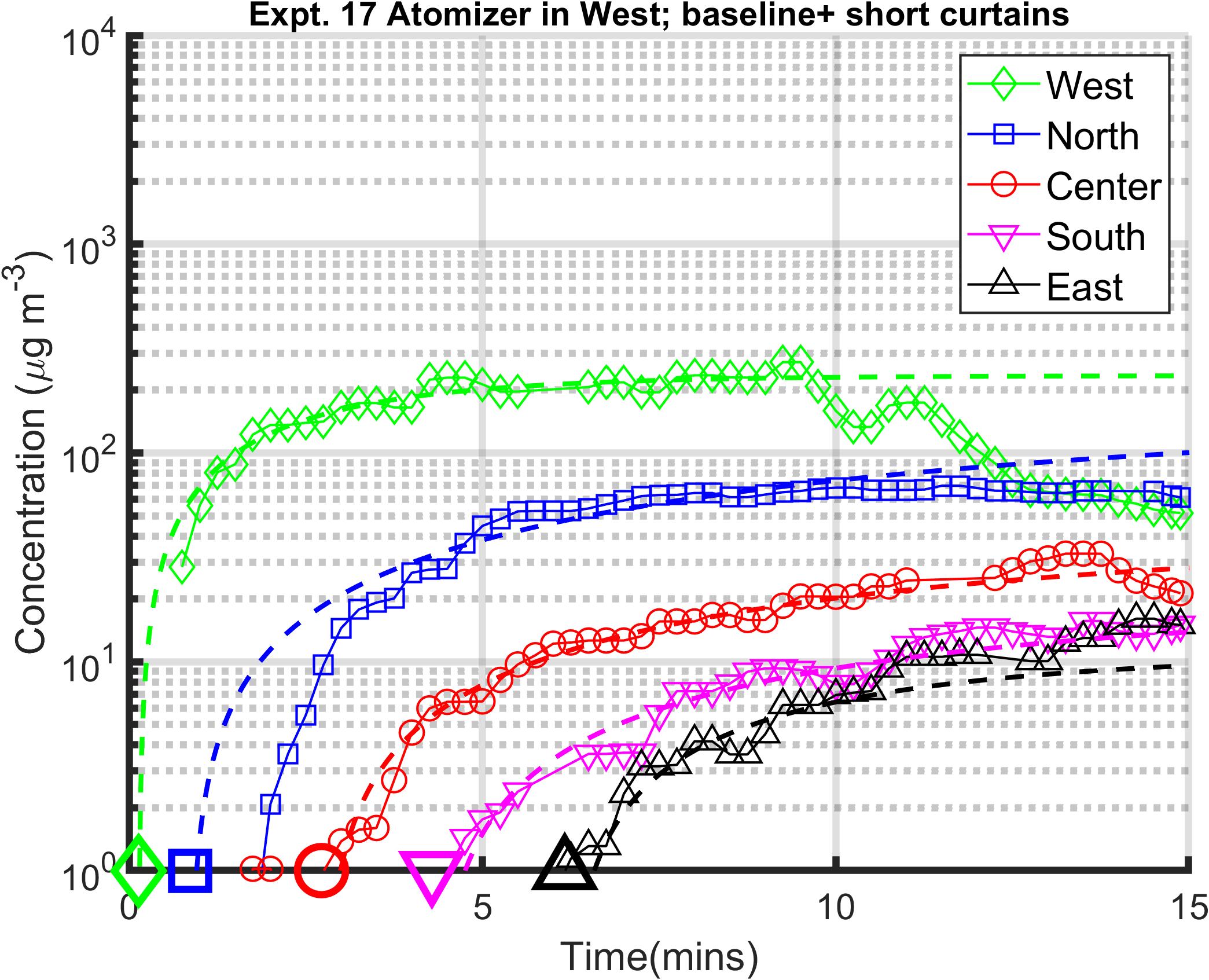}
\caption{Experiment 17. Day 2 Atomizer on West bed, no purifiers; short curtains.}
\label{Expt17}
\end{figure}

\begin{figure}[h!]
\centering
\includegraphics[width=10cm]{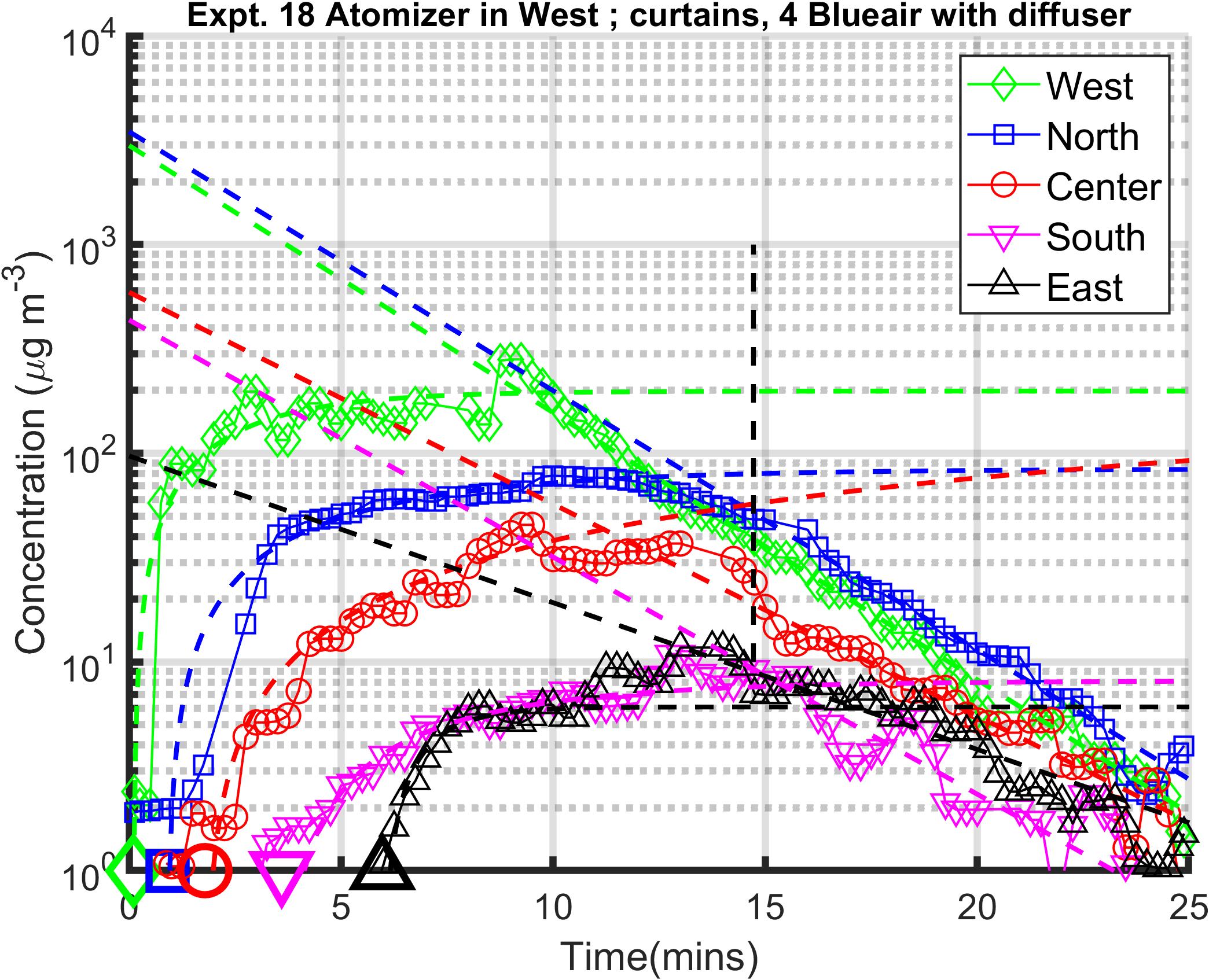}
\caption{Experiment 18. Day 2 Atomizer on West bed, corner purifiers.}
\label{Expt18}
\end{figure}

\begin{figure}[h!]
\centering
\includegraphics[width=10cm]{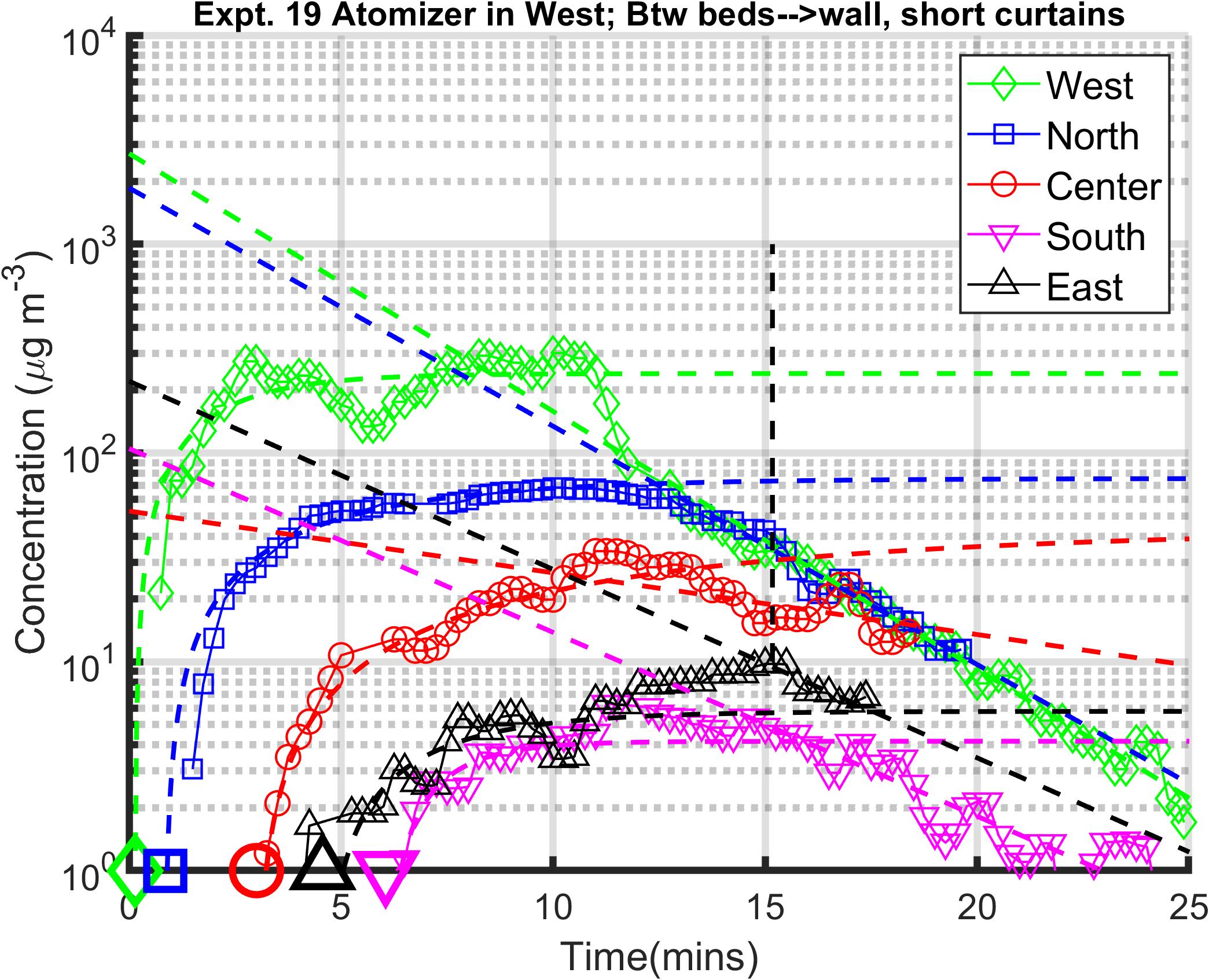}
\caption{Experiment 19. Day 2}
\label{Expt19}
\end{figure}

\begin{figure}[h!]
\centering
\includegraphics[width=10cm]{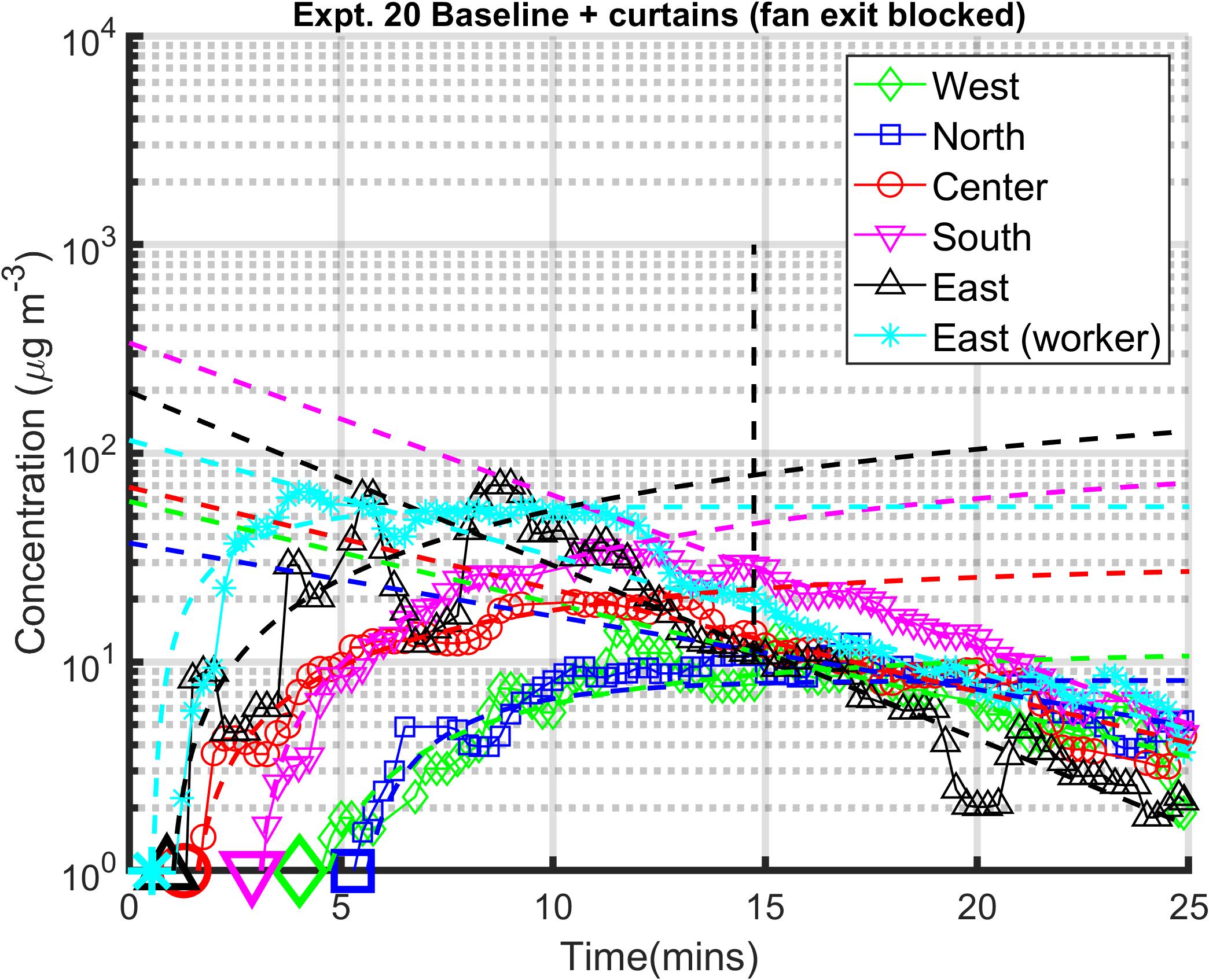}
\caption{Experiment 20. Day 3 Magenta marker is for a RAMP located at 1.5 m elevation just at the South edge of the East bed.  Hereafter, exhaust fan is sealed.}
\label{Expt20}
\end{figure}

\begin{figure}[h!]
\centering
\includegraphics[width=10cm]{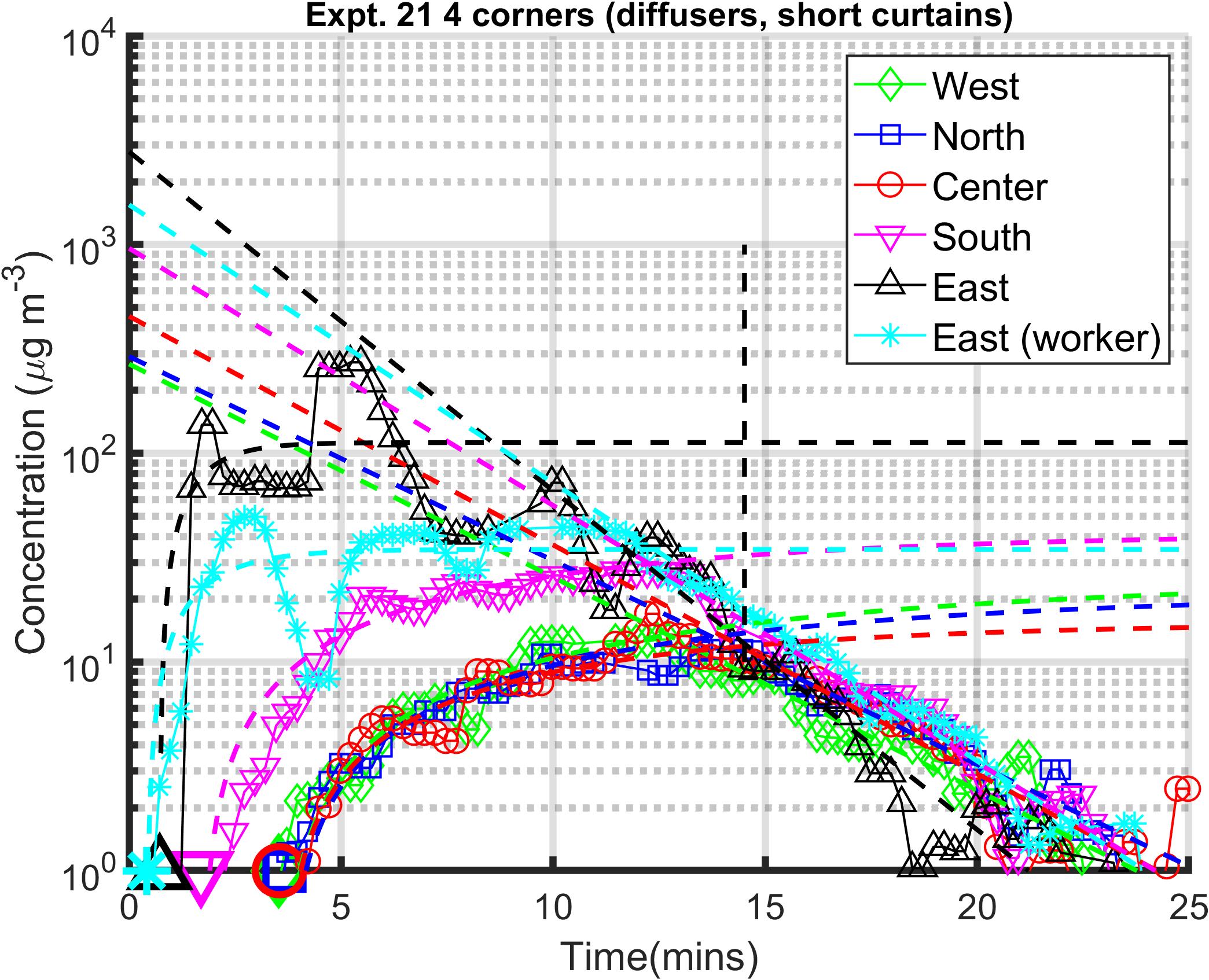}
\caption{Experiment 21. Day 3}
\label{Expt21}
\end{figure}

\begin{figure}[h!]
\centering
\includegraphics[width=10cm]{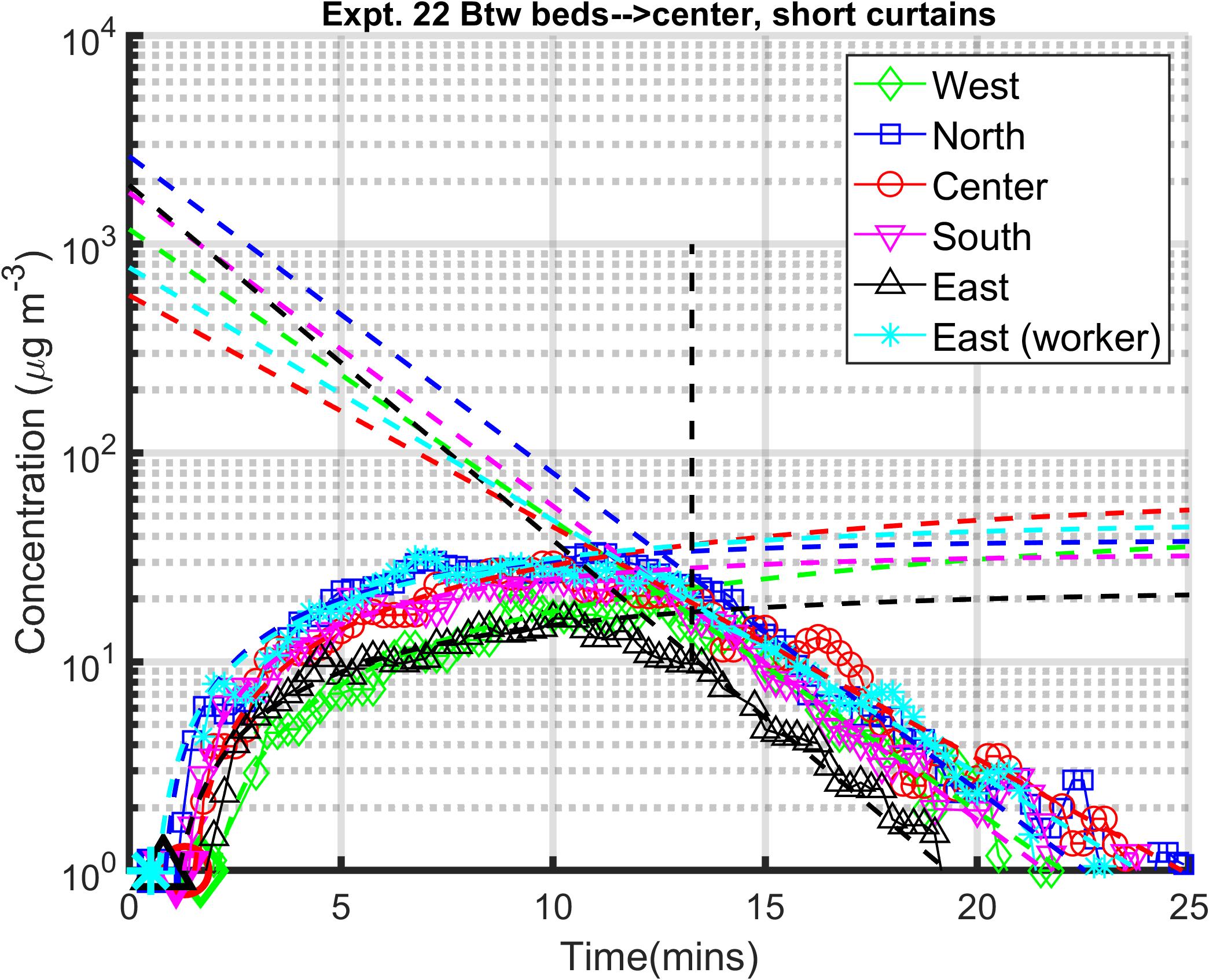}
\caption{Experiment 22 Day 3}
\label{Expt22}
\end{figure}

\begin{figure}[h!]
\centering
\includegraphics[width=10cm]{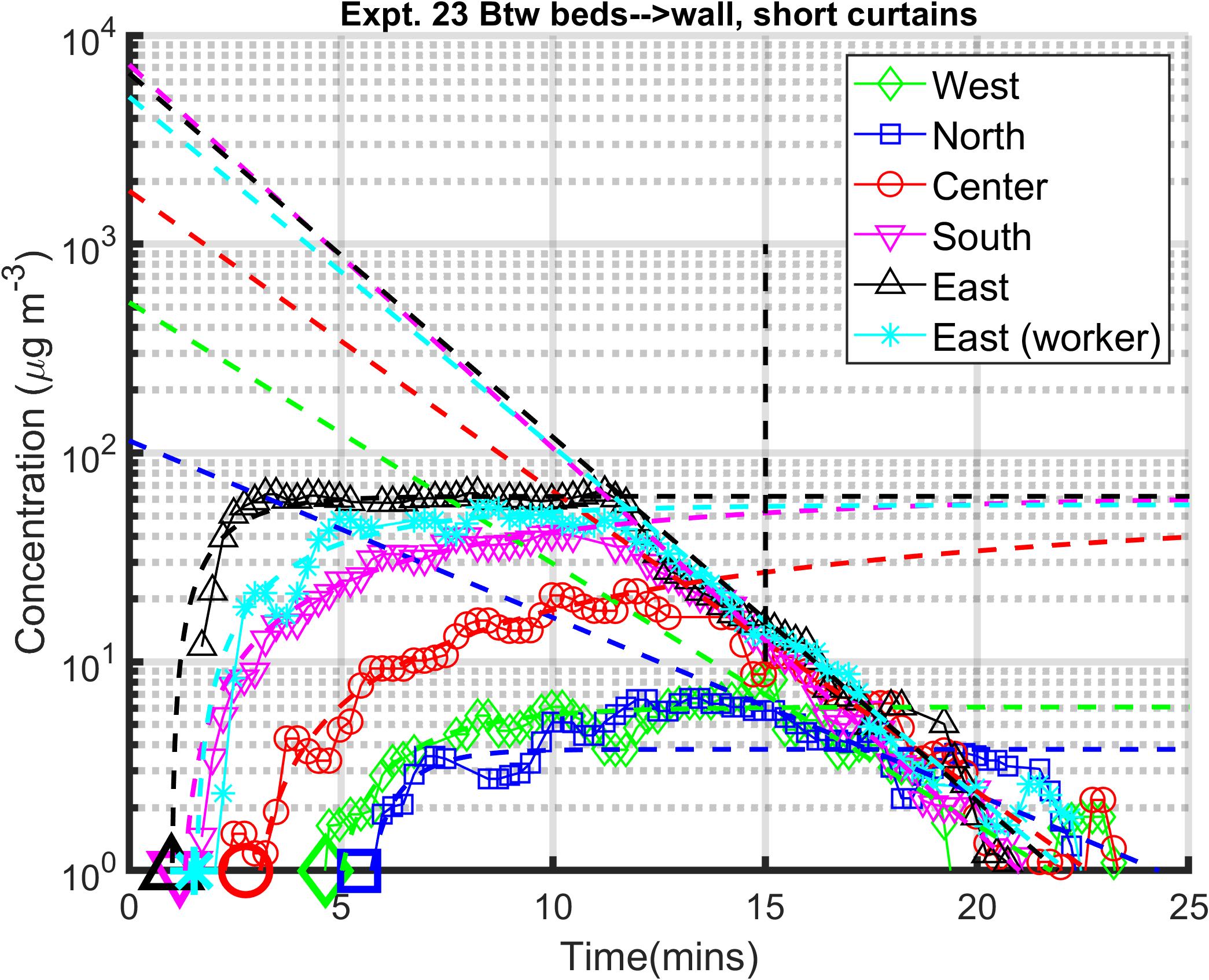}
\caption{Experiment 23 Day 3}
\label{Expt23}
\end{figure}

\begin{figure}[h!]
\centering
\includegraphics[width=10cm]{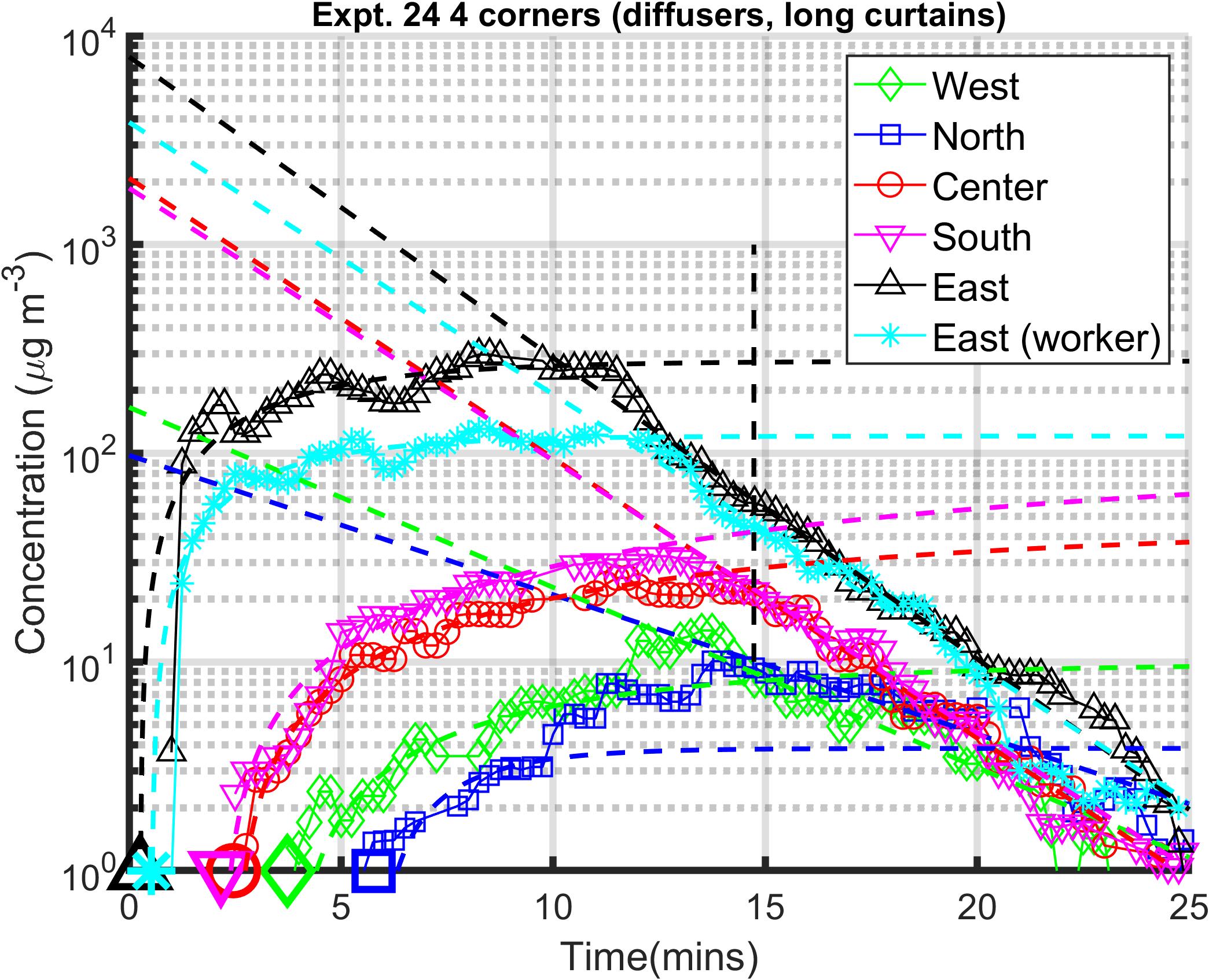}
\caption{Experiment 24. Day 3}
\label{Expt24}
\end{figure}

\begin{figure}[h!]
\centering
\includegraphics[width=10cm]{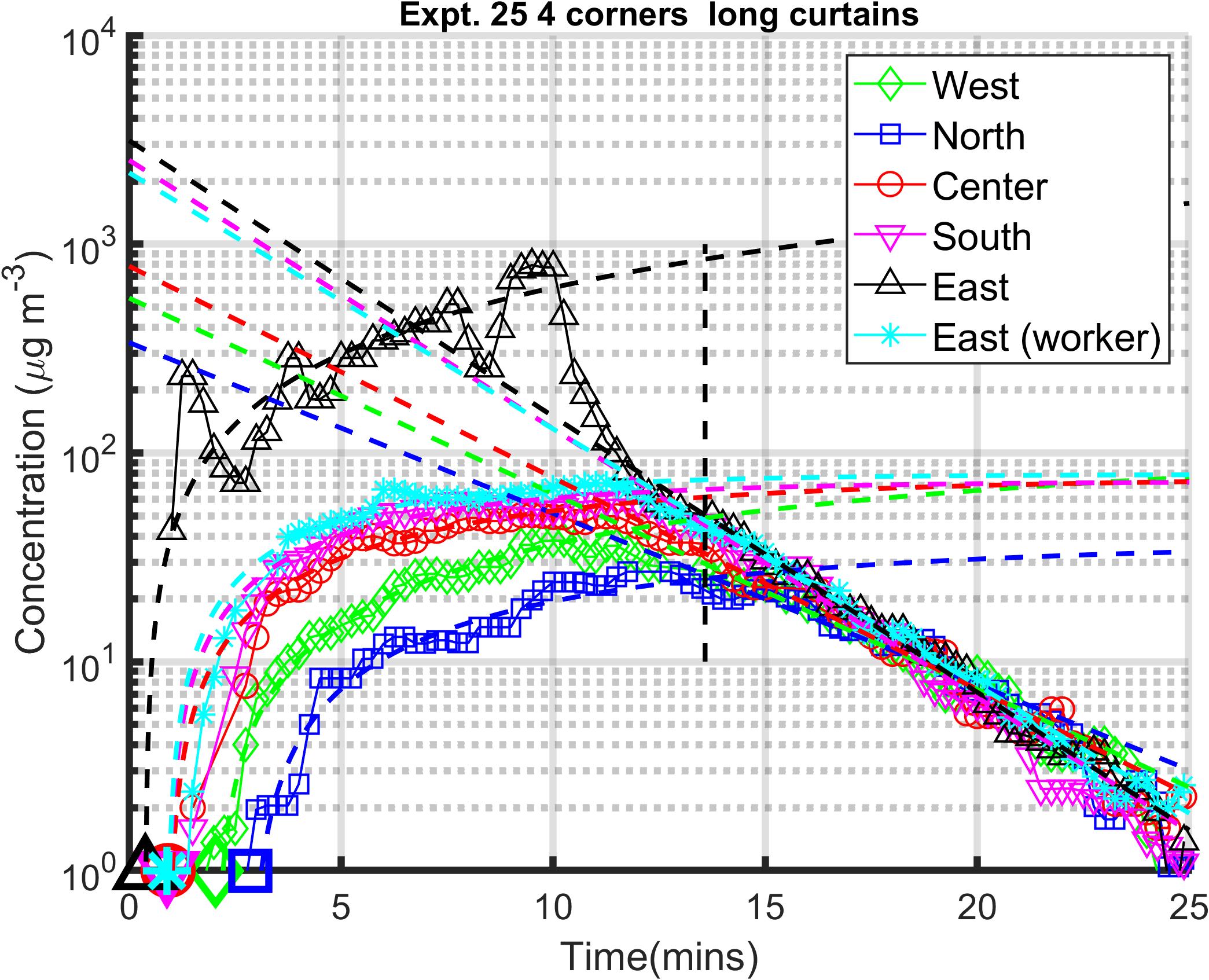}
\caption{Experiment 25. Day 3}
\label{Expt25}
\end{figure}

\begin{figure}[h!]
\centering
\includegraphics[width=10cm]{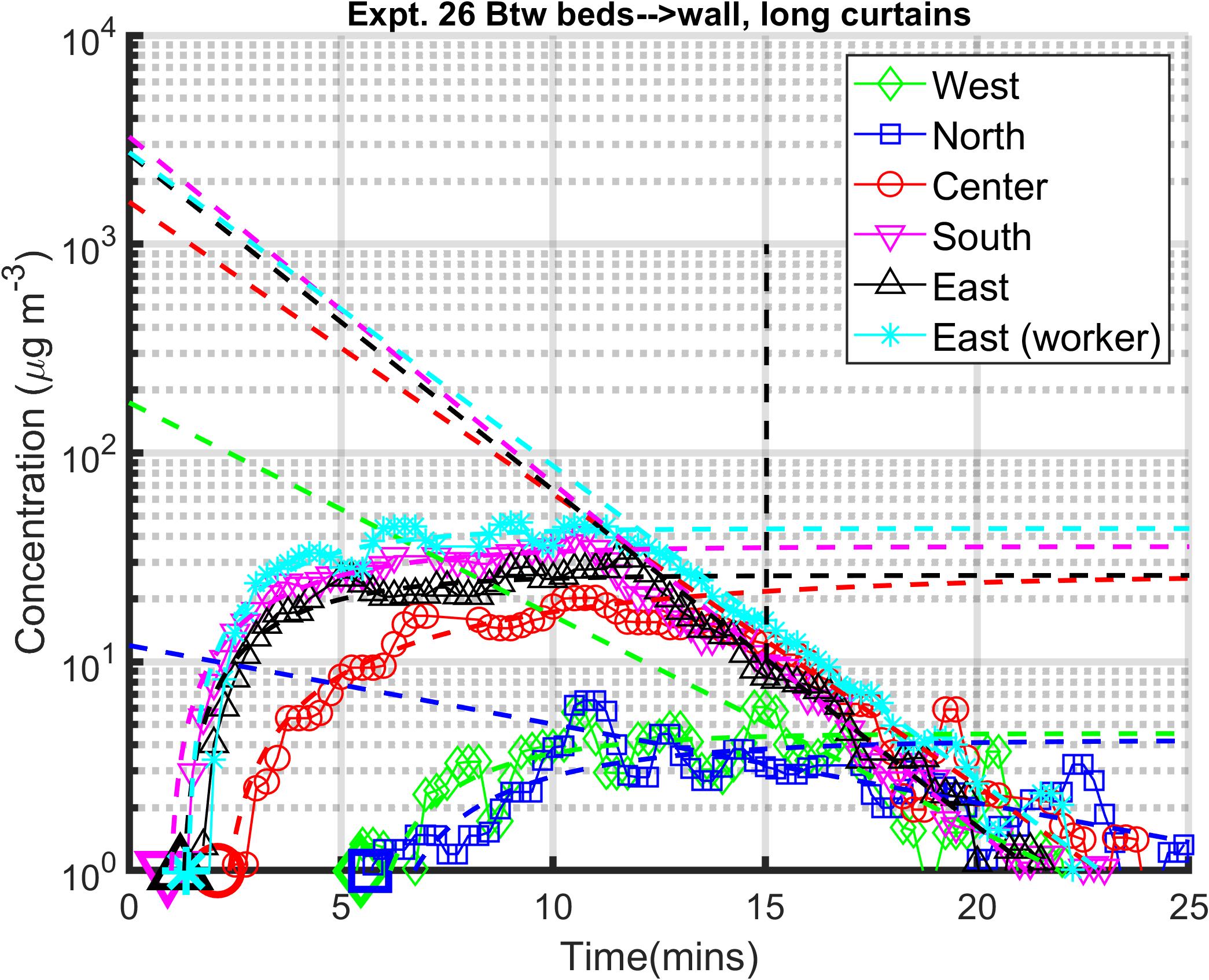}
\caption{Experiment 26. Day 3}
\label{Expt26}
\end{figure}

\begin{figure}[h!]
\centering
\includegraphics[width=10cm]{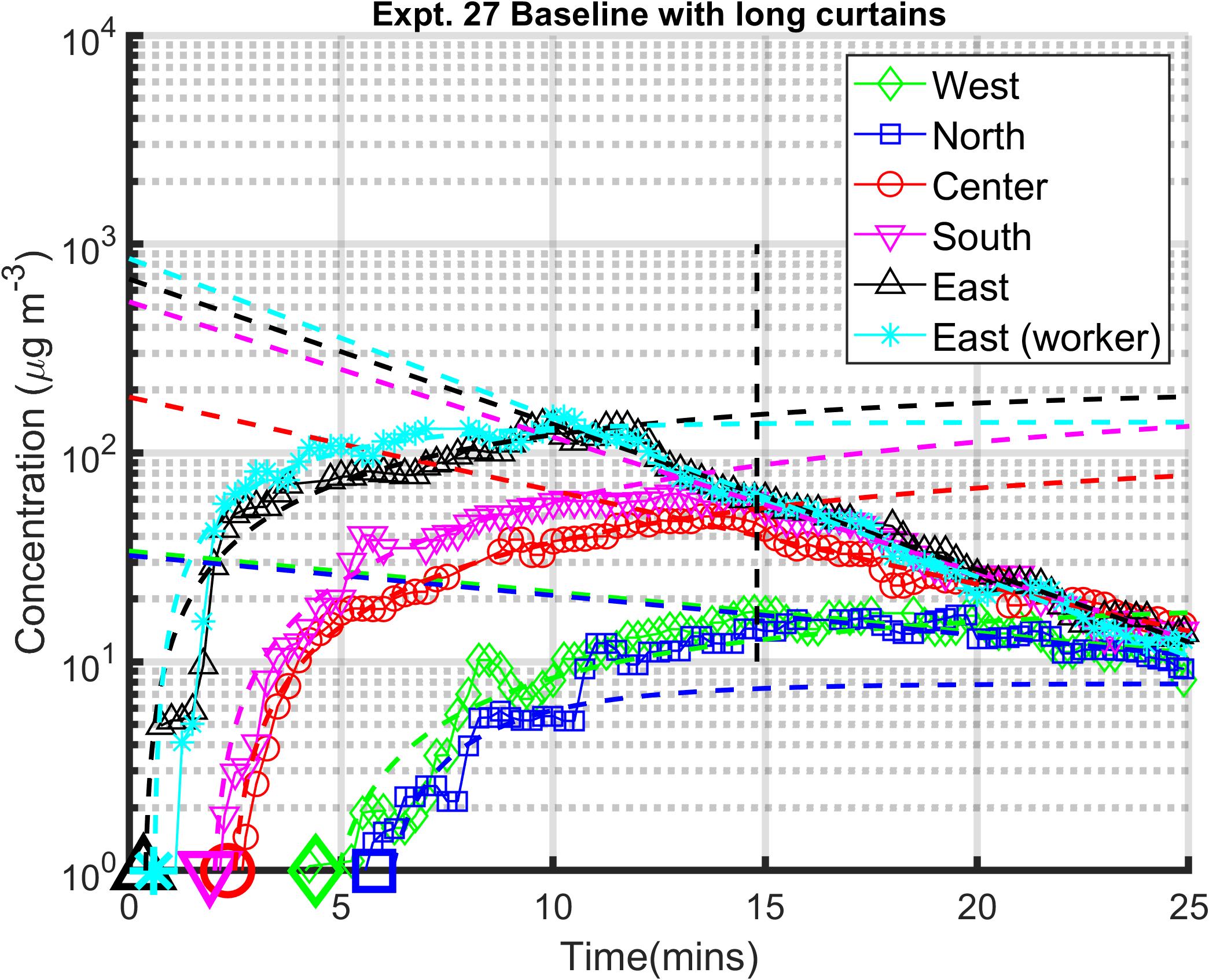}
\caption{Experiment 27 Day3 }
\label{Expt27}
\end{figure}

\begin{figure}[h!]
\centering
\includegraphics[width=10cm]{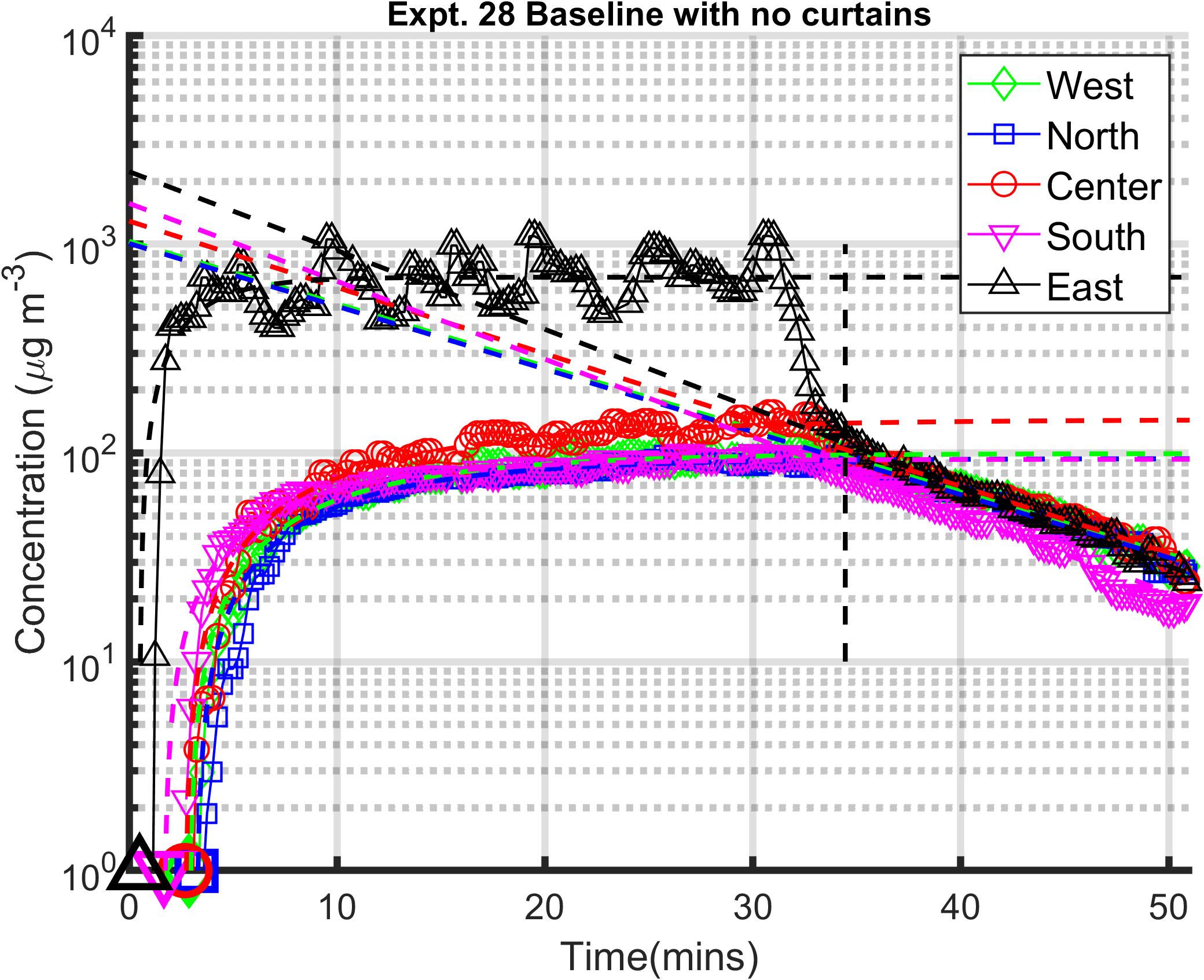}
\caption{Experiment 28. Day 4  Hereafter, experiments use doubled particle generation duration.}
\label{Expt28}
\end{figure}

\begin{figure}[h!]
\centering
\includegraphics[width=10cm]{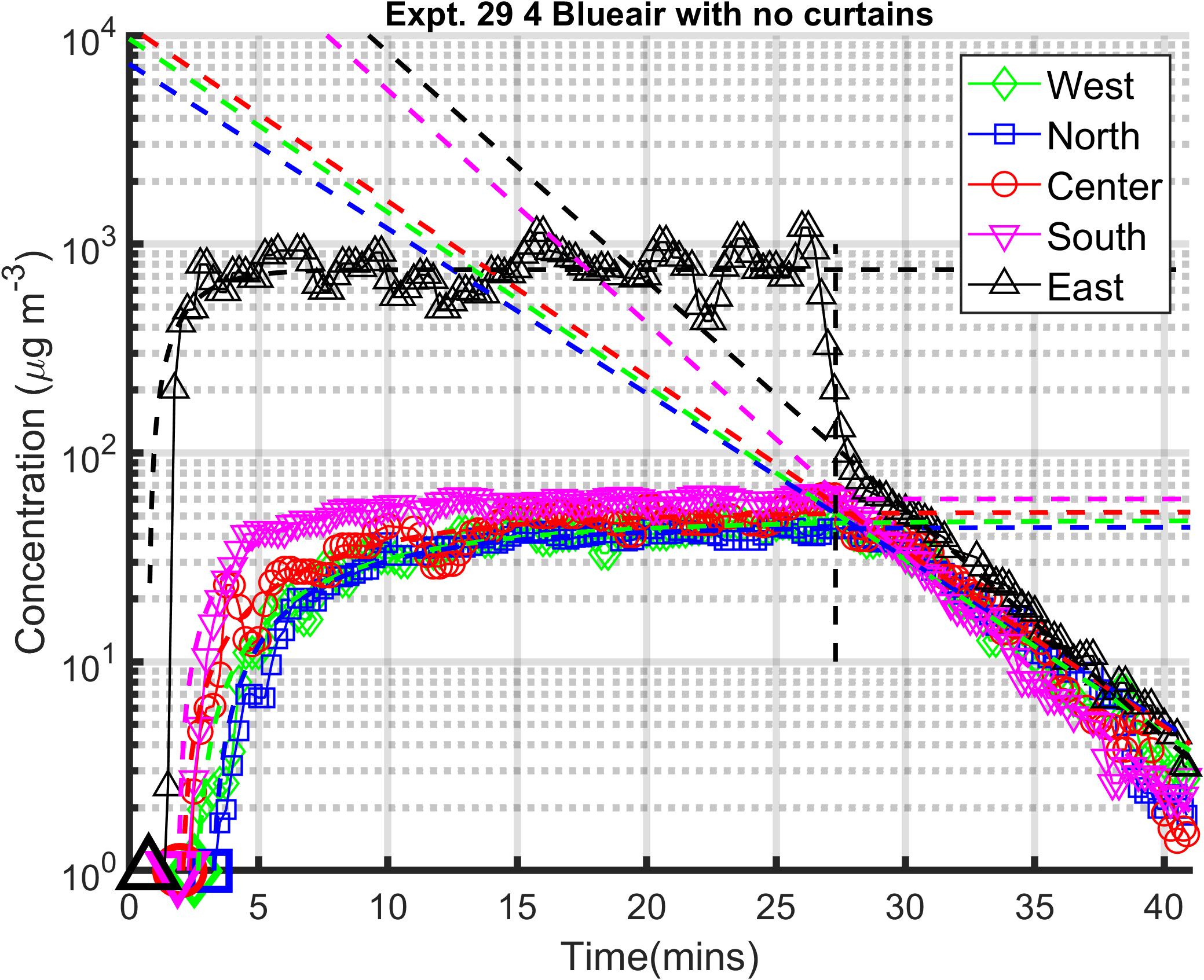}
\caption{Experiment 29. Day 4 Blue Air 411 purifiers in corner, as usual for this configuration.}
\label{Expt29}
\end{figure}

\begin{figure}[h!]
\centering
\includegraphics[width=10cm]{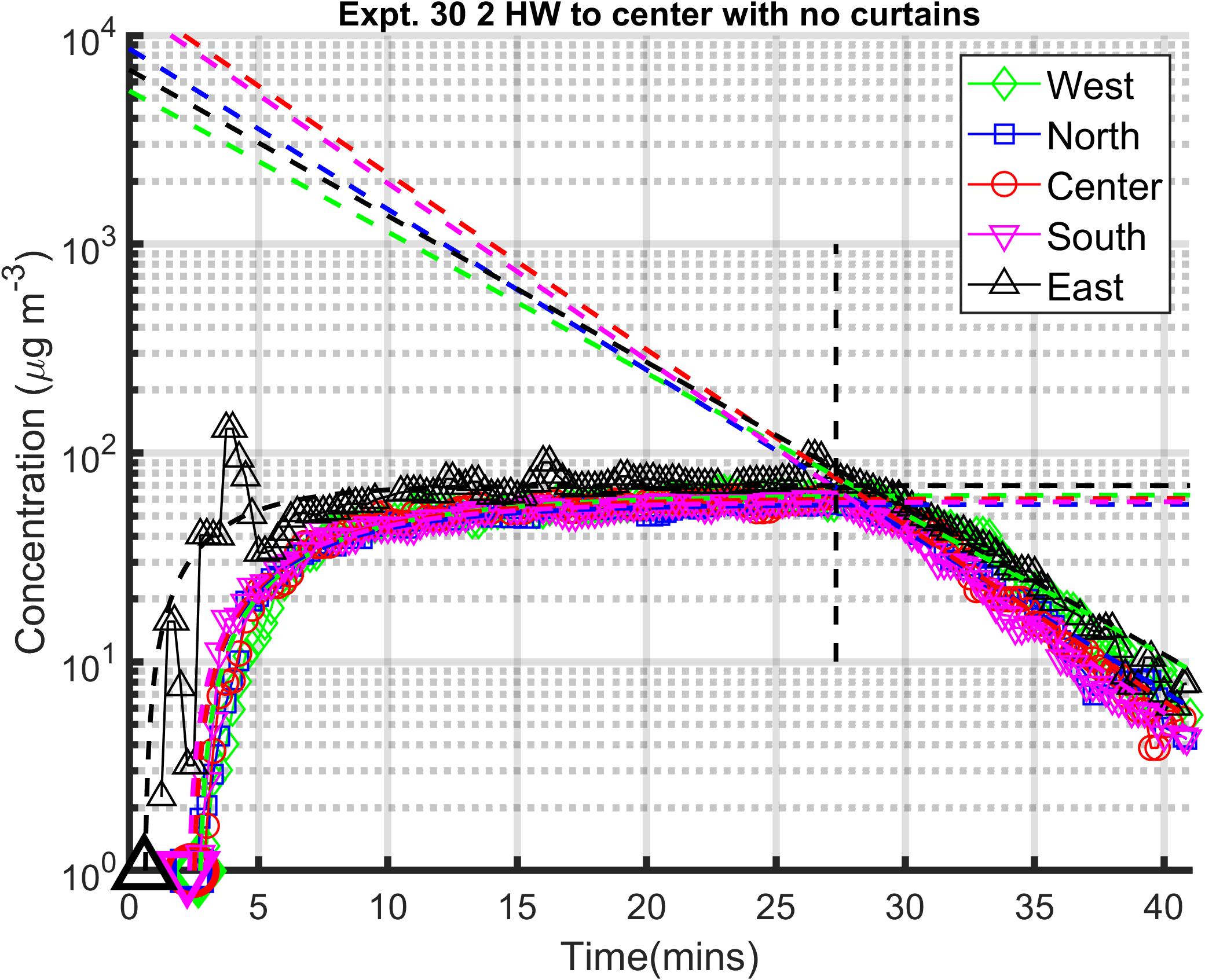}
\caption{Experiment 30. Day 4 HW=Honeywell tower purifier, as for other between-bed tests.}
\label{Expt30}
\end{figure}

\begin{figure}[h!]
\centering
\includegraphics[width=10cm]{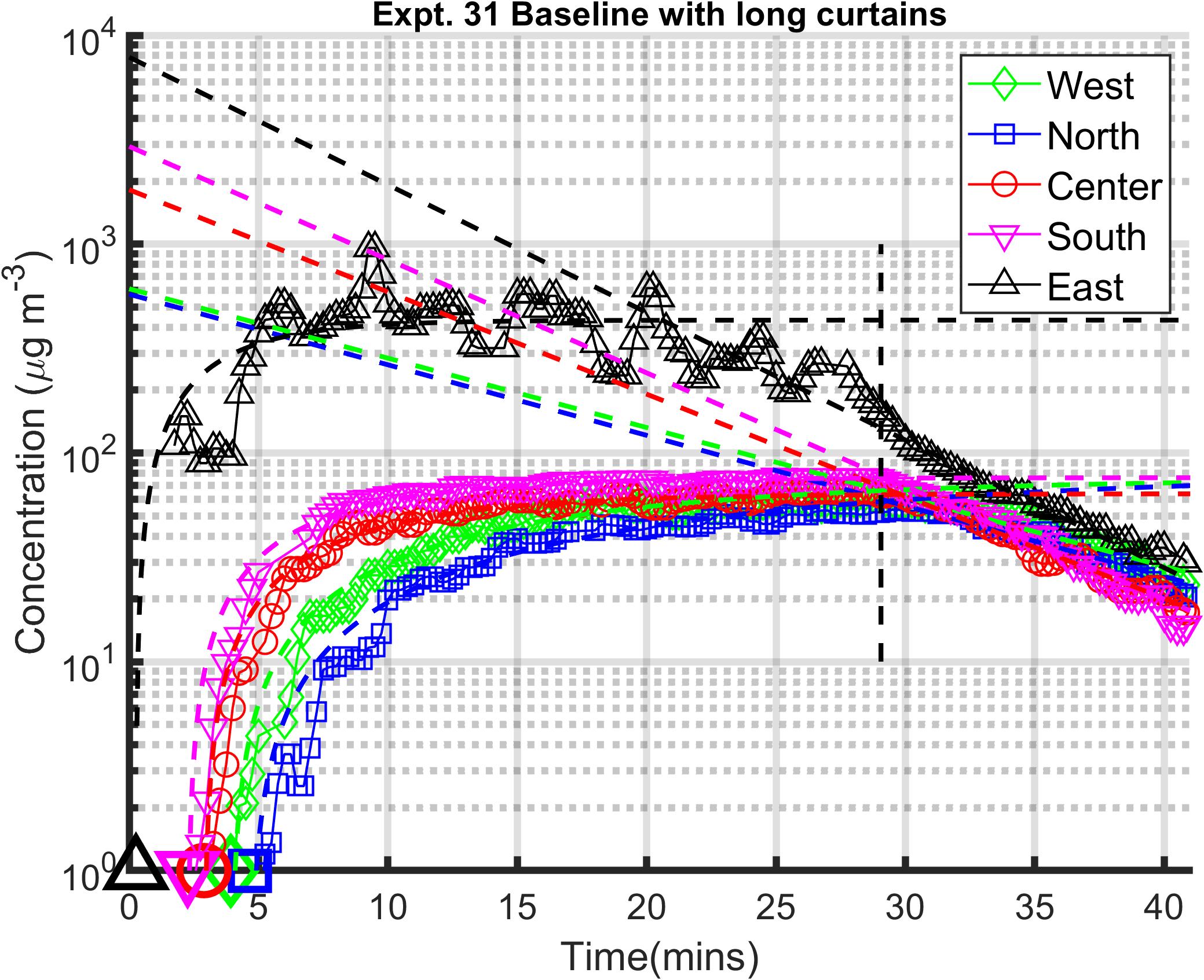}
\caption{Experiment 31. Day 4}
\label{Expt31}
\end{figure}

\begin{figure}[h!]
\centering
\includegraphics[width=10cm]{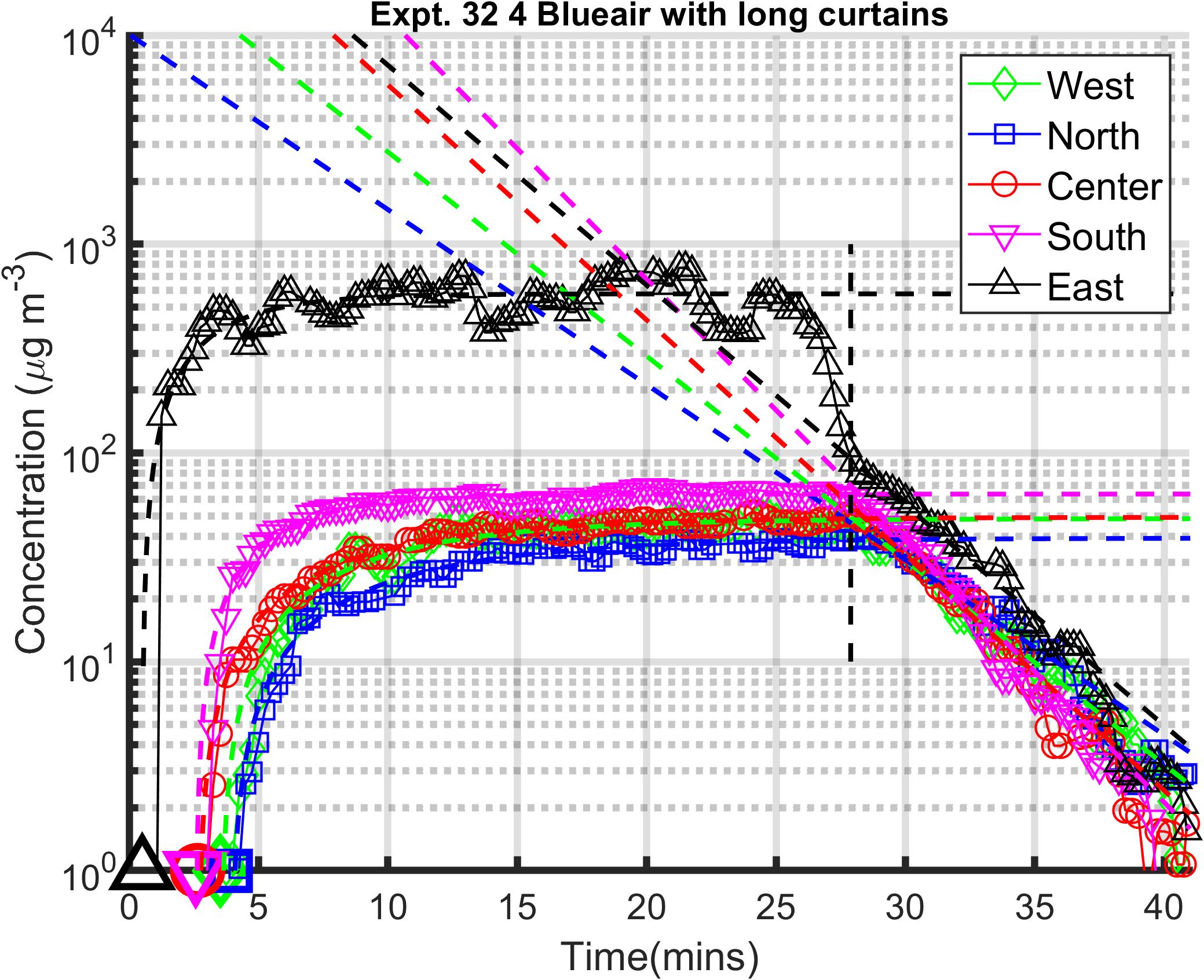}
\caption{Experiment 32. Day 4}
\label{Expt32}
\end{figure}

\begin{figure}[h!]
\centering
\includegraphics[width=10cm]{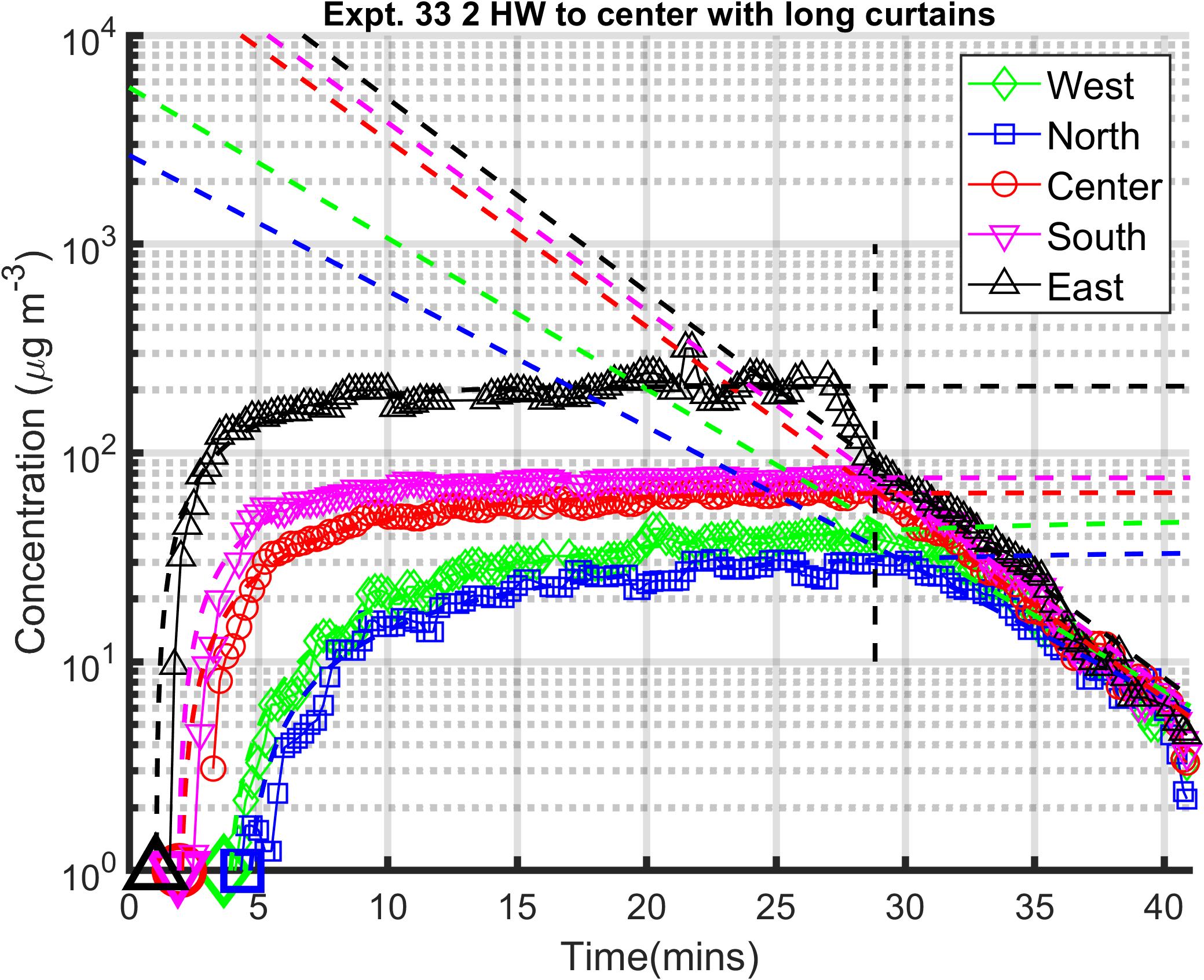}
\caption{Experiment 33. Day 4}
\label{Expt33}
\end{figure}

\begin{figure}[h!]
\centering
\includegraphics[width=10cm]{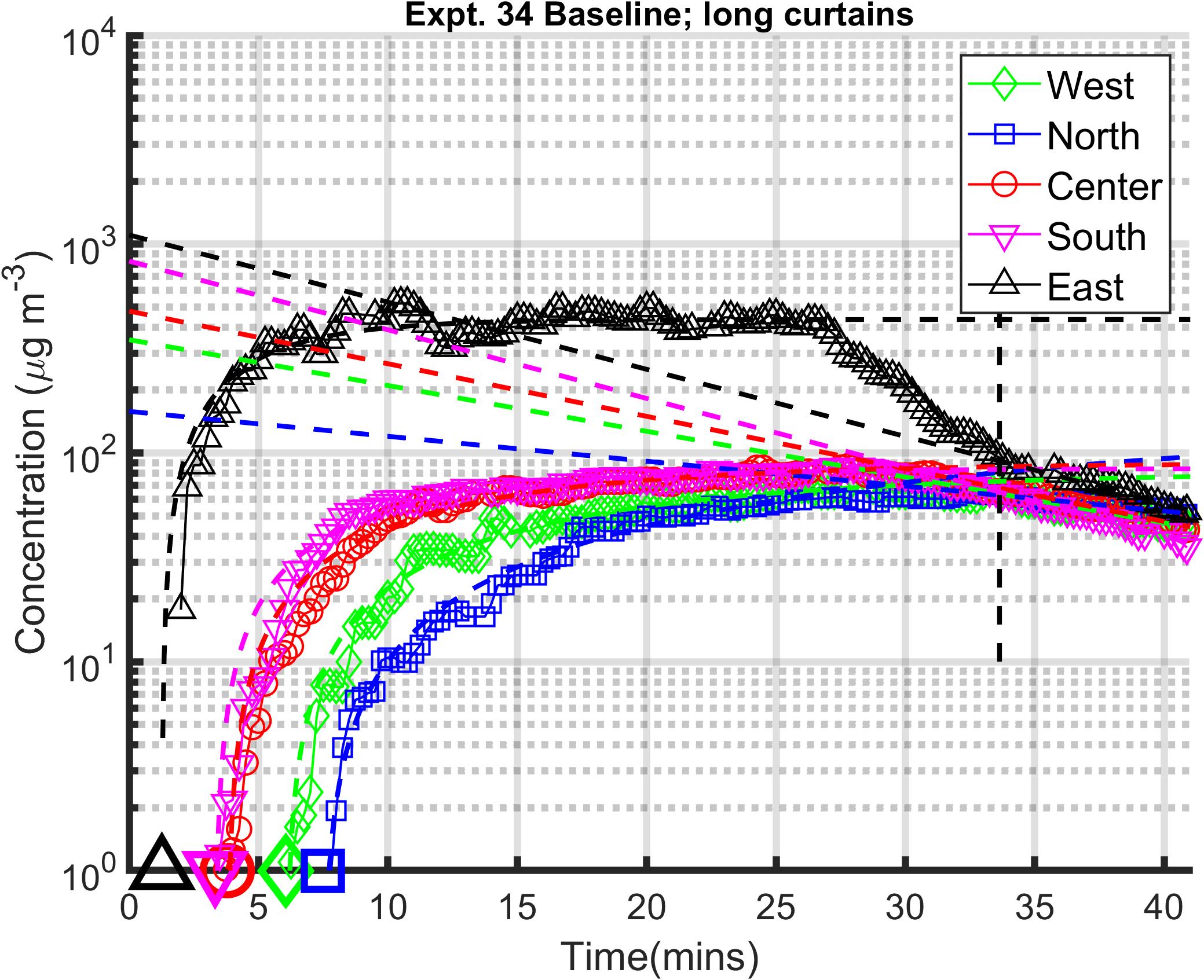}
\caption{Experiment 34. Day 5}
\label{Expt34}
\end{figure}

\begin{figure}[h!]
\centering
\includegraphics[width=10cm]{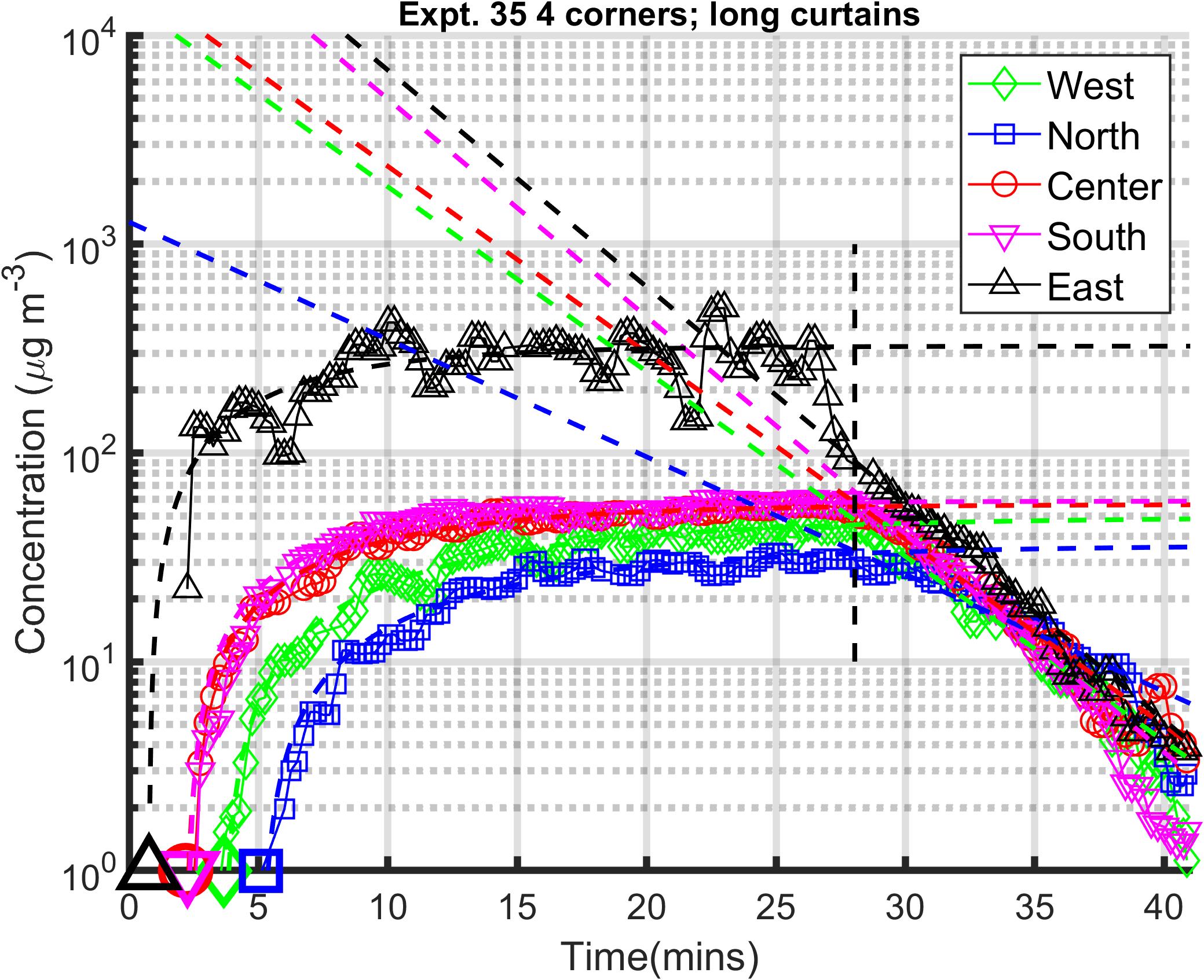}
\caption{Experiment 35. Day 5 }
\label{Expt35}
\end{figure}

\begin{figure}[h!]
\centering
\includegraphics[width=10cm]{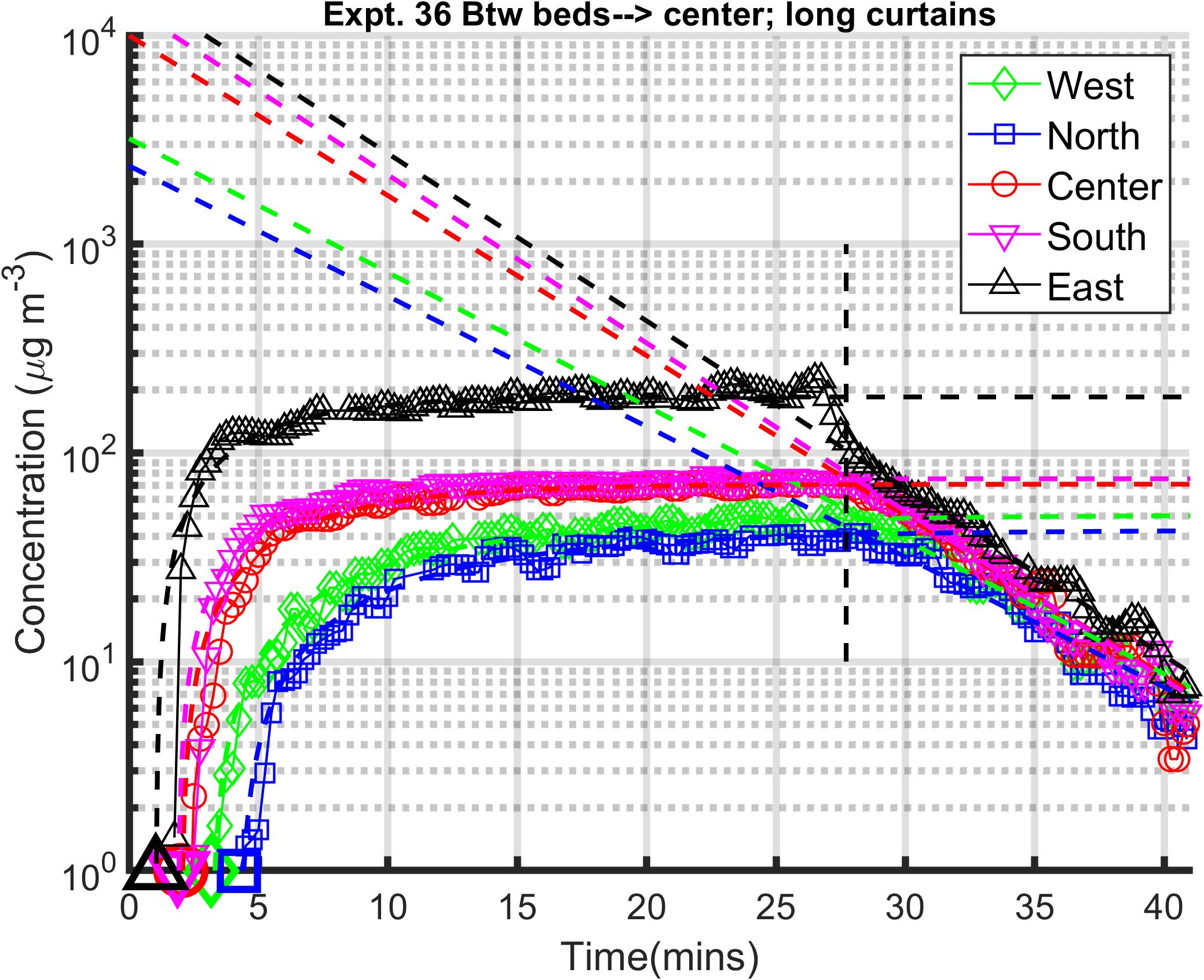}
\caption{Experiment 36. Day 5}
\label{Expt36}
\end{figure}

\begin{figure}[h!]
\centering
\includegraphics[width=10cm]{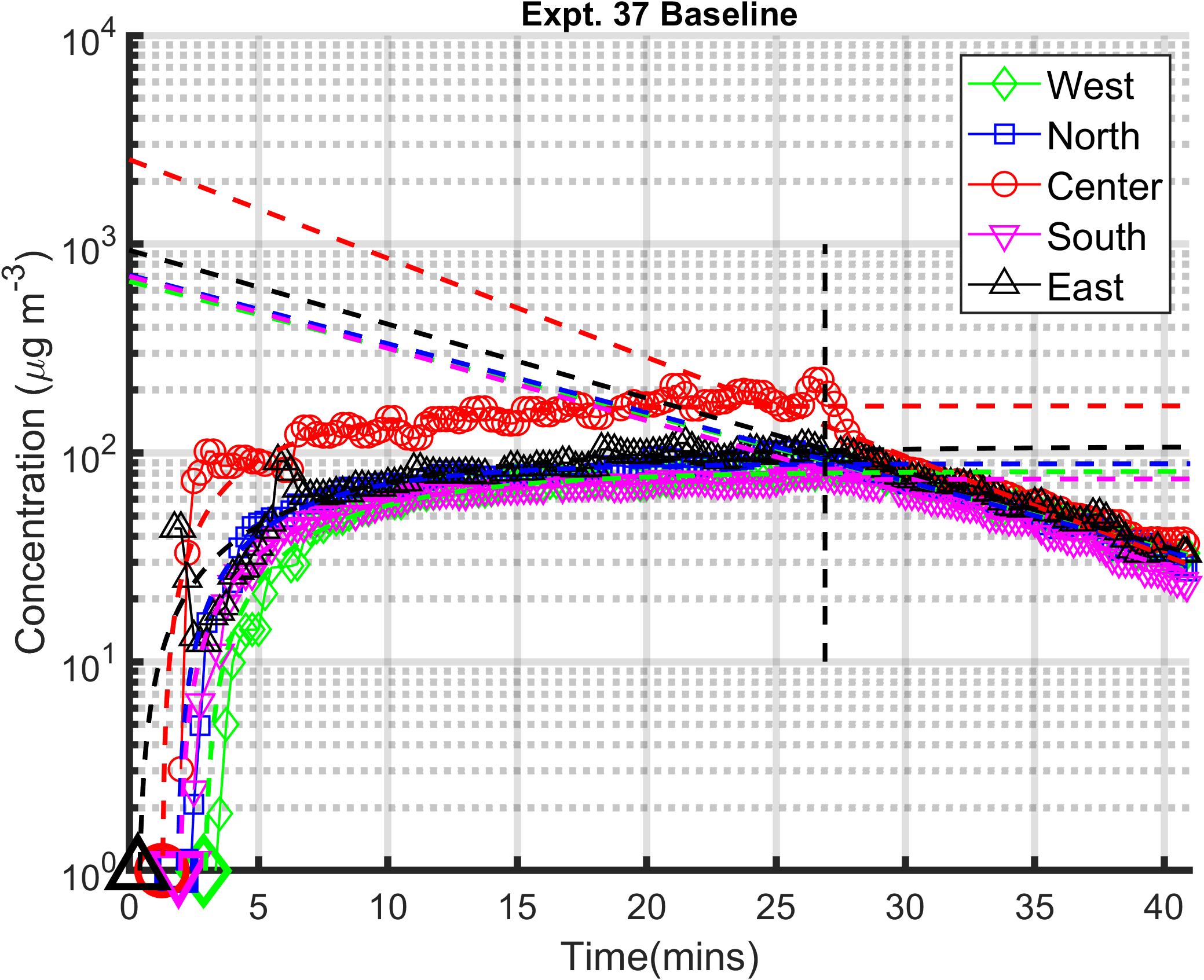}
\caption{Experiment 37. Day 5}
\label{Expt37}
\end{figure}

\begin{figure}[h!]
\centering
\includegraphics[width=10cm]{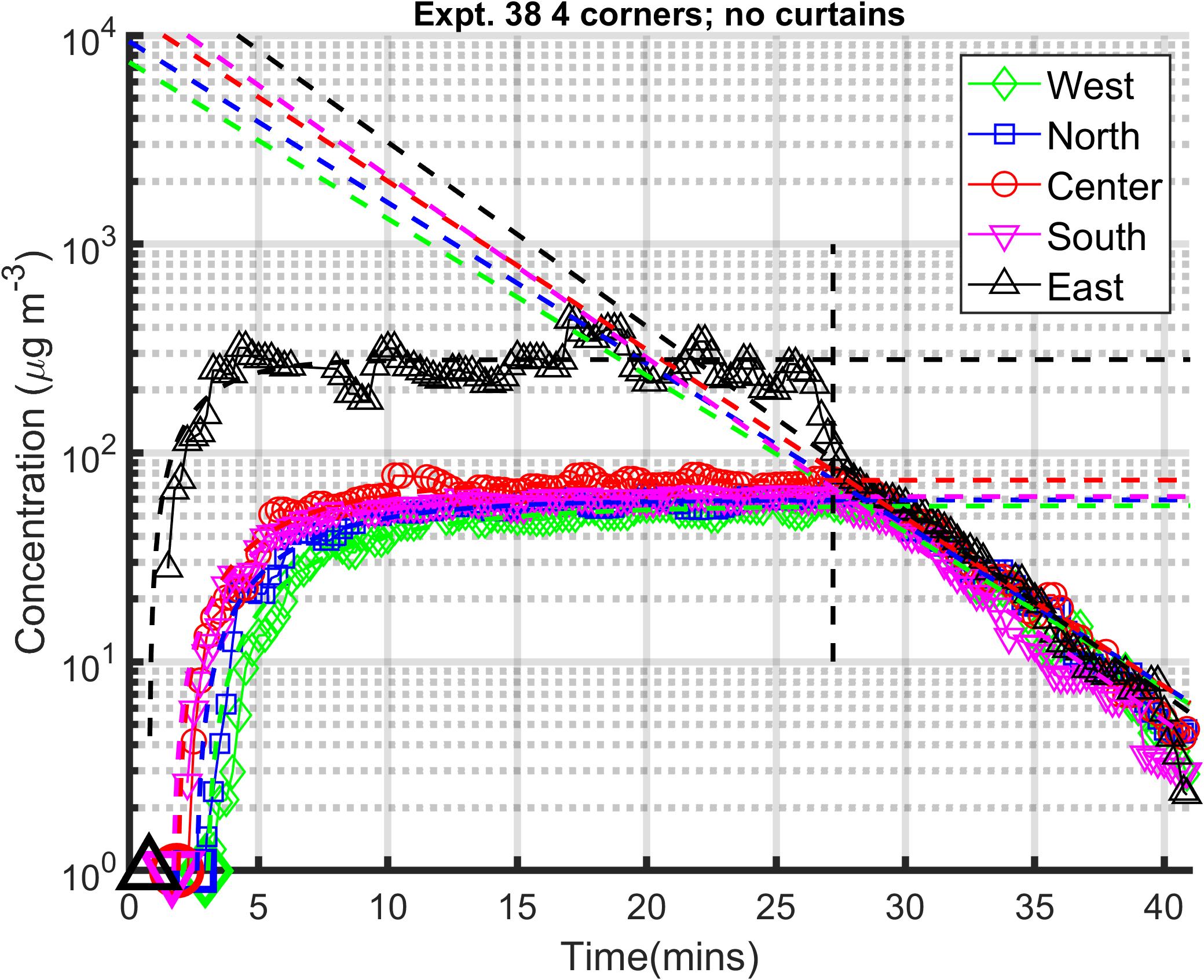}
\caption{Experiment 38. Day 5}
\label{Expt38}
\end{figure}

\begin{figure}[h!]
\centering
\includegraphics[width=10cm]{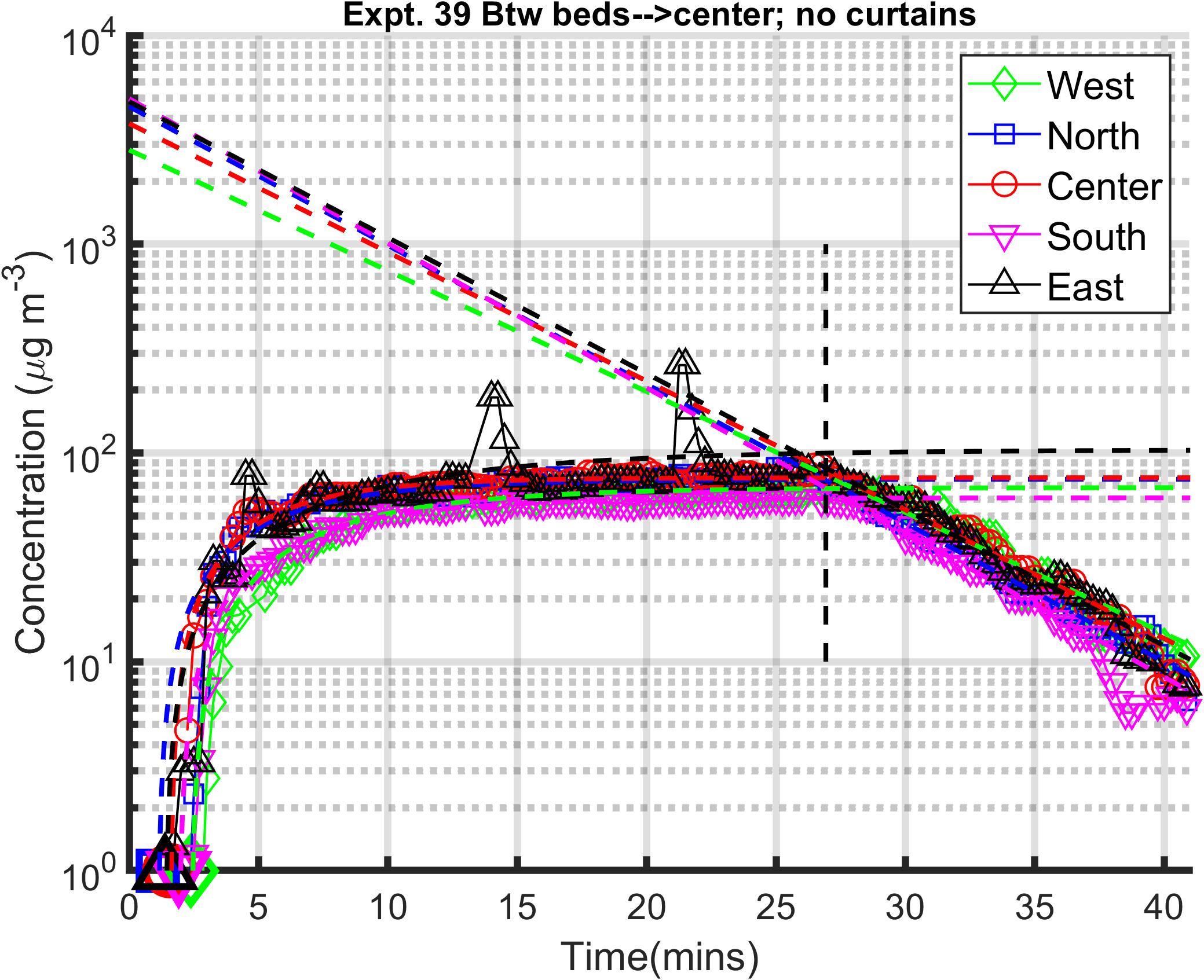}
\caption{Experiment 39. Day 5}
\label{Expt39}
\end{figure}

\begin{figure}[h!]
\centering
\includegraphics[width=10cm]{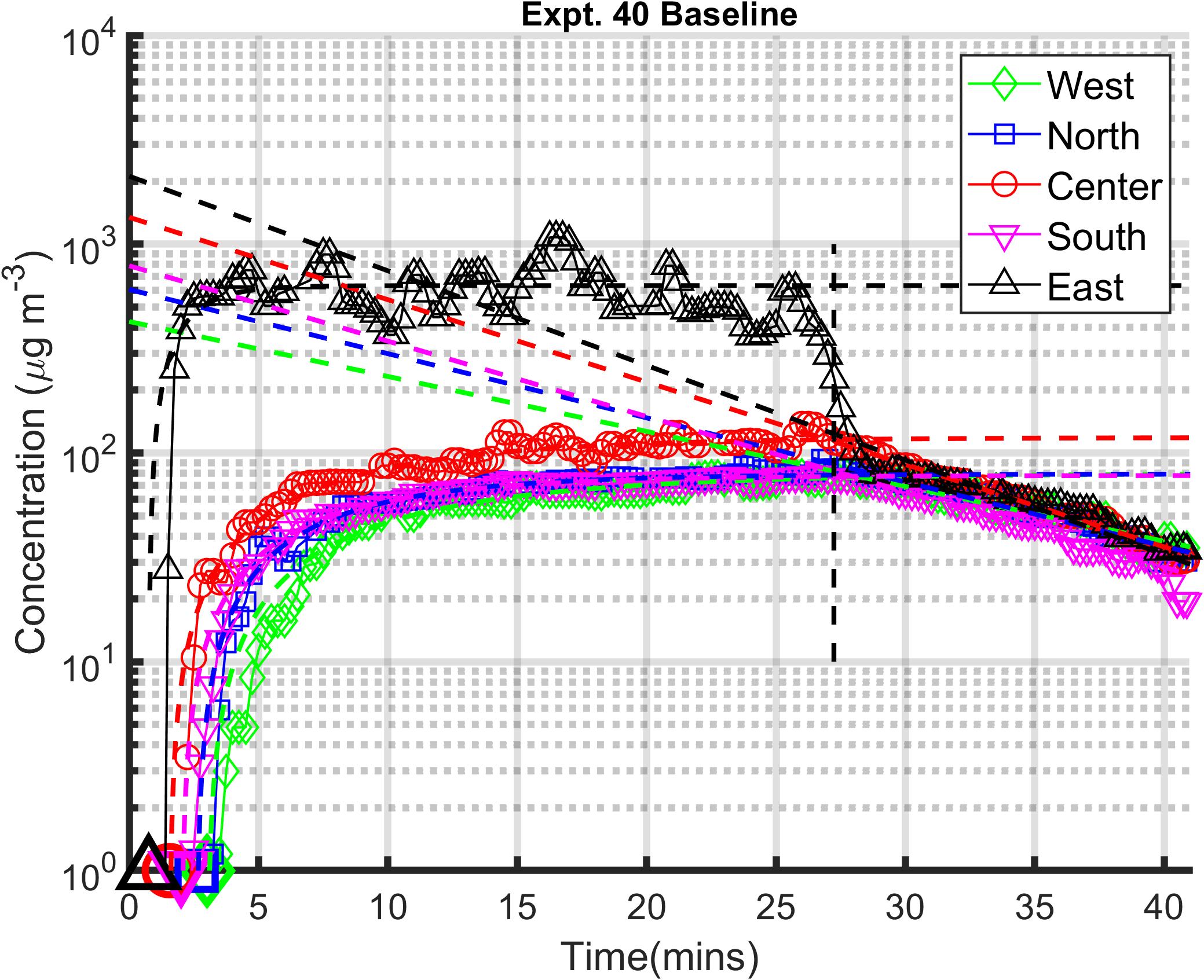}
\caption{Experiment 40. Day 5}
\label{Expt40}
\end{figure}

\begin{figure}[h!]
\centering
\includegraphics[width=10cm]{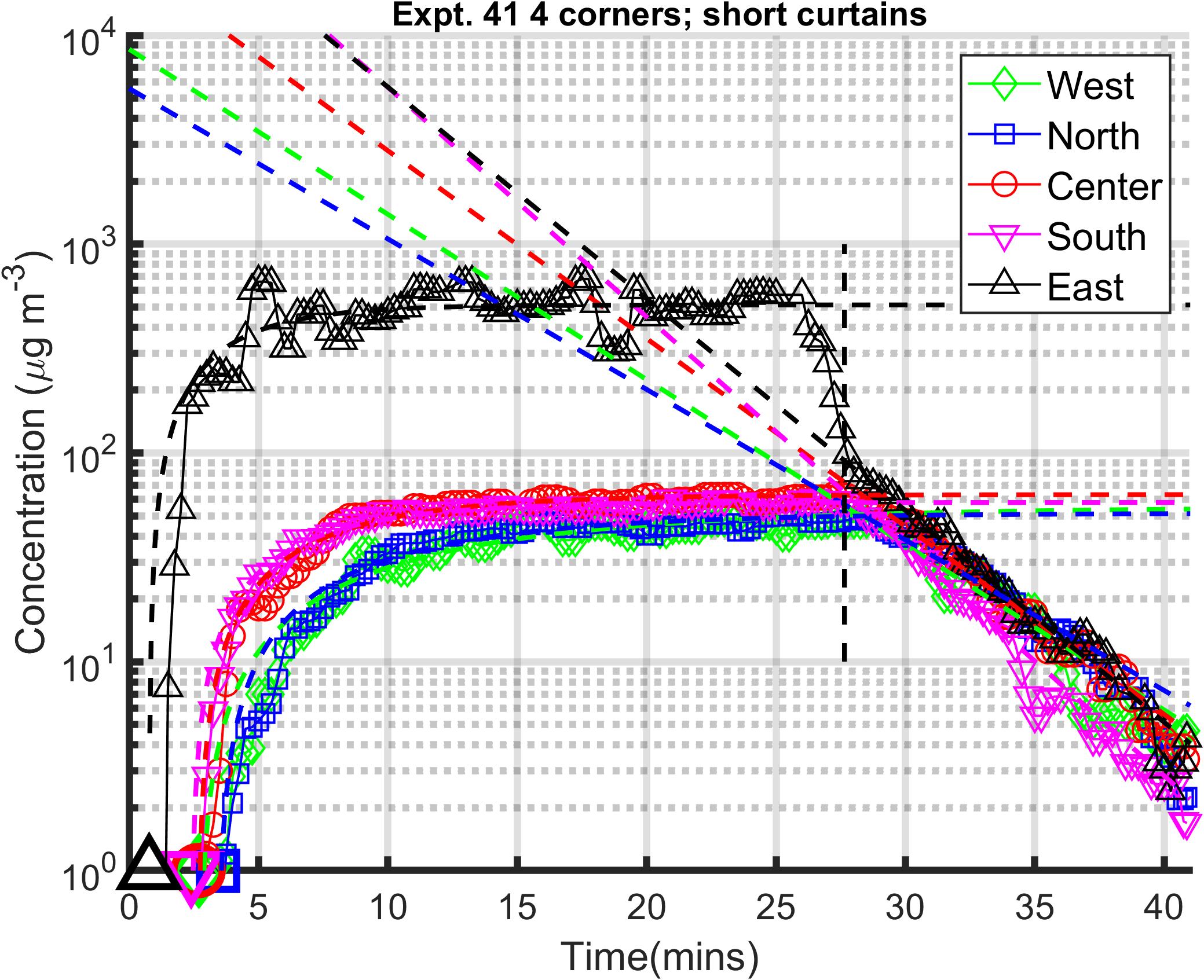}
\caption{Experiment 41. Day 5}
\label{Expt41}
\end{figure}


\end{document}